\documentclass[fleqn,10pt]{wlscirep}
\usepackage{graphicx}
\usepackage{epsfig}
\usepackage{url}
\usepackage{hyperref}
\usepackage{caption}
\usepackage[ruled, vlined, linesnumbered]{algorithm2e}
\usepackage{xspace}
\usepackage{multirow} 
\usepackage{hyperref}

\usepackage[utf8]{inputenc}
\usepackage[T1]{fontenc}

\hypersetup{
    bookmarks=true,         % show bookmarks bar?
    unicode=false,          % non-Latin characters in Acrobats bookmarks
    pdftoolbar=true,        % show Acrobats toolbar?
    pdfmenubar=true,        % show Acrobats menu?
    pdffitwindow=false,     % window fit to page when opened
    pdfstartview={FitH},    % fits the width of the page to the window
    pdftitle={My title},    % title
    pdfauthor={Author},     % author
    pdfsubject={Subject},   % subject of the document
    pdfcreator={Creator},   % creator of the document
    pdfproducer={Producer}, % producer of the document
    pdfnewwindow=true,      % links in new window
    colorlinks=true,       % false: boxed links; true: colored links
    linkcolor=black,          % color of internal links
    citecolor=black,        % color of links to bibliography
    filecolor=black,      % color of file links
    urlcolor=black           % color of external links
}

\newtheorem{mydefinition}{Definition}

\newcommand{\core}{C_{k, \Delta}}

\newcommand{\spancore}{span-core\xspace}

\newcommand{\tdeg}{\mbox{\ensuremath{d}}}

\newcommand{\squishlist}{
 \begin{list}{$\bullet$}
  {  \setlength{\itemsep}{0pt}
     \setlength{\parsep}{3pt}
     \setlength{\topsep}{3pt}
     \setlength{\partopsep}{0pt}
     \setlength{\leftmargin}{2em}
     \setlength{\labelwidth}{1.5em}
     \setlength{\labelsep}{0.5em}
} }
\newcommand{\squishlisttight}{
 \begin{list}{$\bullet$}
  { \setlength{\itemsep}{0pt}
    \setlength{\parsep}{0pt}
    \setlength{\topsep}{0pt}
    \setlength{\partopsep}{0pt}
    \setlength{\leftmargin}{2em}
    \setlength{\labelwidth}{1.5em}
    \setlength{\labelsep}{0.5em}
} }

\newcommand{\squishdesc}{
 \begin{list}{}
  {  \setlength{\itemsep}{0pt}
     \setlength{\parsep}{3pt}
     \setlength{\topsep}{3pt}
     \setlength{\partopsep}{0pt}
     \setlength{\leftmargin}{1em}
     \setlength{\labelwidth}{1.5em}
     \setlength{\labelsep}{0.5em}
} }

\newcommand{\squishend}{
  \end{list}
}

\usepackage{xcolor}

\newcommand{\add}[1]{{{#1}}}

\title{Relevance of temporal cores for epidemic spread in temporal networks}

\author[1]{Martino Ciaperoni}
\author[2]{Edoardo Galimberti}
\author[2]{Francesco Bonchi}
\author[2,3]{Ciro Cattuto}
\author[4]{Francesco Gullo}
\author[5,6,2,*]{Alain Barrat}

\affil[1]{Aalto University, Finland}
\affil[2]{ISI Foundation, 10126 Torino, Italy}
\affil[3]{University of Turin, Turin, Italy}
\affil[4]{UniCredit, R\&D Dept., Italy}
\affil[5]{Aix Marseille Univ, Universit\'e de Toulon, CNRS, CPT, Turing Center for Living Systems, 13288 Marseille, France}
\affil[6]{Tokyo Tech World Research Hub Initiative (WRHI), Institute of Innovative Research, Tokyo Institute of Technology, Japan.}
\affil[*]{alain.barrat@cpt.univ-mrs.fr}

\begin{abstract}  % 200 words max
Temporal networks are widely used to represent a vast diversity of systems, including in particular
social interactions, and the spreading processes unfolding on top of them.
The identification of structures playing important roles in such processes remains largely 
an open question, despite recent progresses in the case of static networks. Here, we consider as candidate structures
the recently introduced concept of span-cores: the span-cores decompose a temporal network into subgraphs of controlled duration and increasing connectivity, generalizing the core-decomposition of static graphs. 
\add{To assess the relevance of such structures,}
we explore the effectiveness of strategies aimed either at containing or maximizing the impact of
a spread, based respectively on removing span-cores of high cohesiveness or duration 
to decrease the epidemic risk, or on seeding the process from such structures.
The effectiveness of such strategies is assessed in a variety of empirical data sets and
\add{compared to} baselines that use only static information on the centrality of nodes
and static concepts of coreness, \add{as well as to a 
baseline based on a temporal centrality measure}. 
Our results show that 
the most stable and cohesive temporal cores \add{play indeed an important role in}
epidemic processes on
temporal networks, and that their nodes are likely to represent influential spreaders.

\end{abstract}
\begin{document}

\flushbottom
\maketitle
\thispagestyle{empty}

\section*{Introduction}

A large variety of systems find a convenient representation as networks of interactions between components.  
Network representations have indeed proved to be useful to understand the structure and dynamics of systems
as diverse as transportation infrastructures or social networks, 
as well as to describe processes occurring on top of them, such as information diffusion, epidemic spread, synchronization, etc \cite{Barrat:2008}. 
A large body of work aims in particular at understanding how the network's features impact the outcome
of processes taking place on top of them, with the ambition of devising control and prediction capabilities. 
For instance, several methods have been put forward to identify
the nodes of a network that play a more important role in a spreading process, either
because they are "influencers" able to widely spread an information, or because they are
"sentinels" with a high probability to be reached by a disease in its early stages and thus 
can give an early warning, or because their immunization is likely to hinder the spread
\cite{Pastor-Satorras:2002,Cohen:2003,Kitsak:2010,Pastor-Satorras:2015,Herrera:2016,Teng:2016,Radicchi:2017,Yuan:2017,Holme:2017,Erkol:2019}.

Such a task becomes even more complex when dealing with temporal networks, in which edges between nodes can appear and disappear
on different time scales. The recent availability of time-resolved data sets of interactions 
has pushed network science beyond the static graph representation and has led to the development of the field of temporal
networks \cite{Holme:2012,Holme:2015}. The temporal dimension can yield non-trivial temporal features such as broad
distributions of interaction times and of inter-event times ("burstiness"), heterogeneous activity distributions, causality constraints, 
and overall a broader diversity 
of activity/connectivity patterns than in static networks \cite{Holme:2012,Holme:2015,Gauvin:2013,Gauvin:2014}. 
This has raised new questions on designing surveillance and control strategies for epidemic processes in temporal networks
\cite{Bajardi:2012,Lee:2012,Starnini:2013,Masuda:2013,Liu:2014,Valdano:2015a,Gemmetto:2014,Gauvin:2015}. 
\add{Identifying single nodes for the design of targeted interventions  (immunization, isolation) can however be challenging in practice: 
indeed, centrality measures can fluctuate strongly over time and depend on details of the system that change in different periods; they might
moreover be difficult to relate to interpretable features. Data with limited resolution can also make it difficult to correctly identify 
the most central nodes \cite{Genois:2018}. 
In particular, several works \cite{Lee:2012,Starnini:2013,Valdano:2015a} have addressed the issue of designing causal interventions, in which 
node properties are measured until a certain time and  determine the nodes to immunize to contain a future spread.
In practice, the impact of such strategies depends strongly on the data set under study and is limited by temporal fluctuations in the centralities
of nodes. Moreover,}
action at the level of individuals is subjected to many informational and operational constraints.

Therefore, strategies \add{based on} intermediate scales have been advocated, targeting structures rather than single nodes,
\add{with the goal of identifying patterns that can be acted upon in practice.
The idea in this case is to first identify potential structures of interest in (static or temporal) networks,
and then verify that they play an important role in spreading processes: this is typically done by thought experiments in which one compares
the outcome of a spreading process on the original network with the outcome if the structure of interest is altered or removed. Alternatively, one can seed
the process from the structure and investigate the spreading power of nodes of the structure. A second important step is to relate the structures
to interpretable features of the system of interest, and the third step is to use this knowledge to propose actionable strategies.}
Examples of such structures include groups or communities in the population, in which
individuals have more contacts with each other than with the rest of the population, 
 \add{leading to the proposal of reactive measures of class closure}  \cite{Gemmetto:2014,Litvinova:2019}, or
sets of links with correlated activity patterns \cite{Gauvin:2015}, \add{which, in a school environment, can be related to the breaks between lectures
and thus help design concrete interventions. We note here that the investigation of structures does not necessarily correspond to causal interventions
in which only past knowledge is used. Rather, one wants to identify a certain type of structures that play an important role and could be interpreted
in practical settings to be acted upon}.

Here, we contribute to \add{the first, most theoretical step of} 
this line of research by proposing a novel type of candidate structures for \add{idealized} targeted interventions: the span-cores of a temporal network.
The span-cores are structures that we recently introduced \cite{Galimberti:2018} 
to decompose a temporal network into hierarchies of subgraphs of controlled duration and increasing connectivity, generalizing 
the core-decomposition of static graphs. \add{They could thus in temporal contact networks be interpreted as long-lasting groups of interacting
individuals.} We recall that the core-decomposition of a static 
network yields a hierarchy of subgraphs that are increasingly densely connected: a 
core with coreness $k$ is defined as the
maximal subgraph such that all nodes within it have degree (number of neighbors in the core) at least $k$
\cite{Batagelj:2011}. The nodes belonging to the
more central cores have been shown to play an important role in \add{standard theoretical models of} epidemic spreading processes
on static networks~\cite{Kitsak:2010,Castellano:2012}.

In temporal networks, the span-cores are defined \cite{Galimberti:2018} as temporal subgraphs
as follows: a span-core ${\cal C}$ of order $k$ is defined 
on an interval $\Delta$ of contiguous timestamps, such that all nodes in ${\cal C}$
have at all timestamps of that interval at least $k$ stable neighbours in ${\cal C}$
(i.e., the links to these nodes 
are present during all timestamps of the interval).
Each span-core is thus characterized by two 
quantities: its order and its duration (the 
length of the interval on which it is defined).
Moreover, a span-core of order $k$ on a temporal interval $[t_1,t_2]$
is said to be maximal if there exist no other span-cores with order at least $k$
and broader temporal interval ($[t'_1,t'_2]$ with $t'_1 \le t_1$ and $t'_2 \ge t_2$). 
In reference \cite{Galimberti:2018}, we have introduced the definition of span-cores and maximal span-cores, devised efficient algorithms to extract them from data, and performed a preliminary analysis of their properties in several data sets.

Maximal span-cores are thus well-connected 
stable groups of nodes, and can be considered as natural candidates when looking for
structures having an important role in spreading processes on temporal networks.
In the present work, we therefore study the role of the maximal span-cores of temporal networks in spreading processes
occurring on these networks. To this aim, \add{we perform two types of thought experiments.} 
We first examine the impact of the removal of maximal 
span-cores with large order and/or duration on spreading processes in a variety of empirical 
temporal networks. We quantify this impact by the resulting decrease in epidemic risk
and compare the removal of span-cores  to random baselines and to 
\add{the isolation of nodes based on properties in the static aggregated network known to be relevant to 
spreading processes~\cite{Kitsak:2010,Castellano:2012}.
For completeness, we also consider a strategy based
on the computation of a centrality measure taking into account the whole temporality of the data, 
namely the temporal PageRank, which assigns a centrality measure to each node at each time~\cite{Rozenshtein:2016}}. We show that the removal of links in the most connected
and stable temporal cores has a particularly strong impact on the epidemic risk. We then
investigate the spreading influence of nodes belonging to maximal span-cores, defined as the final size of a spreading process
originating in those nodes. Our results show that processes seeded in cohesive 
span-cores yield in general large epidemic sizes, 
\add{even compared with several baseline seeding strategies based on properties known to be relevant such as static 
centralities\footnote{\add{We note here that we do not attempt to classify strategies based on all potential 
centrality measures nor to claim that span-cores yield necessarily the best strategy: first, one can expect the classification to be highly 
dependent on specific data set properties; second, centrality measures identify nodes rather than structures. For our goal here, it is thus
enough to check whether strategies based on span-cores yield reliably large effects.}}.} 
All these results confirm the important role of the span-cores structures in diffusion processes on temporal networks.
\add{We discuss some limitations and perspectives in the Discussion section, in particular concerning the 
need to know the temporal network in advance to find the span-cores, showing that one has to be able to interpret the span-core structures in order
to translate this theoretical knowledge into actionable strategies.}

\section*{Results}

\subsection*{Procedure}

Before presenting our results, we describe here our general methodology, and we refer to the Methods section for more details on each step of our procedure.
We consider a series of publicly available data sets describing temporal networks of particular relevance for epidemic spreading scenarios, 
namely high-resolution social network data on face-to-face interactions in a variety of settings, collected independently by two different collaborations
\cite{Sociopatterns,Toth:2015}: schools, conferences, offices, hospital (see Methods for details). This allows us to study data with a broad variety of 
structural and temporal patterns. The data sets we consider are all represented as temporal networks on discrete timestamps: nodes represent individuals,
and a temporal edge $(i,j,t)$ between two nodes $i$ and
$j$ in time window $t$ corresponds to the fact that these two nodes have been in contact during that time window.

We consider standard \add{schematic models of} epidemic spreading processes occurring on top of these temporal networks, namely the
Susceptible-Infected-Susceptible (SIS) and Susceptible-Infected-Recovered (SIR) models.
In the SIS case, nodes can only be in  
either of two states: susceptible (S) or infected (I). A susceptible node in contact with an infected one becomes infected with a fixed probability 
$\beta$ per unit time. Infected nodes recover with a constant rate $\nu$ and become susceptible again. The system can thus either reach a steady state in which there
is constantly a finite fraction of the nodes in the I state, or the epidemic can die out if all nodes recover. 
%The interplay between the parameters $\beta$ and $\mu$ and the temporal network dynamics determines 
In the SIR model, nodes that recover enter the 
Recovered (R) class and can no longer take part in the epidemic process. Therefore, no steady state can be reached: the epidemic process ends when no
more infected nodes are present, and all nodes are then either susceptible (those who have not been reached by the disease) or recovered.

To evaluate the outcome of these processes, we resort to two standard measures of the epidemic risk.
We first take advantage of the theoretical framework developed by \cite{Valdano:2015b,Valdano:2015c} to compute the epidemic 
threshold for both SIS and SIR models
in arbitrary temporal networks
%, based on a Markov chain approach that neglects the temporal correlations between the states of neighboring nodes
(see Methods and \cite{Valdano:2015b,Valdano:2015c} for details). The epidemic threshold represents a crucial way 
of quantifying the epidemic risk in a system,
as it gives the critical value of disease transmissibility above which the simulated pathogen is able to reach a large fraction of the population. 
In the SIR case, we moreover 
perform direct numerical simulations of the process and measure the epidemic size, i.e., the final fraction of recovered nodes, averaged
over $1000$ realizations of the process. 
Note that the process might not have ended at the end of the data set, meaning that there might still be nodes in the I state.
We then replicate the temporal network  until the process has ended. The epidemic size can be interpreted both 
as a further quantification of the epidemic risk and as the spreading power of the seed, i.e., the initial source of the spread.

As mentioned in the introduction, two complementary questions \add{are typically investigated in order to test the role of nodes or structures
in spreading processes}: how to best contain or mitigate a spread, and which seed(s) have the largest
spreading power? Containment and mitigation aim at increasing the epidemic threshold and/or at decreasing the final epidemic size, and various 
strategies can be considered and compared. Here we focus on the role of the maximal span-cores, and thus consider \add{an idealized} 
strategy based on removing
from the temporal network a fraction of these structures. We thus first determine in each data set its 
maximal span-cores
%: We recall that each span-core of a temporal network is characterized by an order $k$ and an interval $\Delta$: it is the maximal set of nodes $C_{k,\Delta}$ such that, for any node $u$ in $C_{k,\Delta}$,  the number of other nodes in $C_{k,\Delta}$ with which $u$ has temporal edges at all timestamps of $\Delta$ is at least $k$ 
(see Methods for the precise definitions).
We then consider an altered version of the temporal network in which a fraction $f$ of the temporal links, taken from the maximal span-cores, 
are removed: for a span-core spanning a temporal interval $[t_1,t_2]$
this amounts to removing the temporal links in which any node of the span-core is involved between $t_1$ and $t_2$ 
(i.e., effectively isolating these nodes during $[t_1,t_2]$). The span-cores on which this strategy is applied are chosen either as the 
ones with the largest order $k$
($k$M maximal span-cores strategy), or the ones with the largest duration $\Delta$ ($\Delta$M maximal span-cores strategy): 
we remove temporal edges starting from the ones of the maximal span-cores ordered by non-increasing
order or duration, until the fraction $f$ is reached. This corresponds to removing all the temporal edges of a certain
number of maximal span-cores, that we denote $n_{msc}$, and potentially some additional temporal edges
taken at random in the next maximal span-core (number $n_{msc}+1$ in the ordered list) to reach exactly $f$. 
For a given data set, the  total number of temporal edges removed is denoted $n_T$ ($n_T = f |E|$ where $E$ is the 
overall set of temporal edges in the data set). Moreover, for a given strategy $s$, we denote by 
$n_t^s$ the number of temporal edges removed at timestep $t$ in this strategy. 
Finally, we compare the epidemic risk in the altered temporal network and in the original one: specifically, we 
measure the relative variation of the epidemic threshold with respect to its value in the original temporal network,
and we compare (for the SIR case) the average final epidemic size in both cases.

Moreover, \add{and to assess the scale of the impact of the removal of maximal span-cores,}
we compare the results 
 to several alternative strategies for removing temporal edges,
\add{and in particular to some strategies based on properties known to be important for spreading processes}.
To perform a sensible comparison, each of these alternative strategies 
must consist in the removal of a globally equivalent number of  temporal edges $n_T$, to have overall the same impact on the global temporal network activity.
We first consider strategies (see details in Methods section) based on static measures of coreness, since static cores have been
shown to play a role in spreading processes: these would be effective strategies if the processes were taking place on temporally aggregated networks.
To this aim, we aggregate the temporal network on the data set temporal window: in the aggregated network, two nodes are connected if they are connected
at least once in the temporal network, and the corresponding static edge has a weight equal to the number of timestamps with a temporal edge between them.
We perform the static $k$-core decomposition as well as its weighted counterpart $s$-core decomposition \cite{Eidsaa:2013} and then consider
sequentially the nodes starting from the cores with highest order (either $k$ or $s$, yielding the SC and SWC strategies) 
and remove all the temporal edges of these nodes, until
$n_T$ temporal edges have been removed (nodes are ordered at random in each core; also, whenever removing all temporal edges of a node
would lead to a total number of removed edges larger than $n_T$, we remove edges of this node at random until we reach exactly $n_T$).
\add{Although our focus is on structures rather than single nodes, we}
 also consider strategies based on \add{the most well-known} node centrality measures in the aggregated network (degree and strength, yielding
the SD and SWD strategies), proceeding in the same manner.
\add{As several centrality measures have been
generalized to temporal networks, we moreover consider
a strategy based on computing the temporal PageRank
\cite{Rozenshtein:2016} of each node at each time, classifying the pairs $(node, time)$ according to this centrality and
removing temporal edges of nodes using this classification
(tPR strategy).}
Finally, we consider three random baselines strategy. In the "random times" (RT) strategy, $n_T$ 
temporal edges are removed totally at random from the temporal network.
In the "random by timestamp", we remove exactly $n_t^s$ temporal edges at random at each timestep $t$: there are thus two
such strategies, kRTT (resp. $\Delta$RTT) removing the same number of edges at each time step than the $k$M 
(resp. $\Delta$M) maximal span-cores strategy. 
This means that these strategies are informed by the temporality of the maximal span-cores, but not by which nodes and edges they contain.

To investigate on the other hand whether nodes belonging to maximal span-cores tend to have a high spreading power, we proceed as follows:
we compute the ratio of the average epidemic sizes obtained (i) when a node of a maximal span-core is chosen as seed of the SIR process
and (ii) when a random node is chosen instead (see Methods). We then compare the epidemic ratio obtained for nodes chosen in maximal span-cores
and for nodes chosen according to aggregated static characteristics (nodes of the static cores with highest order, nodes with high degree, nodes with high 
strength).

While we perform these investigations for $8$ different data sets (see Methods), 
we show in the main text the results for two data sets corresponding to: a high school (where the aggregated contact network
displays a clear community structure, and contacts between classes are observed only during the breaks, giving rise
to interesting correlations between structure and temporal patterns \cite{mastrandrea2015contact,Gauvin:2015}) and a workplace 
(where group structure is much weaker and individuals mix with no time constraints
\cite{Genois:2015}).
The results for the other data sets are shown in the Supplementary Material.

\begin{figure}[thb]
\centerline{\includegraphics[width=0.9\columnwidth]{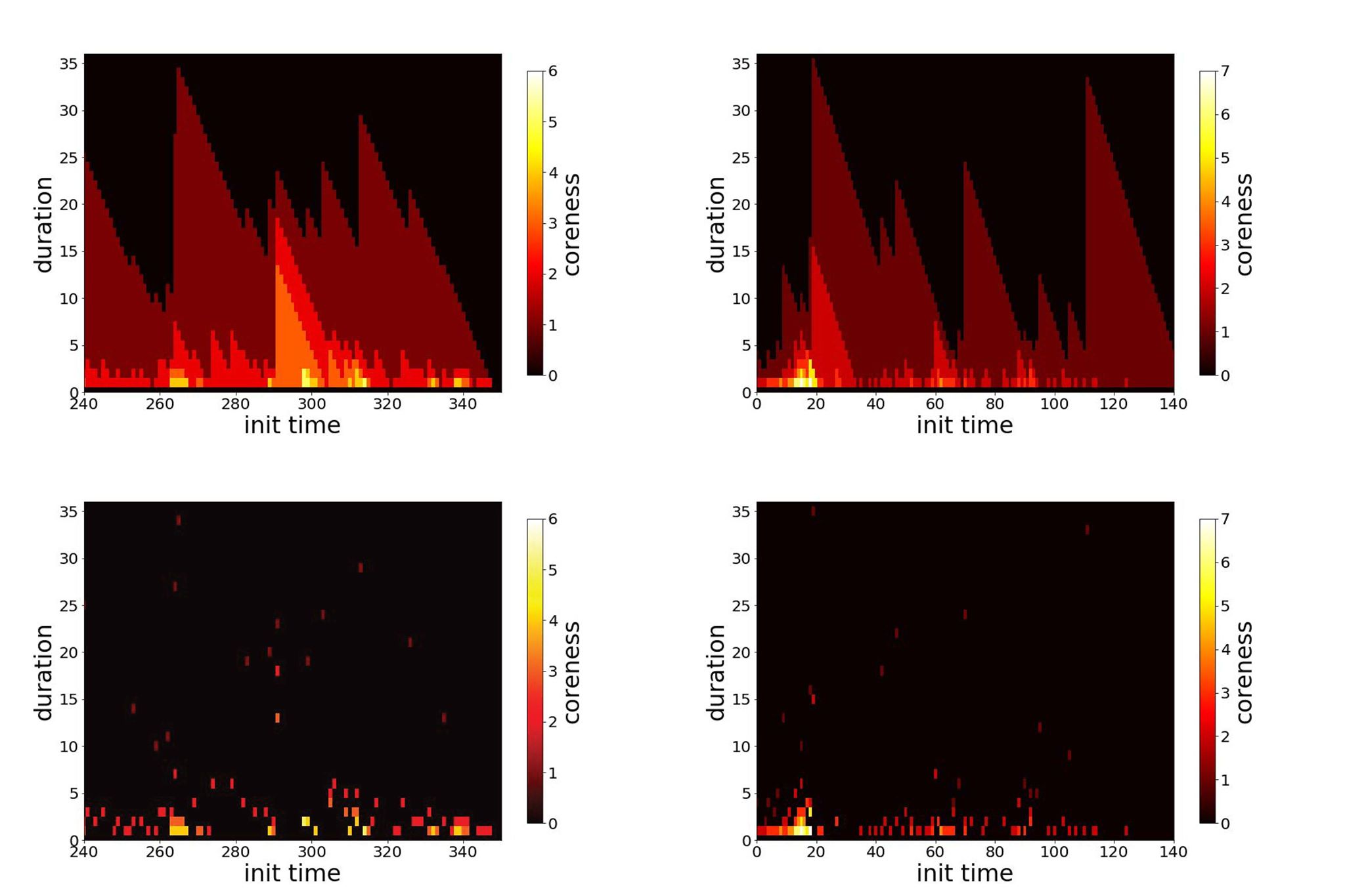}}
\caption{Colorplot of the span-cores temporal activity for two data sets. Left column: High School. Right column: Workplace.
Top plots: all span-cores. Bottom plots: maximal cores. 
We restrict in each case to one day of data for a better visibility.
In each plot, the x axis reports the timestamp 
at which the span of a span-core starts, 
the y axis specifies the size of the span (in minutes), and the color scale shows its order (coreness) k. }
\label{fig:cores_timeline}
\end{figure}

\begin{figure}[thb]
\centerline{\includegraphics[width=\columnwidth]{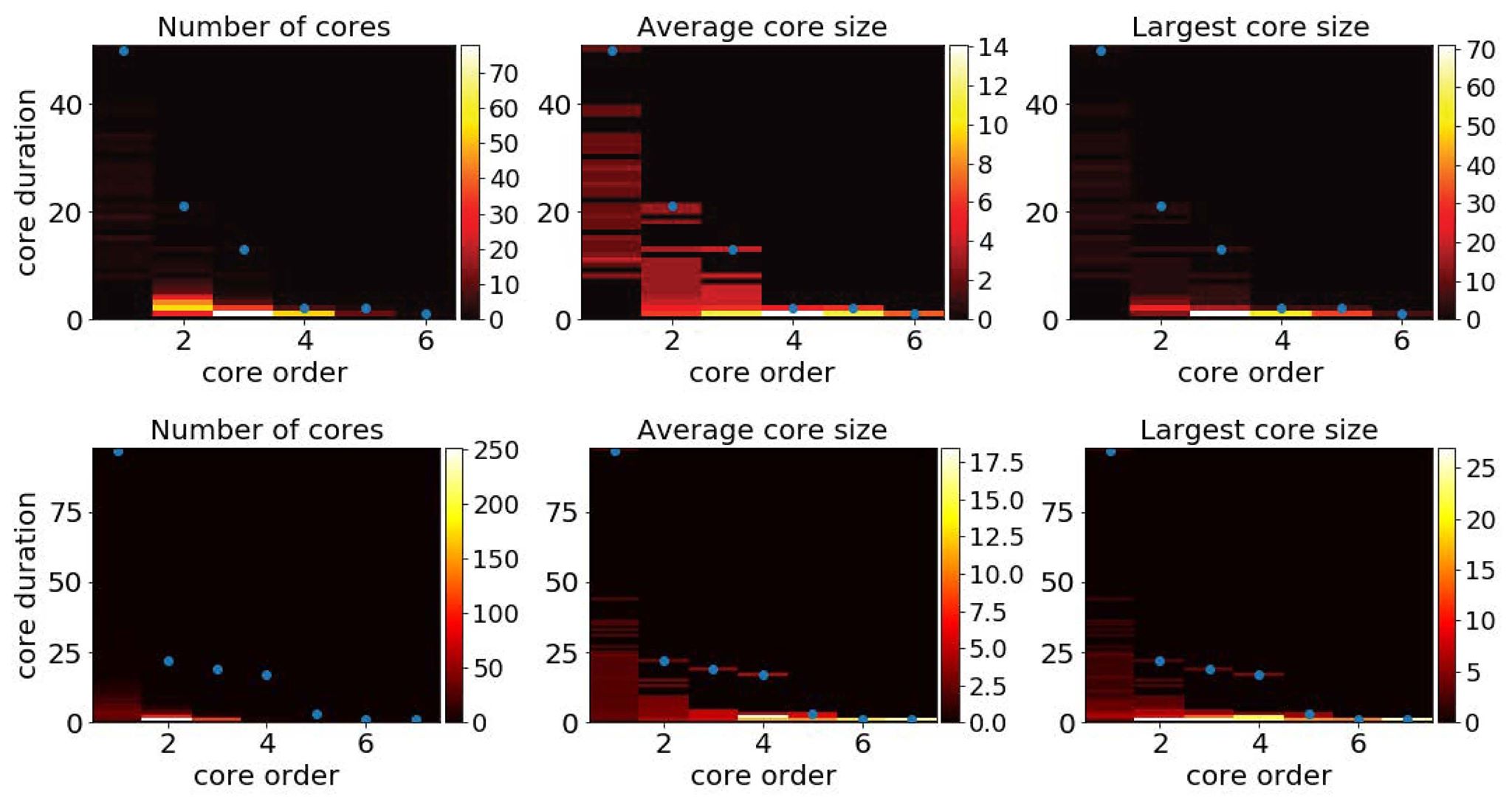}}
\caption{Aggregated statistics of the maximal span-cores for two data sets. Top row: High School. Bottom row: Workplace.
Each panel shows with a color scale the property  of cores with given core order (x-axis) and core duration (y-axis). 
Left plots: number of cores. Middle plots: average size (number of nodes) of these cores.
Right plots: size (number of nodes) of the largest of these cores.
The blue dots give, for each order, the maximal duration of cores with that order.
}
\label{fig:cores_flattened}
\end{figure}

\subsection*{Statistics of the data's (maximal) span-cores}

Let us first discuss some statistics regarding the maximal span-cores found in the temporal
networks, which are here the structures of interest targeted by the intervention strategies.
Figure \ref{fig:cores_timeline}  presents the timeline of the span-cores of the High School and Workplace 
data sets. As each span-core is characterized 
by a temporal interval (its span) and an order, we resort to a colorplot for their visualization: in each plot, the x-axis corresponds to the starting time of the interval, the y-axis
to the duration of the core, and the color encodes its order. This allows to highlight periods in which cores with long durations and/or high order are observed. Note that the triangular shapes
observed in the timelines showing all span-cores are a direct consequence of the span-cores definition: if a span-core of order $k$ is present on an interval $\Delta = [t_1,t_2]$, then there is also a span-core of the same order on all the intervals $[t_1+1,t_2],\cdots,[t_2-1,t_2], [t_2,t_2]$,
with respective durations $|\Delta|-1$, \dots, $2$, $1$.
Note also that it was shown in Ref \cite{Galimberti:2018}, that the observed timeline is not trivially linked to the activity timeline: 
reshuffled data sets in which the 
number of temporal edges in each timestamp is conserved, as well as the instantaneous degree of each node, yield indeed span-cores with 
durations and orders systematically much lower than in the original data.

Figure \ref{fig:cores_flattened}
%, Table \ref{tab:dataset_properties} 
and Table S1 in the Supplementary Material present some aggregated 
statistics of the maximal span-cores, aggregated over the whole temporal network (see also Methods). 
We show in  
particular, as a function of core order and duration, the number of maximal span-cores, their average number of nodes and the largest
number of nodes of a maximal span-core with these order and duration values.
We also highlight the largest core duration found at a given core order: for larger values of the order, the largest duration decreases,
and the largest order cores have duration $1$, corresponding to cohesive groups 
that last only briefly. On the other hand, some cores of order $1$ can last for many timestamps: they correspond to the most stable
contacts between two nodes. 

While these statistical properties do not provide a detailed temporal information as in Figure \ref{fig:cores_timeline}, they nonetheless
give an idea of the richness of the temporal patterns present in a data set. To highlight this point, we show in the Supplementary Material how the results of Figure \ref{fig:cores_flattened} are changed when the data is temporally 
reshuffled using several null models: as there are many possible null models for temporal networks~\cite{Gauvin:2018},
we consider several reshuffling possibilities, which all preserve the global activity timeline. In all cases, the reshuffled data 
show a clearly less rich core structure, with in particular smaller maximal order at a given core duration.

Finally, we mention that span-cores might be used to give a time-dependent measure of centrality of each node in the temporal network: one can for instance
at each timestamp take into account the span-cores to which a node belongs, and consider as centrality measure the highest order of these cores. One can
also measure the duration of these cores, their size, etc. One can then aggregate these measures over the whole temporal window, to obtain  
the maximal core order to which a node belongs, or the average order, etc, yielding global centrality measures for each node 
based on the temporal structures to which it contributes. 
In the Supplementary Material we show that these measures are statistically correlated with static coreness measures in the static network obtained by temporal aggregation of the temporal network, but that the correlation is weak and many 
outliers are observed: this is
expected as a given node's static connectivity properties in the aggregated network might come from the results of non-simultaneous interactions,
while span-cores are defined by cohesiveness properties that are local in time.

\subsection*{Impact of the removal strategies on activity timelines}

As mentioned above and detailed in the Methods section, \add{in order to assess the relevance of these structures,} 
we consider as \add{idealized} intervention strategies to mitigate 
spreading processes the removal of maximal span-cores, as well as other strategies based on static centrality measures that are known
to be relevant for spreading processes
\cite{Pastor-Satorras:2002,Kitsak:2010}. 

\begin{figure}[h]
\centerline{\includegraphics[width=0.8\columnwidth]{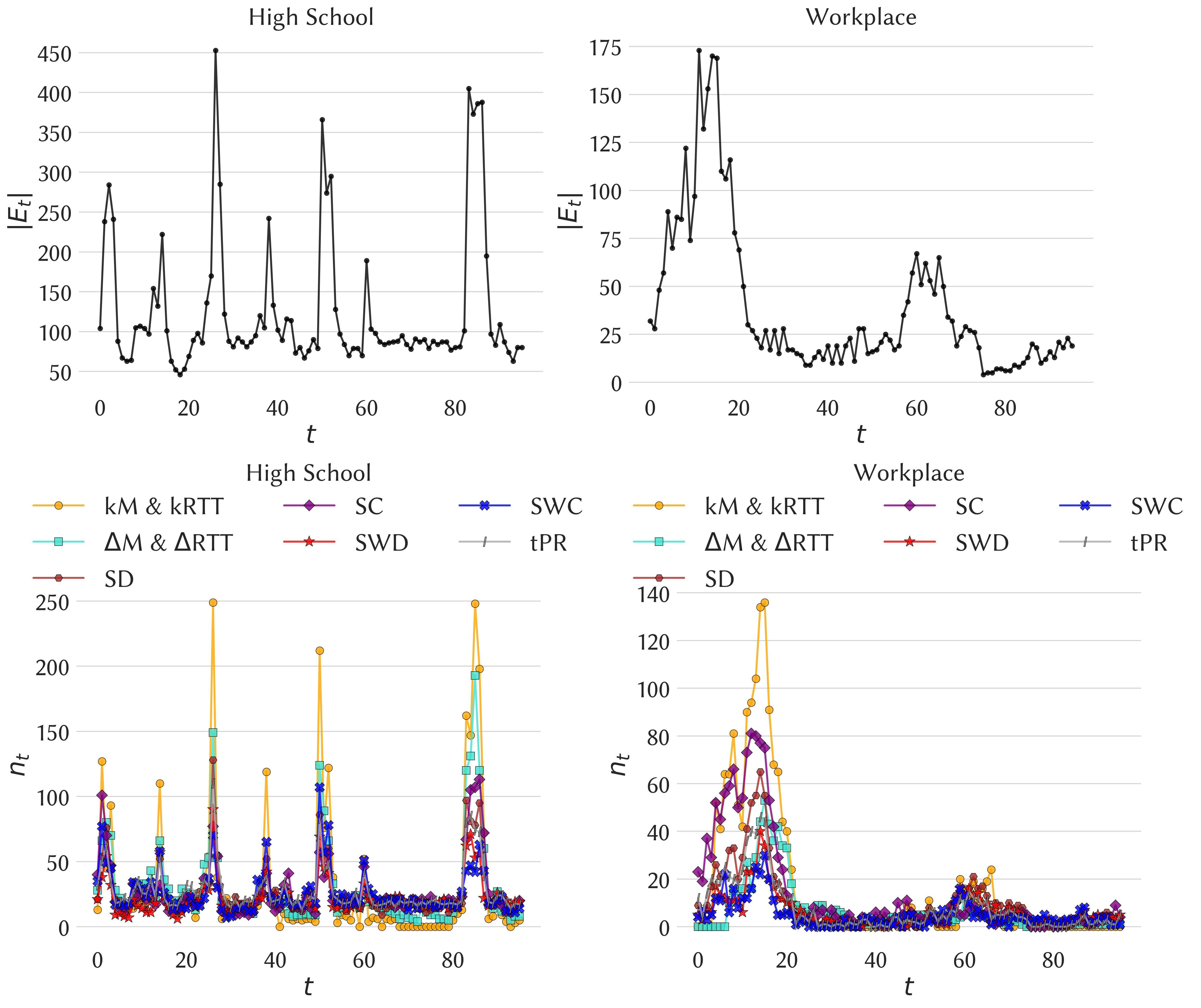}}
\caption{
Original activity timelines (number of temporal edges per timestamp, top plots) and number of temporal edges removed per timestamp
for each strategy (bottom plots), for one
day of data. Here the overall fraction of removed temporal edges is $f=20\%$.
The { kM} and { $\Delta$M} strategies target maximal span-cores (respectively with largest order and largest duration); the 
kRTT and $\Delta$RTT remove the same number
of temporal edges at each time as the {kM} and { $\Delta$M} strategies, respectively, 
but choosing the edges at random; the SD, SWD, SC and SWC are based on static quantities measured on the aggregated network, namely degree, 
strength, coreness and weighted coreness, respectively (See Methods for the detailed list of strategies). Left column: High School. Right column: Workplace.
The values of the total numbers of removed span-cores and temporal edges are given in Table \ref{tab:core_properties_interventions}.}
\label{fig:act_timelines_interventions}
\end{figure}

Table \ref{tab:core_properties_interventions} gives, for each strategy based on span-cores, some properties of the span-cores targeted. As we consider
the overall removal of a fixed fraction of the temporal edges (here $f=20\%$), the number of span-cores targeted
depends on the strategy. Moreover, targeting the cores with largest order ({kM} strategy) leads to the removal of cores with smaller duration but larger size than the $\Delta$M strategy.
Figure \ref{fig:act_timelines_interventions}
moreover shows the impact on the activity timelines of the various strategies: each curve gives the number of temporal edges removed at each time step
(i.e., $n_t$ vs $t$) for a specific strategy.
The figure shows that all strategies tend to remove more temporal edges in periods of high activity. This effect is however much stronger for the strategies based on span-cores, especially for the  { kM} strategy: this is linked to the fact that large order cores appear during these periods. 
On the other hand, strategies based on static measures or on temporal PageRank remove
also a substantial number of temporal edges during low-activity periods. Note that the {$\Delta$M} strategy strikes a balance, in particular in the
high school case, removing more temporal
edges than the static strategies in the large activity periods but also more than the {kM}
strategy in low activity periods (this is due to the
fact that some cores with large duration have order $1$, corresponding for instance to long-lasting links that extend beyond large activity periods, 
see Fig.s~\ref{fig:cores_timeline} and \ref{fig:cores_flattened}).

\begin{table}[htb]
\centering
\begin{tabular}{|p{2cm}|p{4cm}|p{1.5cm}|p{1.5cm}|p{1.5cm}|p{1.5cm}|p{1.5cm}|}
\hline
Data set & strategy & $n_{msc}$  & $\langle k \rangle$   & $\langle |\Delta| \rangle$  & $\langle n \rangle$ &  $\langle e \rangle$ \\
\hline
High School   % & first cores in chronological order & 300 & 2.53 & 5.16 & 6.98 & 43.98 \\
%\cline{2-7}
& largest order cores & 300 & 2.98 & 2.72 & 7.97 & 40.56 \\
\cline{2-7}
& largest duration cores & 289 & 2.13 & 7.17 & 4 & 43.24  \\
\hline
Workplace % & first cores in chronological order & 287 & 2.08 & 4.02 & 4.70 & 26.38 \\ \cline{2-7}
& largest order cores & 232  & 2.74 & 2.46 & 5.20 & 27.46 \\
\cline{2-7}
& largest duration cores & 179 & 1.13 & 10.50 & 2.16 & 30.71 \\
\hline
\end{tabular}
\caption{\label{tab:core_properties_interventions}
Basic properties of the maximal span-cores removed in each of the targeted strategies (with removal of $f=20\%$ of the temporal edges).
$n_{msc}$: number of maximal span-cores with all temporal edges removed;
$\langle k \rangle$: average order of the targeted maximal span-cores;
$\langle |\Delta| \rangle$: average duration of the targeted maximal span-cores;
$\langle n \rangle$: average number of nodes in the targeted maximal span-cores;
$\langle e \rangle$: average number of temporal edges removed per time step impacted by the strategy.}
\end{table}

%\clearpage
%\newpage

\subsection*{Impact of the removal strategies on the epidemic risk}

\begin{figure}[h]
\centerline{\includegraphics[width=1\columnwidth]{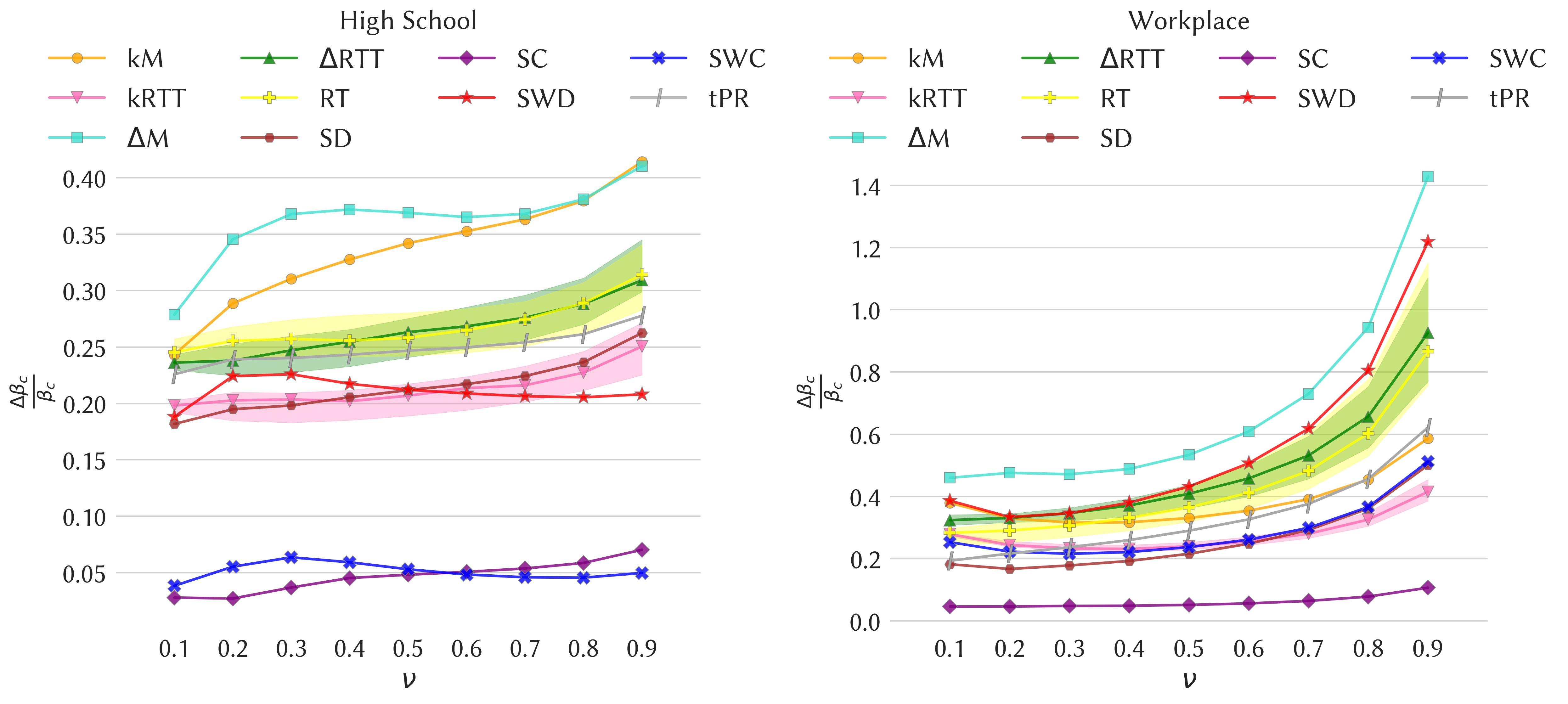}}
\caption{Impact of the various intervention strategies as measured by the change in the epidemic threshold of SIS processes.
Left plot: High School. 
Right plot: Workplace.
In each panel we plot for the various strategies the relative change $\Delta \beta_c / \beta_c$
in the epidemic threshold of an SIS process on the temporal network, computed using the method of Valdano et al. \cite{Valdano:2015b}, as a function of the recovery rate $\nu$. For each strategy
based on random choices, we show the confidence
interval between the $5^{th}$ and $95^{th}$ percentiles as a shaded area (computed on $30$ samples).}
\label{fig:delta_lambdac}
\end{figure}

Figure~\ref{fig:delta_lambdac} shows the impact of the various removal strategies on the epidemic threshold of SIS and SIR spreading processes for the data sets High School and Workplace (results for the other data sets are shown in the Supplementary Material). The figure shows, as a function of the
recovery rate $\nu$, the relative variation $\Delta \beta_c / \beta_c$ between the threshold computed on the temporal network 
after removal of temporal edges and the threshold in the original temporal network. In all cases the variation is positive, meaning as expected that the
removal of edges increases the epidemic threshold and thus decreases the epidemic risk.

\begin{figure}[h]
\centerline{\includegraphics[width=0.8\columnwidth]{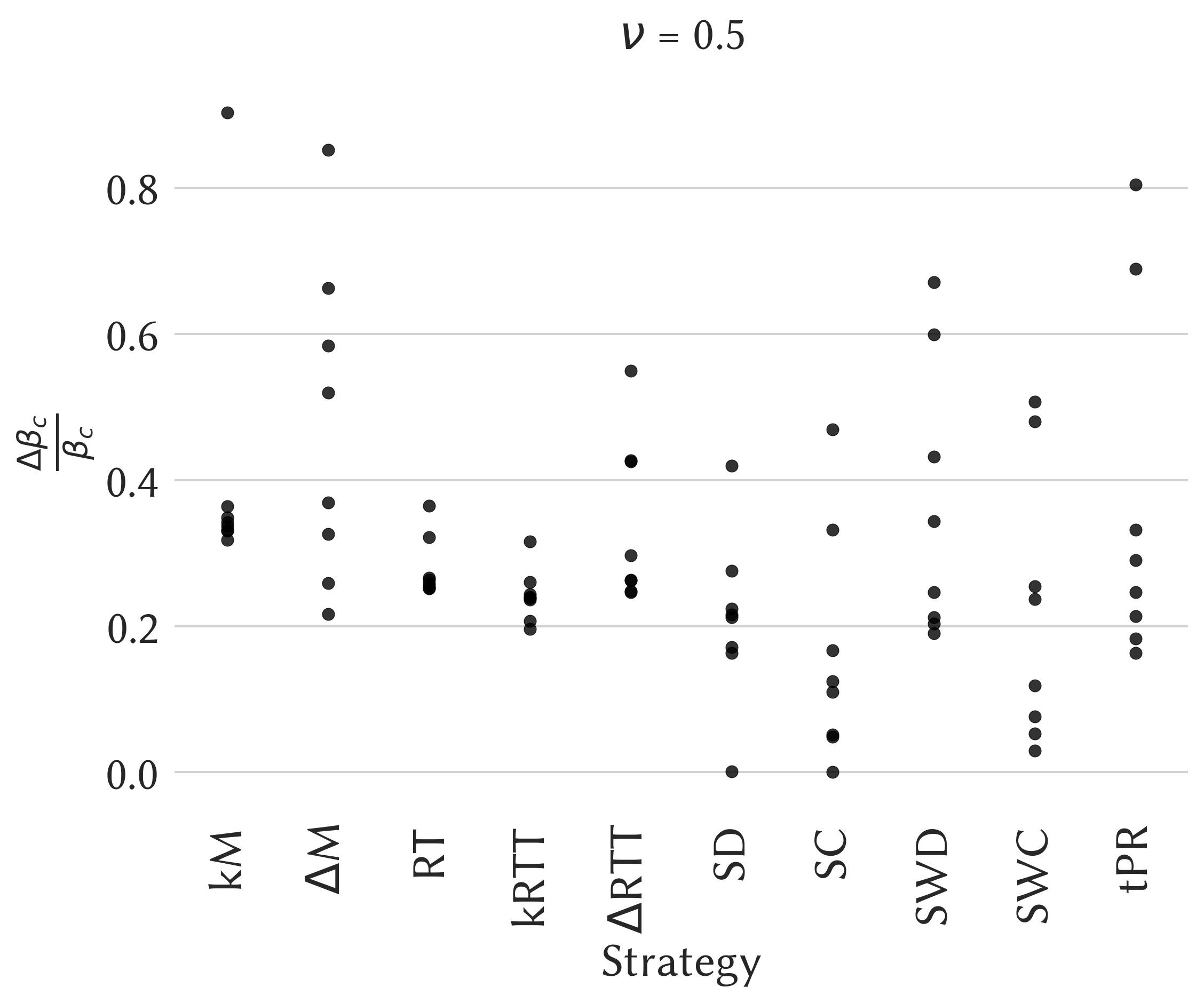}}
\caption{Epidemic threshold relative variation $\Delta \beta_c / \beta_c$ for $\nu = 0.5$ over all the $8$ datasets and candidate strategies. For a given strategy, indicated on the $x-$ axis, each point corresponds to one data set and its $y-$axis value is given by $\Delta \beta_c / \beta_c$. }
\label{fig:th_all}
\end{figure}

We first observe that the size of the effect
varies a lot for different strategies and also between data sets: some strategies lead only to a small shift of the epidemic threshold, while in some cases its value can be doubled. % (we recall that here $f=20\%$ of the temporal edges are removed in each strategy).
We also note that the best strategy, and more generally the classification of strategies according to the size of the epidemic threshold shift, depend
on the data set under consideration. However, several points are worth highlighting.
First, the best strategy is almost always the { $\Delta$M} or the {kM} one, i.e., strategies based on maximal span-cores.
Moreover, in school contexts the {kM} strategy is either best or second-best, and in three cases out of four the two span-cores based 
strategies are  the best performing ones. 
Finally, even when they are not the two best ones, the strategies based on maximal span-cores always have a large impact on the epidemic threshold, while the effect of strategies based on static centrality measures is more variable across data sets. This is illustrated in Fig.~\ref{fig:th_all} that shows, for a specific value of $\nu$, the eight values of $\Delta \beta_c / \beta_c$ for each strategy (one value per data set). The results are consistently large for the span-core strategies. Strategies based instead on static coreness (either weighted or not) and on static degree have in some cases a very small impact, while the static strength strategy, that targets nodes according to their total number of contacts, also yields a strong impact for all data sets, \add{as well as the
tPR strategy that in addition integrates temporal
information}. We however note that \add{both these strategies target} single nodes \add{(and the tPR strategy even isolates
single nodes at specific times)}
and not whole structures as the span-core strategies do \add{and hence carry an advantage in the comparison, despite
being typically less actionable in practice}. 
We moreover show in the Supplementary Material that these results are robust with respect to a change in the fraction of removed temporal edges.

As a further investigation, we have considered the epidemic size reached by the SIR process on temporal networks altered by the various
intervention strategies. While all strategies tend to decrease the epidemic size, we show in the Supplementary Material that there is no clear
optimal strategy: which strategy performs best depends both on the spreading parameters and on the data set, and in many cases the 
epidemic size mitigation is of similar amplitude for the various strategies. \add{In particular, the removal of span-cores has thus an impact as important as other
well-known strategies using single node centralities.}

\clearpage
\newpage

\subsection*{Impact of the seeding strategies on the epidemic sizes of SIR processes}

\begin{figure}[h]
\centering
\includegraphics[width=0.9\columnwidth]{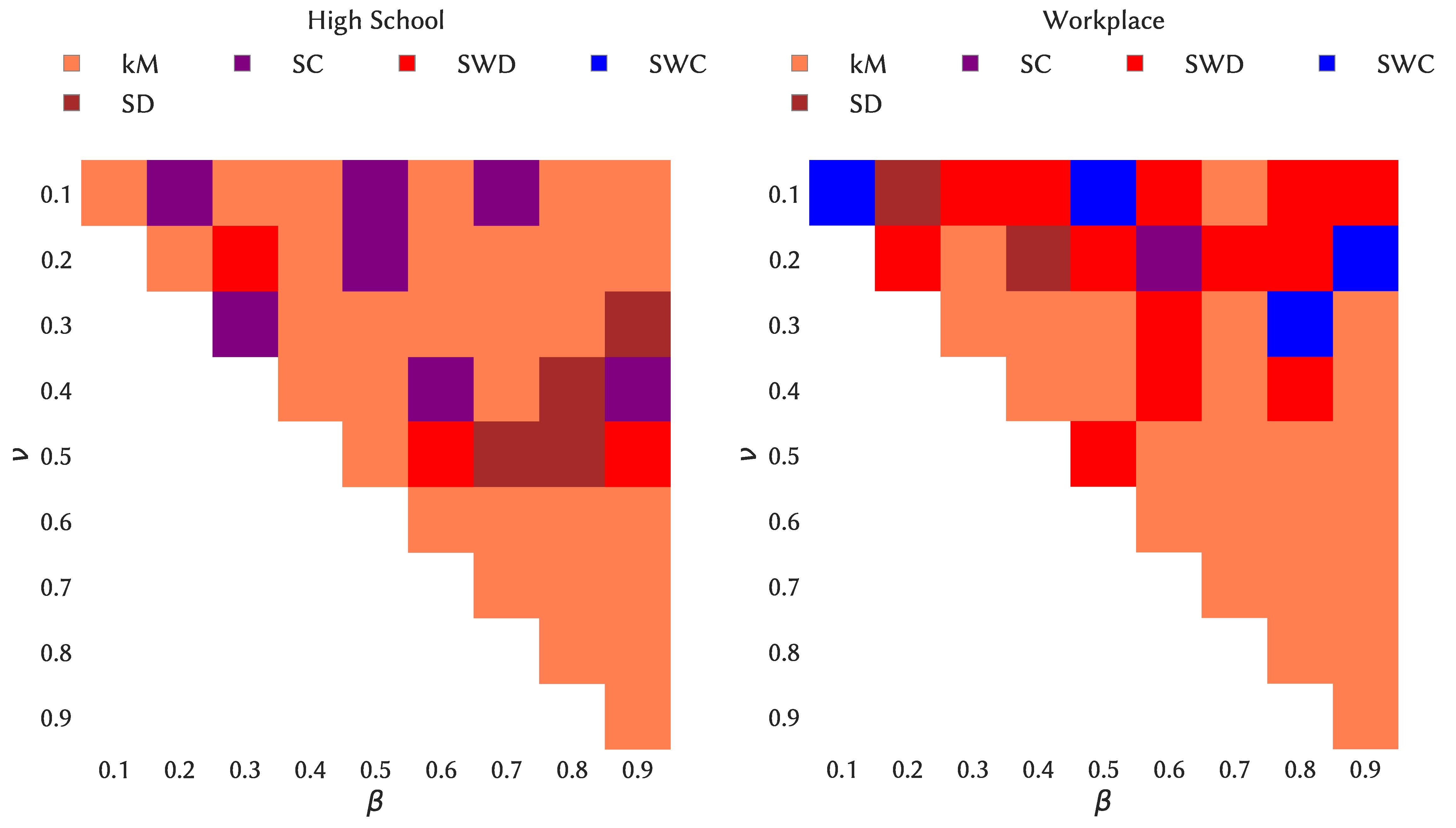} 
\caption{Heatmap indicating the seeding strategy leading to the largest ratio of average final sizes, for each combination of spreading parameter values.
Left column: High school. Right column: Workplace. }
\label{fig:SIR_heatmaps}
\end{figure}

We now consider the issue of spreading maximization, namely, which choice of the initial seed and time of the 
spread leads to the largest spread. 
Here, we consider as possible seeding strategy an initial seed belonging  
to a span-core of high order, with an initial time of the spread
at a time in the temporal interval of this span-core (kM strategy, see Methods). Moreover, as nodes with large degree or strength and nodes belonging to cores of high order 
are known to have a large spreading power in static networks, we also consider processes originating in such nodes (see Methods).
Figure \ref{fig:SIR_heatmaps} shows, for the various combinations of spreading parameters, which strategy leads to the largest average
epidemic size (see also Fig. \ref{fig:SIR_distributions} and the Supplementary Material for the other data sets). 
While some dependency on the parameter values is observed, seeds chosen in 
maximal span-cores of high order have \add{consistently a large spreading power, and in fact} in most cases the largest 
\add{among the considered baselines}. Moreover, we show in the Supplementary Material 
that the spreading power tends to increase with the order of the span-core to which the seed belongs.

\begin{figure}[thb]
\centering
\includegraphics[width=0.9\columnwidth]{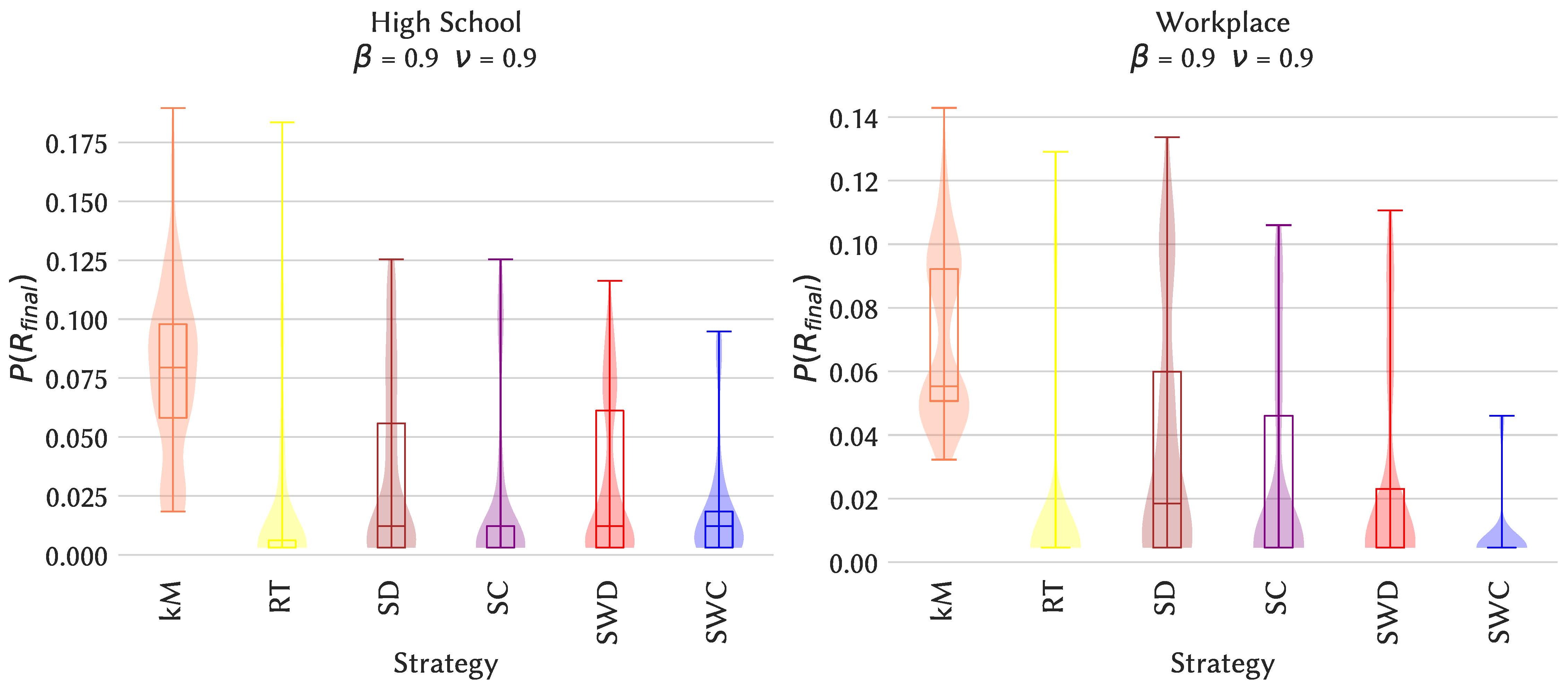} 
\caption{Distributions of final epidemic sizes for some illustrative cases, for an SIR process on the original temporal network
and for several seeding strategies. The RT
case corresponds to the baseline of randomly chosen
initial time and seed.
Left column: High School. Right column: Workplace.
}
\label{fig:SIR_distributions}
\end{figure}

\section*{Discussion}

In this work, we have investigated the relevance of specific temporal structures, namely the span-cores put forward in \cite{Galimberti:2018}, 
for spreading processes on temporal networks. The span-core decomposition generalizes to temporal networks the $k$-core decomposition of 
static graphs, by extracting in each time interval subgraphs of increasing internal connectivity ("order"). Given their definition as lasting well-connected structures, it is rather intuitive that span-cores of high order and/or duration might 
play an important role in propagation processes on temporal networks, just as static cores of large coreness are relevant for 
such processes on static networks~\cite{Kitsak:2010,Castellano:2012}. We therefore focused here on the maximal span-cores: 
a span-core on an interval $\Delta$ is defined as maximal if no other span-core can be found with higher order 
and on a broader temporal interval than $\Delta$.

To test the relevance of the maximal span-cores on propagation processes, we have investigated how two standard \add{simplified models for}
epidemic spreading processes, the SIS and SIR models, are impacted by the removal of a fraction of the temporal edges forming these span-cores, starting with the cores of highest order or of longest duration: this corresponds to temporary "quarantining" the nodes involved in these cohesive structures. 
We have compared the impact of such mitigation strategies to several 
baselines strategies of temporal edge removal targeting nodes chosen according to their centrality (degree or coreness, both weighted or not) in the static network obtained by aggregation of the temporal network, \add{as these properties are known to impact spreading processes. We have also considered a strategy based on the generalization of PageRank to temporal
networks, which allows to define a centrality that is local in time for each node}.

We have quantified the impact of these strategies on the one hand by measuring the induced relative shift in the epidemic threshold of the process, and on the other hand, in the SIR case, by the change in the resulting epidemic size (i.e., the fraction of the population affected by the spread). We have performed these numerical experiments using
several empirical temporal networks of  relevance for epidemic spread,
namely data describing human interactions in a variety of contexts. 

The results show that the shift in epidemic threshold resulting from the removal of temporal edges in maximal span-cores is consistently large; the
strategies of removing edges in the span-cores of large orders or durations figure moreover among the 
mitigation strategies leading to the strongest impact \add{among the considered strategies},
while the impact of strategies based on static measures depends more on 
the specific data set. 
The good performance of the strategies based on maximal span-cores in school environments is particularly interesting as
such contexts are of special relevance for containing infectious disease spread \cite{Gemmetto:2014} \add{and span-cores
might be related to interpretable events in such settings, such as groups forming during the breaks.}
\add{The fact that strategies based on temporal structures perform well in such environment} could be related to the fact that such environments 
present at the same time a strong community structure and also specific temporal patterns of correlated edges that cannot be exposed by purely static measures \cite{Gauvin:2013,Gauvin:2014}.
Overall, span-cores are thus more able to detect structures whose targeting leads to an increase 
of the epidemic threshold than static quantities. In terms of final epidemic size, we note 
that all strategies \add{considered} lead to a \add{large} decrease of this epidemic risk measure of comparable size,
\add{confirming the consistently important role of span-cores}.

%%%%SPREADING MAXIMIZATION%%%%%%
We have also investigated the spreading power of nodes
belonging to span-cores of high order, comparing the final size of SIR processes seeded at these nodes (and with initial time in the span of the core) and 
at nodes chosen with other seeding strategies. The results indicate that seeds chosen in span-cores
of the highest order have \add{consistently a large} spreading power, 
which tends to increase with the order of the span-core considered.

Overall, our findings confirm that maximal span-cores in temporal networks play an important role in  propagation processes on these networks. 
These results should in particular stimulate the design of models of temporal networks with non-trivial span-core structures, in addition to the usually considered statistics of contact durations and inter-contact times. 

\add{Our work has several limitations worth discussing. First, we have limited our investigations to the schematic SIS and SIR models, while
real diseases are most often described a more complicate description with a larger number of possible states for each individual (such as latent, 
asymptomatic,...). However, despite their simplicity, SIR and SIS models  
capture the main phenomenology of spreading processes, and these paradigmatic processes therefore constitute the first natural testing ground for
understanding whether a certain type or node or structure is relevant in spreading processes
\cite{Kitsak:2010,Radicchi:2017,Holme:2017,Erkol:2019,Bajardi:2012,Starnini:2013,Liu:2014}. 
It should nevertheless be kept in mind, and future work should also concern more realistic
models of spread, in particular when addressing more concrete applications. 
Second, we consider here that the whole temporal network is known ``in advance'', and the thought experiments we perform 
by removing structures are not causal. As described in the introduction, such causal interventions have been discussed in other works
\cite{Lee:2012,Starnini:2013,Valdano:2015a} and correspond to a different type of studies. Here, our goal was to assert the importance of
a certain type of temporal network structures in the unfolding of spreading processes. The next steps towards the effective design of applied
strategies would be to relate the span-cores to interpretable events or features in a context of interest and to use such knowledge to devise
actionable strategies, as done for instance in several works dealing with strategies towards
controlling outbreaks in schools \cite{Gemmetto:2014,Gauvin:2015,Smieszek:2013}.
}

Further work includes \add{thus} more realistic design and investigation of strategies, \add{in particular under the hypothesis that}
the temporal network is not fully known \add{but for instance only its main statistical properties}
\cite{Genois:2015,Genois:2018}. 
Moreover, it would be interesting to investigate the role of span-cores in other types 
of dynamical processes on networks, such as
opinion formation, complex contagion or 
synchronization processes.

\clearpage
\newpage

\section*{Methods}

\subsection*{Span-core decomposition and maximal span-cores}

In a recent work, Galimberti \textit{et al.}~\cite{Galimberti:2018} proposed an extension of core decomposition to temporal networks, whereby cores are
associated with their temporal spans, i.e., intervals of contiguous timestamps for which the coreness property (i.e., of minimal
connectivity) holds. Such cohesive temporal structures are named span-cores.

%In order to formally define them, some notation is introduced. 
Let us consider a temporal graph $G = (V,T,\tau)$, where
$V$ is a set of nodes,
 $T = [0, 1, \ldots, t_{max}]$ is a discrete time domain, and $\tau: V  \times V \times  T\rightarrow \{0,1\}$ is a function defining for each pair of vertices $u,v \in V$ and each timestamp $t \in T$ whether the 
 edge $(u,v)$ exists in $t$. We denote $E = \{(u,v,t) \mid \tau(u,v,t) = 1 \}$ the set of all temporal edges. Given a timestamp $t \in T$, $E_t = \{(u,v) \mid \tau(u,v,t) = 1 \}$ is the set of edges existing at time $t$.
Given a subset of nodes $S \subseteq V$, let $E_{\Delta}[S]$ be
the set of edges connecting the nodes of $S$ that exist in {\em all} timestamps $t \in \Delta$.
We then define the temporal degree of a node $u$ within the subgraph $G_{\Delta}[S] = (S , E_{\Delta}[S])$ as
$\tdeg_\Delta(S,u) = |\{v \in S \mid (u,v) \in E_\Delta[S] \}|$. In other words, the temporal degree of $u$ is the number
of other nodes to which $u$ is linked in all the timestamps of $\Delta$, without interruption.

\begin{mydefinition}[$(k,\Delta)$-core] \label{def:core}
        The  $(k,\Delta)$\emph{-core} of a temporal graph $G = (V,T,\tau)$ is (when it exists) a maximal and non-empty set of nodes $ \emptyset \neq C_{k,\Delta} \subseteq V$, such that $\forall u \in C_{k,\Delta} : \tdeg_\Delta(C_{k,\Delta},u) \geq k$, where
        $\Delta \sqsubseteq T$ is a temporal interval and $k \in \mathbb{N}^+$.
\end{mydefinition}

The interval $\Delta$ and the integer $k$ are referred to as span and order of the span-core, respectively.
As it is well known, the cores of a static graph define a hierarchy. On the other hand, span-cores are not all nested into each other.
Nonetheless, they exhibit containment properties.
In particular, we say that a  $(k,\Delta)-$core is contained into another $(k^{'},\Delta^{'})-$core if $k^{'} <  k$ and $\Delta \sqsubseteq \Delta^{'}$.

The number of span-cores is quadratic in $|T|$. This is definitely not desirable when human inspection is of interest.
Therefore, it is useful to focus only on the most relevant ones.
Thus, Galimberti \textit{et al.}~\cite{Galimberti:2018} introduced the concept of maximal span-core.

\begin{mydefinition}[Maximal \spancore] \label{def:maximal}
        A \spancore $\core$ of a temporal graph $G$ is said \emph{maximal} if there does not exist any other \spancore $C_{k',\Delta'}$ of $G$ such that $k \leq k'$ and $\Delta \sqsubseteq \Delta'$.
\end{mydefinition}

A span-core is thus identified as maximal if it is not dominated by any other span-core in terms of both order and span.
Clearly, maximal span-cores resemble the idea of innermost core, i.e., the core of highest order, in the core decomposition of a static graph.
However, maximal span-cores are not unique. Instead, there is at most one maximal span-core for every temporal interval.

We also recall here some basic ideas of the efficient algorithm to compute all span-cores in a temporal
network~\cite{Galimberti:2018}. A naive approach would involve executing a static core decomposition routine independently for each temporal
interval $\Delta$. A more efficient procedure exploits the containment property in both its dimensions,
coreness and temporal intervals.
Indeed, given a temporal graph $G$, and a temporal interval 
$\Delta = [t_s,t_e] \sqsubseteq T$, let $\Delta_+ = [\min\{t_s+1,t_e\}, t_e]$ and $\Delta_- = [t_s, \max\{t_e-1,t_s\}]$.
It holds that
$$
C_{1,\Delta} \ \subseteq \ (C_{1,\Delta_+} \cap C_{1,\Delta_-}) \ = \ \bigcap_{\Delta' \sqsubseteq \Delta} C_{1, \Delta'} .
$$
The algorithm~\cite{Galimberti:2018} takes advantage of this simple property by processing temporal intervals of increasing size
(starting from size one) and, for each interval $\Delta$ of width larger than one, the core decomposition is initiated 
from $(C_{1,\Delta_+} \cap C_{1,\Delta_-})$,  the smallest intersection of cores containing $C_{1,\Delta}$.
This expedient produces a speed-up of orders of magnitude in the obtention of all span-cores.

The problem of computing the maximal span-cores of a temporal graph can also be addressed by the simple approach of extracting all span-cores
and then filtering out those which are not recognized as maximal.
However, theoretical properties that relate the maximal span-cores to each other prove that it is not required to compute the overall temporal core decomposition but it is possible to extract only the maximal span-cores.
Note that this is a challenging design principle, as it contrasts
the idea that a core of order $k$ is typically computed from the core of order $k-1$.
Theoretical findings provide bounds on the order of a maximal span-core and suggest a top-down algorithm that
processes temporal intervals starting from the larger ones, in opposition to the method used to extract the entire span-core decomposition.
The procedure does not search the whole span-core space and it has been empirically shown to be markedly more efficient compared to the approach based on filtering out non-maximal span-cores~\cite{Galimberti:2018}.

The algorithms are detailed in Ref. \cite{Galimberti:2018} and the code is publicly available on
\verb+https://github.com/egalimberti/span_cores+

\subsection*{Datasets}
\label{subsec:datasets}

We consider {\bf $8$} data sets describing human interactions with high spatial and temporal resolutions in a 
variety of contexts: schools (at different levels and in different countries), conferences, workplace (office building) and hospital.
These data are publicly available thanks to two independent collaborations. {SocioPatterns} \cite{Sociopatterns} 
gathers longitudinal data on physical proximity and face-to-face 
contacts of individuals in different contexts using a sensing platform based on wearable badges equipped with 
radiofrequency identification devices (RFIDs). 
Contact data are collected with a temporal resolution of $20$ seconds.  Further data describing human proximity
are provided by Toth \textit{et al.} \cite{Toth:2015}, 
who deployed a platform composed of wireless ranging enabled nodes (WRENs) in several schools in the USA. 
Each WREN collects signals from other WRENs in proximity at intervals of approximately $20$ seconds. 
Signal strength criteria are used to select pairs of individuals wearing these WRENs
located at distance lower than or equal to $1$ meter: this is the practical definition used to define a contact between individuals at each time.
For each data set, we aggregate all interactions in successive time-windows (timestamps) of $300$ seconds. We thus obtain from each data set
a temporal network in which nodes represent individuals, and a temporal edge is drawn in a timestamp $t$ between two nodes
if the two corresponding individuals have been in contact in this time-window.

The specific data sets we use are the following. 
The first $6$ are provided by the {SocioPatterns} \cite{Sociopatterns} collaboration, and the last $2$ by Toth \textit{et al.} \cite{Toth:2015}. 
\begin{itemize}
	
	\item The \emph{Primary School} data set contains the contact events between $242$ individuals ($232$ children and $10$ teachers) in a primary school in Lyon, during two days in October $2009$ \cite{Stehle:2011b}.
	
	\item The \emph{High School} data set gives the interactions between $327$ students
	of nine classes within a high school in Marseille, during five days in December $2013$ \cite{mastrandrea2015contact}.
	
	\item The \emph{Hospital} data set describes the face-to-face interactions of 
	patients and health-care workers (HCWs) in a hospital ward in Lyon, France during one week in December 2010. The study included 46 HCWs and 29 patients \cite{Vanhems2013estimating}. 
	
	\item The \emph{Conference} data set was collected during the ACM Hypertext $2009$ conference, which took place between June $6$, $2009$ and July $1$, $2009$ in Turin, Italy. 
	The data cover a period of two days and a half \cite{Isella:2011}. 
	
	\item The \emph{SFHH conference} data set describes the face-to-face interactions of $405$ participants to the $2009$ SFHH conference in Nice, France
	(June $4-5$, $2009$) cite{Genois:2018}. 

	\item The \emph{Workplace} data set contains the temporal network of contacts between individuals recorded in an office building in France in $2015$ \cite{Genois:2018}. 

	\item The \emph{Elementary School} data set contains the contact data  associated with the $476$ students in the $21$ classes of a suburban elementary school in Utah (USA) on January $31$, $2013$ and February $1$, $2013$ \cite{Toth:2015}. 
	
	\item The \emph{Middle School} data set describes the proximity interactions occurred on November $28$ and $29$, $2012$ in an urban public middle school in Utah (USA) \cite{Toth:2015}. 

\end{itemize}
Some more details are given in Table \ref{tab:dataset_properties}.

\begin{table}[ht]
\centering
\begin{tabular}{|p{4cm}|p{2cm}|p{2cm}|p{2cm}|p{2cm}|p{2cm}|}
\hline
Data set  and reference  & number of nodes &  number of temporal edges & total duration (in number of timestamps) & core of largest order: (order, duration, number of nodes) & core of max duration: (order, duration, number of nodes)  \\
\hline
Primary school & 242 & 55k & 208 & (7, 1, 10) & (1, 27, 2) \\
\hline
High school & 327 & 47k & 492 & (6, 1, 7) & (1, 50, 2) \\
\hline
Elementary school  & 339 & 41k & 152 & (9, 1, 13) & (1, 49, 5) \\
\hline
Middle School & 590 & 68k & 168 & (10, 1, 11) & (1, 23, 2)  \\
\hline
SFHH & 403 & 21k & 247 & (9, 1, 10) & (1, 83, 2)  \\
\hline
ACM Hypertext Conference & 113 & 7k & 424 & (7, 1, 9) & (1, 38, 2) \\
\hline
Workplace & 217 & 21k & 1k & (7,  1, 24) & (1, 97, 2) \\
\hline
Hospital & 75 & 9218 & 627 & (6,1,7) & (1,55,2) \\ 
\hline

\end{tabular}
\caption{\label{tab:dataset_properties} Properties of the data sets considered here and of the span-cores with largest order or duration.
}
\end{table}

\subsection*{Spreading processes and impact evaluation}

We consider the two paradigmatic spreading processes Susceptible-Infectious-Susceptible (SIS) and 
Susceptible-Infectious-Recovered SIR, with parameters $\beta$ and $\nu$, as described in the main text: $\beta$ is the probability per 
unit time that a susceptible node in contact with an infectious one becomes
infectious, and $\nu$ is the probability per unit time that an infectious node recovers spontaneously (becoming again susceptible
in the SIS case, and recovered in the SIR case).
These spreading processes are considered on the temporal network data: contagion can occur only along the temporal edges.

{In the spreading mitigation scenario}, our aim is to compare the unfolding and impact of 
these processes, for each data set, on the original temporal network and on temporal networks
modified according to several intervention strategies.
To quantify the differences between processes in original and altered networks, we resort to two
common measures of 
the epidemic risk and study how its value
is modified by the intervention.
We first consider the value of the epidemic threshold, i.e., the critical value of disease transmissibility 
$\beta$ above which the spread is able to reach a large fraction of the population.  The analytical method developed by Valdano et al.
\cite{Valdano:2015b}, based on the approximation of the process by a Markov chain, allows to express the epidemic threshold for both processes
in terms of the spectral radius of a matrix that encodes both network structure and disease dynamics. We use the Python package publicly available at
\verb+http://github.com/eugenio-valdano/threshold+ to compute the threshold as a function of the recovery
parameter $\nu$ for the various data sets and the various intervention strategies, and measure the impact
of each strategy through the relative change in the threshold value.

We  moreover 
consider the final size of the epidemic, i.e., the fraction of nodes that have been reached by the
process when it ends. At the end of the SIR process, no infectious nodes are left and the epidemic size is given by the fraction
of nodes in the R state. Note that, as the SIR epidemic might not end within the finite span of the
data, the temporal network is repeated if needed until the process ends.
For the original data, for each strategy and for each set of parameters $(\beta,\nu)$
we simulate $N_{sim} = 1000$ SIR processes. We then compute the epidemic size ratio for each strategy and parameter values, defined as the ratio between the
average epidemic size for processes on the reduced temporal network obtained according to a strategy and the average epidemic size for processes on the original temporal network. 

$$
\rho_{strategy}(\beta,\nu) = 
\frac{\langle R_{final} \rangle_{strategy}}
{\langle R_{final}\rangle_{original}}  \ .
$$

{In the spreading maximization scenario, the goal is to assess the performance of the maximal span-cores as a tool to identify the nodes that,when infected, lead to wide propagation. 
We consider numerical simulations of the SIR process and we adopt the final size of the epidemics as a performance metric for a seeding strategy. For each seed, we simulate $N_{sim} = 100$ SIR processes starting at different timestamps. Again, if required, the temporal domain of a temporal network is repeated until the process end. 
We then compute the epidemic size ratio, in which
the numerator and denominator are given by the average epidemic sizes for processes seeded according to a strategy and for randomly seeded processes, respectively.}

$$
\rho_{strategy}(\beta,\nu) = 
\frac{\langle R_{final} \rangle_{strategy}}
{\langle R_{final}\rangle_{random}}  \ .
$$

{Clearly, although in the spreading mitigation and maximization frameworks the epidemic size ratios are defined similarly, in the former smaller values are desiderable while in the latter larger values are to be preferred.}

\subsection*{Mitigation strategies}

We describe here in detail the  targeted intervention strategies aimed at mitigating the spread of an epidemic
process  unfolding in a host population described by a temporal network. 
These interventions consist in removing temporal edges in the network, and different strategies consider
different ways of choosing the edges to be removed.
The strategies we put forward target the maximal span-cores of the given temporal network $G = (V, T, \tau)$.
One possibility is to choose an a priori number $n_{msc}$ of maximal span-cores and remove all the corresponding
edges. This can be interpreted as temporary isolation: 
a node $u$ belonging to one of the chosen maximal span-cores $C_{k,\Delta}$ is isolated over the time interval $\Delta$. 
As different cores have different sizes, fixing $n_{msc}$ would however lead to different fractions of removed temporal
edges for the different strategies and the fraction $f$ of temporal edges to be removed and remove them starting
with the maximal span-cores taken in a chosen order. We consider here $f=20\%$ and show in the supplementary material the results of using $f=10\%$.

As the maximal cores can be classified along two properties, their order and their durations, we consider in fact two separate strategies:
\begin{itemize}

\item 
The top-$k$ maximal span-cores strategy (for short {kM}) removes temporal edges starting from the maximal span-cores
with highest order, independently of their duration. We denote by $G_{kM}$ and $E_{kM}$ the  temporal network and set
of temporal edges remaining after the intervention, respectively.
%network (for short \emph{kM}) 
%$G_{kM} = (V, T, \tau_{kM})$ has edge set $E_{kM} = E \setminus \{(u,v,t) \mid \exists C_{k,\Delta} \in S_k \ s.t. \ (u \in C_{k,\Delta} \lor v \in C_{k,\Delta}) \land t \in \Delta \}$.
\item The top-$\Delta$ maximal span-cores strategy (for short { $\Delta$M}) removes temporal edges starting from the maximal span-cores
with longest duration, independently of their order. We denote by $G_{\Delta M}$ and $E_{\Delta M}$ the  temporal network and set
of temporal edges remaining after the intervention, respectively.
%Let $S_{\Delta}$ be the set of the largest-span $n_{msc}$ maximal span-cores.
%The no-top-$\Delta$ maximal span-cores network (for short \emph{$\Delta$M}) $G_{\Delta M}$ = $(V, T, \tau_{\Delta  M})$ has edge set $E_{\Delta M} = E \setminus \{(u,v,t) \mid  \exists C_{k,\Delta} \in S_{\Delta} \ s.t. \ (u \in C_{k,\Delta} \lor v \in C_{k,\Delta}) \land t \in \Delta \}$.

\end{itemize}
We denote by $n_T$ the total number of temporal edges removed: $n_T = f |E|$ where |E| is the set of temporal edges in the
temporal network considered. For each strategy, we precise that: (i) at given duration or order, the cores are ordered randomly; 
(ii) if removing all the edges of a span-core would lead to removing more than $n_T$ edges,  edges are removed at random 
from that core until reaching exactly $n_T$ removed edges.

Table \ref{tab:core_properties_interventions} gives for each strategy the properties of the span-cores removed for the two data sets
considered in the main text. These properties are given in the Supplementary material for the other data sets. 

The  $n_T$ temporal edges removed are not removed uniformly along the timeline of the temporal network, and we denote by
$n_t^{s}$ the number of temporal edges removed at timestep $t$ for strategy $s$ ($s=kM$ or $s=\Delta M$).
We evaluate the effectiveness of these intervention strategies by comparing their impact to the one of several baselines. 
Each baseline consists in removing the same number $n_T$ of temporal edges from the temporal network. The two
simplest baselines consist in removing these edges at random:
\begin{itemize} 

\item Randomly trimmed network (for short {RT}): the simplest benchmark is the intervention in which $n_T$ edges are randomly 
removed over the temporal domain. 

\item Randomly trimmed by timestamp network (for short {RTT}): this strategy uses the knowledge of the timestamps in which the targeted
span-cores are active, by removing  exactly $n_{t}^s$ edges randomly at each timestamp $t$. 
There are therefore two such strategies, { kRTT} that removes  $n_{t}^{kM}$ random edges at time $t$ and 
{$\Delta$RTT} that removes  $n_{t}^{\Delta M}$ random edges at time $t$. 

\end{itemize} 

Note that the two {RTT} strategies exploit the temporal information provided by the temporal core decomposition while the 
{RT} one does not.  

We consider as well more sophisticated baselines based on node centrality measures computed on the time-aggregated network. 
Indeed, it is known for static networks that nodes with large degree or static coreness play important roles in spreading processes.
We also consider the weighted counterparts of degree and coreness, strength and weighted coreness \cite{Eidsaa:2013}.
We recall that in the time-aggregated network, the degree of a node $u$ is equal to the number of distinct nodes with 
whom $u$ has been in contact, and the weight of an edge $(u,v)$ gives the number of temporal edges between $u$ and $v$ in the
temporal network. 

For each centrality measure, the strategy works as follows:
\begin{itemize} 
\item we first sort the nodes in decreasing order of their centrality; 

\item we then consider the nodes one by one, starting by the most central ones, and removing all the temporal edges to which it belongs 
over the entire temporal domain, until the stopping criterion is met (i.e., until $n_{T}$ edges have been discarded).
If discarding all the interactions of a node would mean exceeding $n_T$, the temporal interactions to be removed 
for this node are chosen at random.
\end{itemize} 
We thus obtain the four following strategies:
\begin{itemize} 
	\item The highest static degree strategy ({SD}) strategy;
	%$G_{SD} = (V, T, \tau_{SD})$ is obtained by sequentially removing all the interactions associated with the nodes 
	%having highest static degrees until the total number of removed interactions equals $n_{T}$.
	\item The highest static coreness ({SC}) strategy;
	%removes sequentially the temporal edges of the nodes
	%network $G_{SC}= (V, T, \tau_{SC})$ is obtained by sequentially removing all the edges incident to nodes
	 %having highest static coreness until the total number of removed interactions equals $n_{T}$.
	\item The highest static strength (weighted degree, {SWD}) strategy;
	%removes sequentially the temporal edges of 
	%network $G_{SWD}= (V, T, \tau_{SWD})$  is obtained by sequentially removing all the contacts adjacent to 
	%the nodes having highest static weighted degree until the total number of removed interactions equals $n_{T}$. 
	\item The highest static s-coreness  (weighted coreness, {SWC}) strategy.
	% removes sequentially the temporal edges of
	% network $G_{SWC}= (V, T, \tau_{SWC})$ is obtained by sequentially removing all the edges associated with 
	%the nodes having highest weighted static s-coreness until the total number of removed interactions equals $n_{T}$. 
\end{itemize} 

\add{Finally, we consider the strategy tPR, which is based on a generalization of PageRank to temporal network
 \cite{Rozenshtein:2016}. Here, a temporal PageRank value is assigned to each node $u$ at each timestamp $t$: tPR$(u,t)$.
The pairs $(u,t)$ are ranked according to these values
and we remove their temporal edges using this ranking
until $n_T$ temporal edges have been removed.}

\subsection*{Seeding strategies}

{In the spreading maximization scenario considered for the SIR model, we consider several seeding strategies, i.e., choice
of the initial seed of the spread (the first node in state I) aimed at favouring the spread. 
The idea underlying the proposed procedure is that nodes that represent influential spreaders are likely to belong to the span-cores of highest order 
at the timestamp in which the spread begins. 
We consider a fraction
$f = 5 \%$ of the nodes of a temporal network:
The top-k maximal span-core seeding strategy (for short $kM$) requires to carry out the following steps:}

\begin{itemize}

    \item we rank the nodes according to the value of the highest order of the span-cores they belong to;
   
   \item in decreasing order, we take each node as seed of $N_{sim}=100$ SIR processes until the fraction $f$ of the total amount of nodes has been considered. For a given node, the spread starts at a timestamp randomly sampled from the union of the spans of the highest order span-cores it belongs to. We then compute the average size of the epidemic, averaged over the $N_{sim}=100$ processes.

\end{itemize}

{Strategies based on randomness are not properly defined in this case since random seeds are exploited in order to construct the performance metric. 
Instead, baseline strategies are based on the same static node centrality measures used in the spread mitigation scenario, namely degree (strategy SD), strength (strategy SWD), coreness (strategy SC) and weighted coreness (strategy SWC).
For each centrality measure, the associated strategy is implemented as follows: }

\begin{itemize}

    \item we rank the nodes according to their centrality; 
    
 %   \item selecting the fraction $f$ of nodes having the largest values;
 
    \item in decreasing order of centrality, we take each node as a seed of $N_{sim}=100$ SIR processes until the fraction $f$ of the total amount of nodes has been considered. In order for the comparison to be fair, the same sequence of initial timestamps considered in the seeding strategy based on the highest order maximal span-cores is considered.

\end{itemize}

\section*{Acknowledgements}

This work was partially supported by
the ANR project DATAREDUX (ANR-19-CE46-0008) to A.B.

\bibliography{sample}

\clearpage
\newpage

\graphicspath{{FIGS_SI/}}

\setcounter{section}{0}
\setcounter{table}{0}
\setcounter{equation}{0}
\setcounter{figure}{0}

\renewcommand{\thetable}{S\arabic{table}}
\renewcommand{\thefigure}{S\arabic{figure}}
\renewcommand{\thesection}{S\arabic{section}}
\renewcommand{\theequation}{S\arabic{equation}}

\begin{center}
{\Large{\textbf{Supplementary Material \\ \bigskip
``Relevance of temporal cores for epidemic spread in temporal networks"}}}\\ \bigskip
\large{Martino  Ciaperoni, Edoardo Galimberti, Francesco Bonchi, Ciro Cattuto, Francesco Gullo, and Alain Barrat}

\end{center}

\section{Aggregated statistics of maximal span-cores for original and reshuffled data sets}

 \begin{figure}[thb]
     \centering
\includegraphics[width=.8\columnwidth]{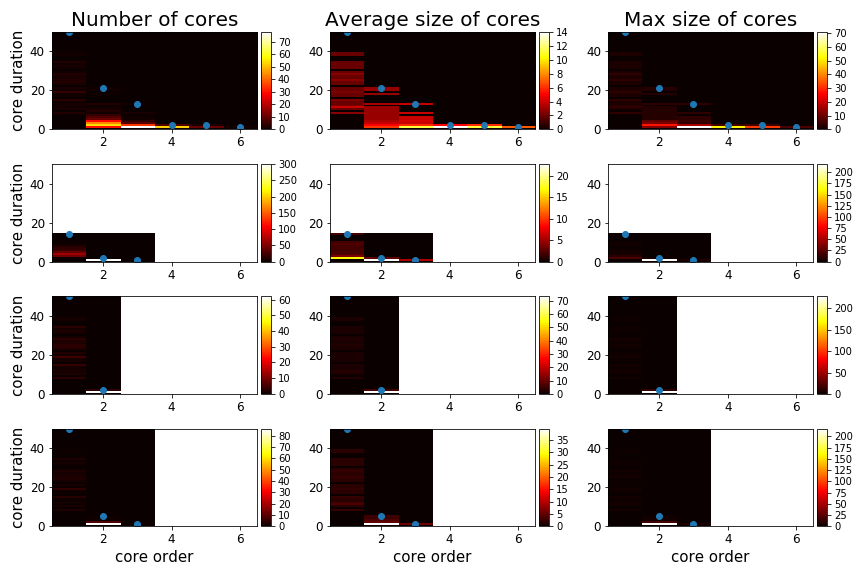}
\caption{Aggregated statistics of the maximal span-cores for the original High School data set and for three 
reshuffled versions of this data set.
Each panel shows with a color scale the property
of cores with given core order (x-axis) and core duration (y-axis).
Left plots: number of cores;
Middle plots: average size of these cores;
Right plots: size (number of nodes) of the largest of these cores.
The blue dots give, for each order, the maximal duration of cores with that order.
First row: original data.
Second row: reshuffling R1, noted $P[w,t]$ in \cite{Gauvin:2018}, which shuffles the timestamps
among the temporal edges, thus preserving aggregated network and its weights and the global 
activity timeline.
Third row: reshuffling R2, noted $P[{\cal L}, p(t,\tau)]$ in \cite{Gauvin:2018}, 
which shuffles the contacts while keeping their starting time and duration, thus preserving
contact duration statistics, global activity timeline and structure of the aggregated network.
Fourth row: reshuffling R3, noted $P(k, p_{\cal L} (\Theta) )$ in \cite{Gauvin:2018},
for which one shuffles the links of the aggregated network according to the procedure of Sneppen \& Maslov
\cite{sneppen-maslov}: this preserves the activity timeline and statistics of contact durations but reshuffles any static structure.
}
         \label{fig:heatmaps_nullmodels_1}
      \end{figure}
      
\begin{figure}[thb]
\centering       
\includegraphics[width=.8\columnwidth]{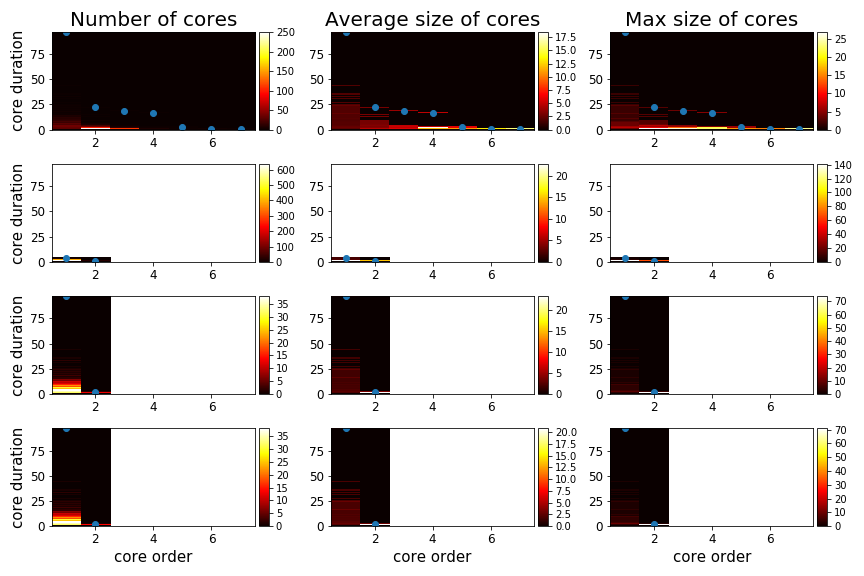}
\caption{Same as Figure \ref{fig:heatmaps_nullmodels_1}, for the Workplace data set.}
\label{fig:heatmaps_nullmodels_2}
\end{figure}
      
 \begin{figure}[thb]
     \centering 
     \includegraphics[width=.8\columnwidth]{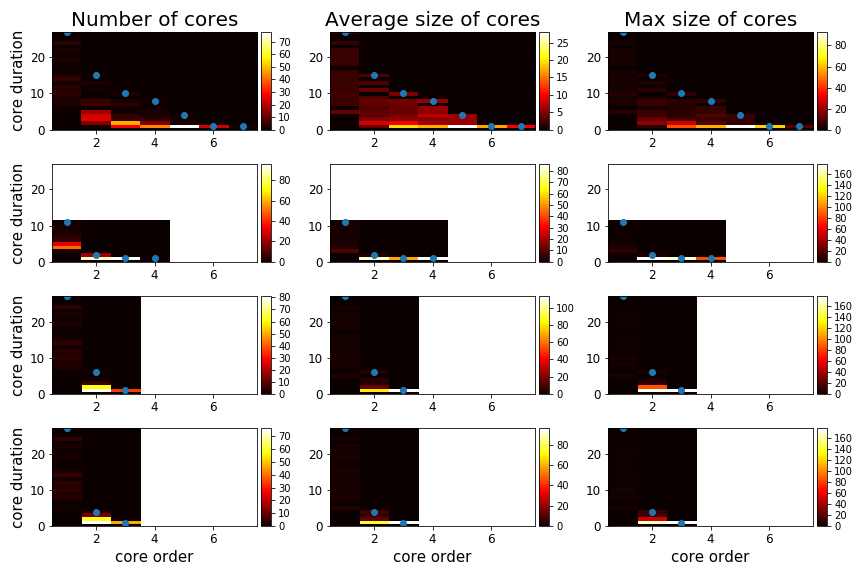}
\caption{Same as Figure \ref{fig:heatmaps_nullmodels_1}, for the primary school  data set.}
\label{fig:heatmaps_nullmodels_3}
\end{figure}

\begin{figure}[thb]
\centering   
 \includegraphics[width=.8\columnwidth]{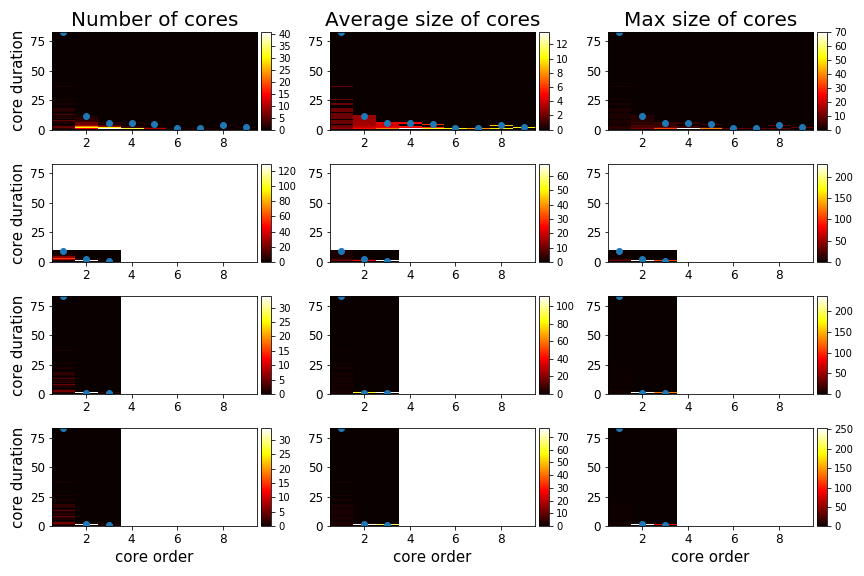}
    \caption{Same as Figure \ref{fig:heatmaps_nullmodels_1}, for the SFHH data set.}
     \label{fig:heatmaps_nullmodels_4}
\end{figure}

\begin{figure}[thb]
\centering   
 \includegraphics[width=.8\columnwidth]{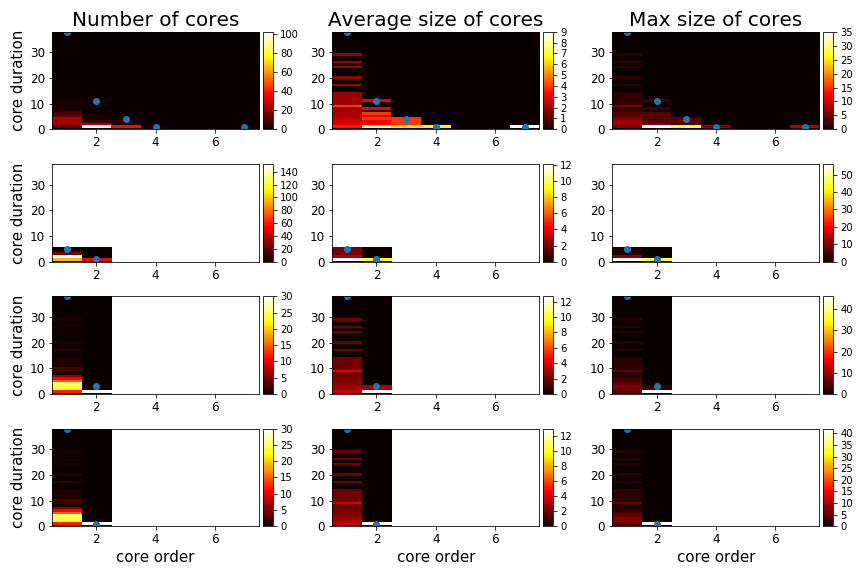}
    \caption{Same as Figure \ref{fig:heatmaps_nullmodels_1}, for the ACM Hypertext data set.}
     \label{fig:heatmaps_nullmodels_5}
\end{figure}

\begin{figure}[thb]
\centering   
 \includegraphics[width=.8\columnwidth]{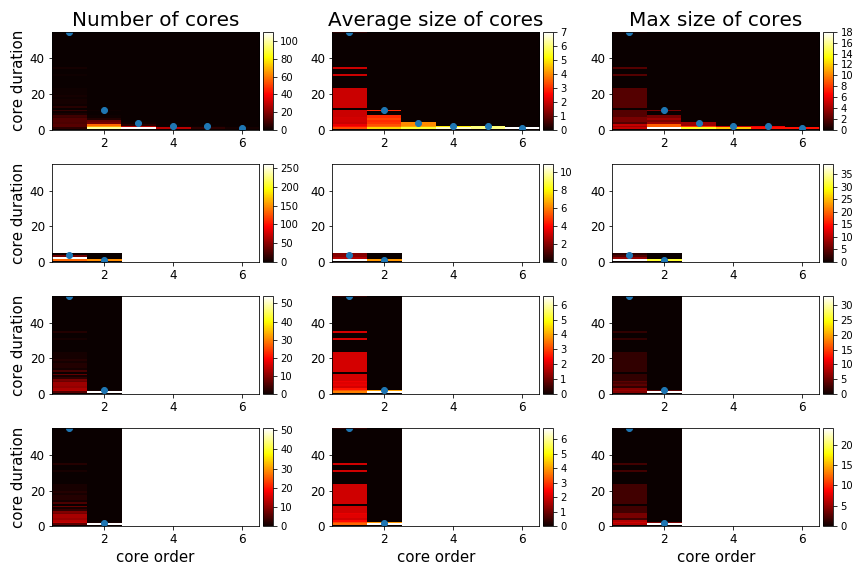}
    \caption{Same as Figure \ref{fig:heatmaps_nullmodels_1}, for the Hospital data set.}
     \label{fig:heatmaps_nullmodels_6}
\end{figure}

\begin{figure}[thb]
\centering   
 \includegraphics[width=.8\columnwidth]{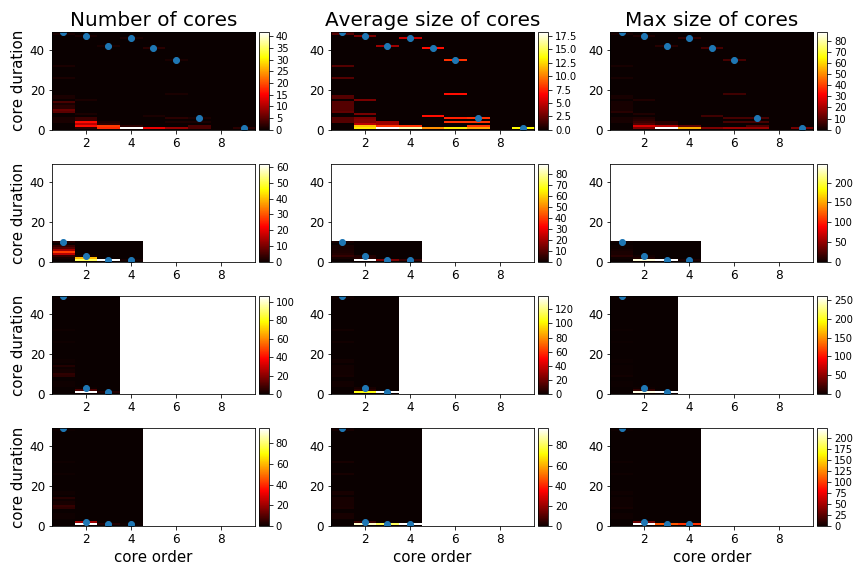}
    \caption{Same as Figure \ref{fig:heatmaps_nullmodels_1}, for the Elementary School  data set.}
     \label{fig:heatmaps_nullmodels_7}
\end{figure}

\begin{figure}[thb]
\centering   
 \includegraphics[width=.8\columnwidth]{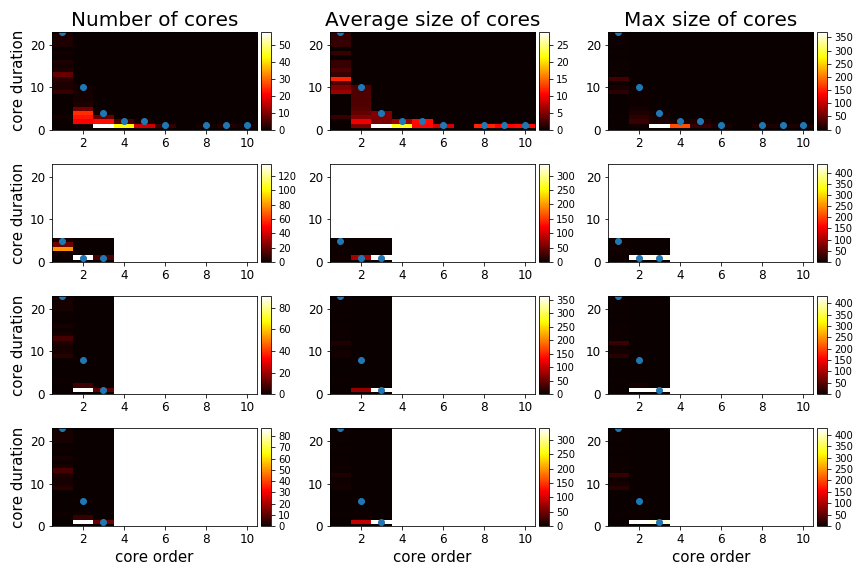}
    \caption{Same as Figure \ref{fig:heatmaps_nullmodels_1}, for the Middle School data set.}
     \label{fig:heatmaps_nullmodels_8}
\end{figure}

\clearpage
\newpage
    
 \begin{figure}[thb]
\centering
\includegraphics[width=.48\columnwidth]{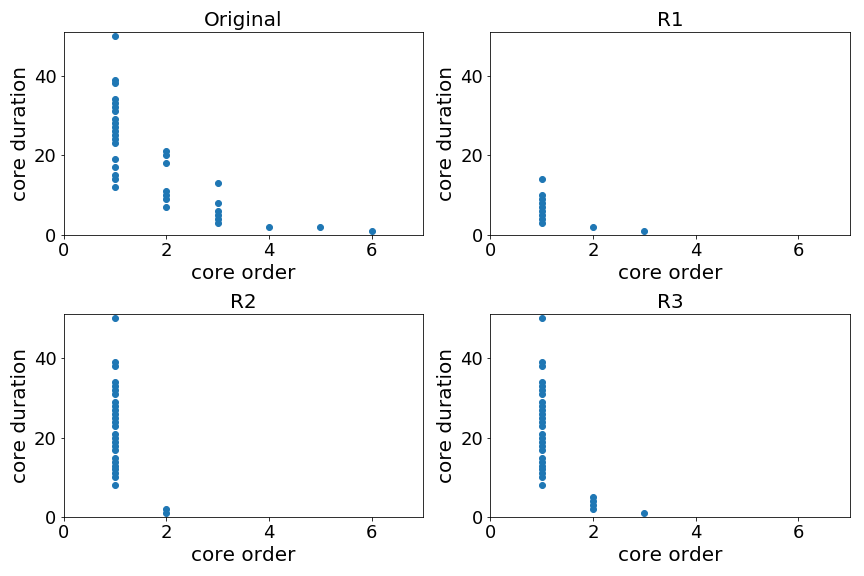}
\hspace{.01cm}
\includegraphics[width=.48\columnwidth]{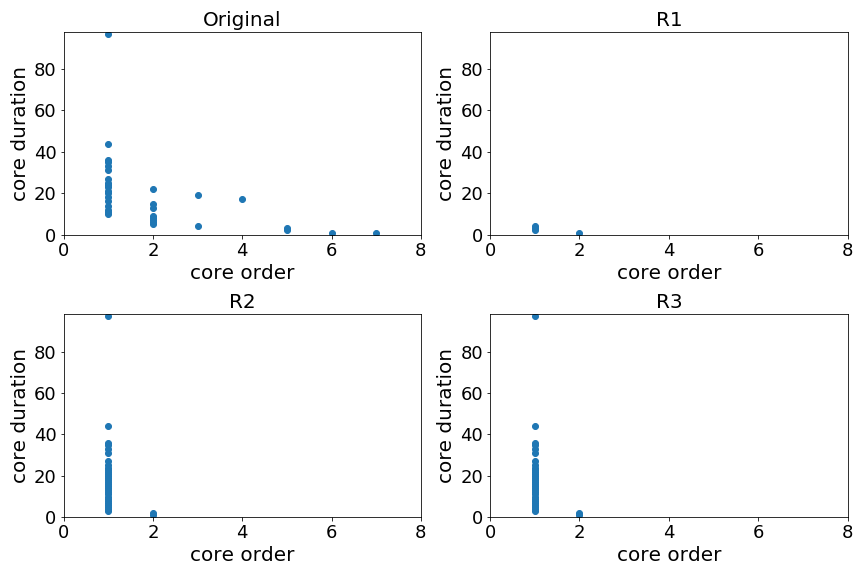}
\caption{Maximal values of the order and duration of the span-cores of the High school data set (left)
and of the Workplace data set (right).
Each blue dot gives either the
largest observed duration of maximal span-cores of a given order, or
the largest observed order of maximal span-cores of a given duration,
for the data set Primary School and for three reshuffled versions of the data set
(See Fig. \ref{fig:heatmaps_nullmodels_1} for the definition of the reshuffling procedures R1, R2, R3.)}
\label{fig:maxcoremaxduration_1}
 \end{figure}
  
  \begin{figure}[thb]
\centering
\includegraphics[width=.48\columnwidth]{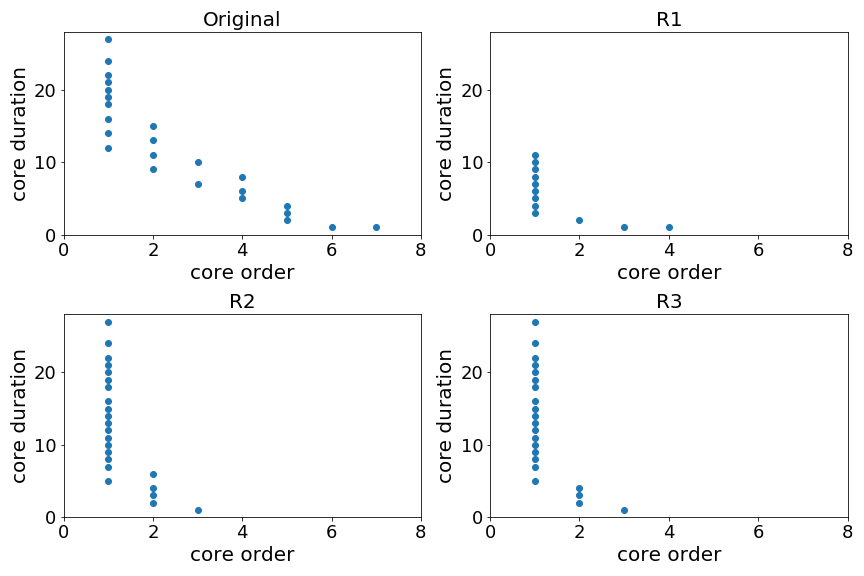}
\hspace{.01cm}
\includegraphics[width=.48\columnwidth]{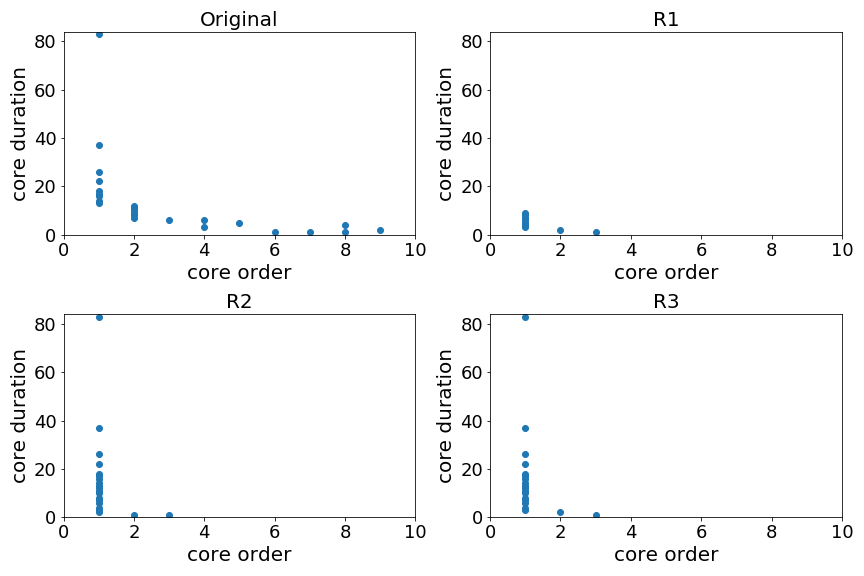}
\caption{Same as Figure \ref{fig:maxcoremaxduration_1}, for the Primary School data set (left) and
 the SFHH data set (right).}
\label{fig:maxcoremaxduration_2}
 \end{figure}
  
   \begin{figure}[thb]
\centering
\includegraphics[width=.48\columnwidth]{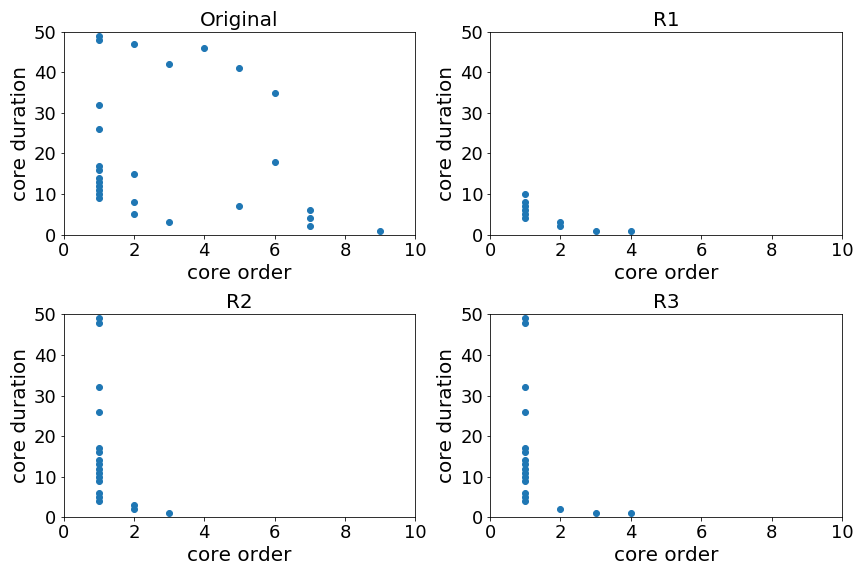}
\hspace{.01cm}
\includegraphics[width=.48\columnwidth]{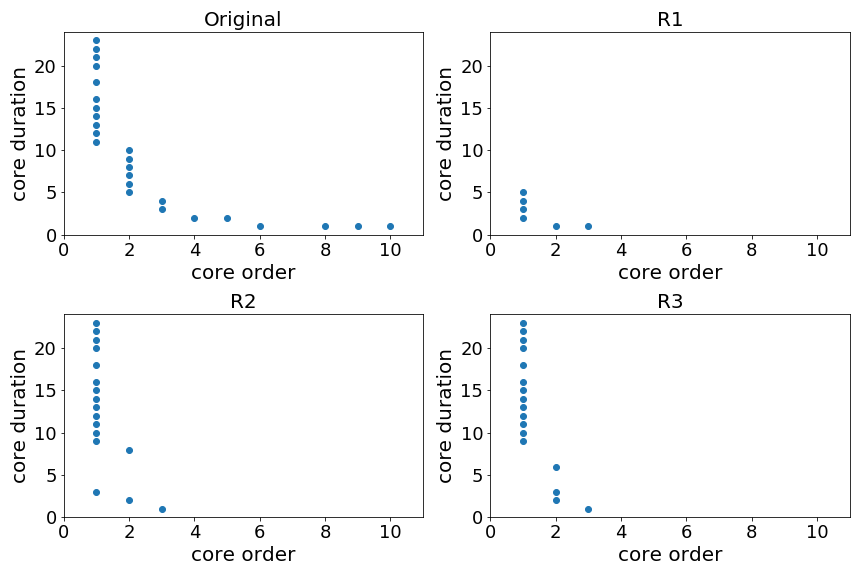}
\caption{Same as Figure \ref{fig:maxcoremaxduration_1}, for the Elementary School data set (left) and
 the Middle School data set (right).}
\label{fig:maxcoremaxduration_3}
 \end{figure}

     \begin{figure}[thb]
\centering
\includegraphics[width=.48\columnwidth]{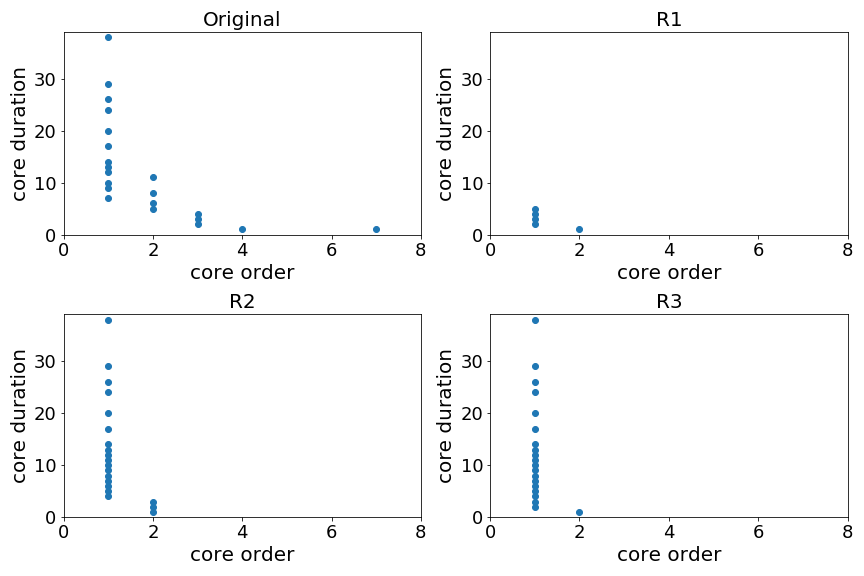}
\hspace{.01cm}
\includegraphics[width=.48\columnwidth]{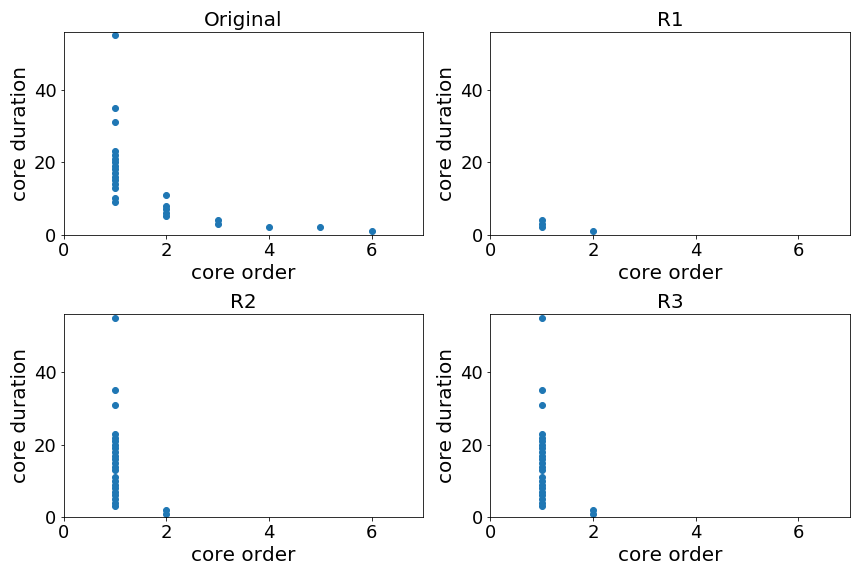}
\caption{Same as Figure \ref{fig:maxcoremaxduration_1}, for the ACM Hypertext data set (left) and
 the Hospital data set (right).}
\label{fig:maxcoremaxduration_4}
 \end{figure}

\clearpage
\newpage
 
 \section{Static vs. dynamic coreness based centrality measures}
 
\begin{figure}[thb]
\centering
\includegraphics[width=.48\columnwidth]{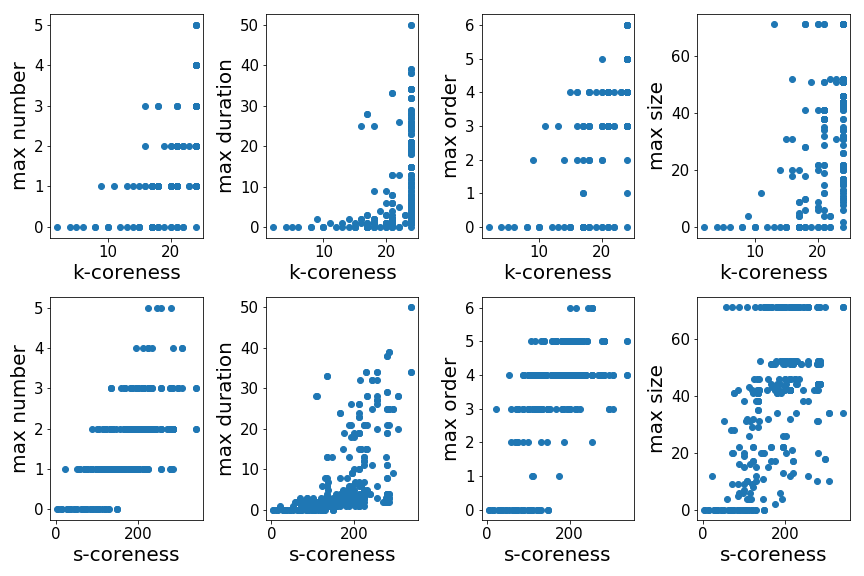}
\hspace{.1cm}
\includegraphics[width=.48\columnwidth]{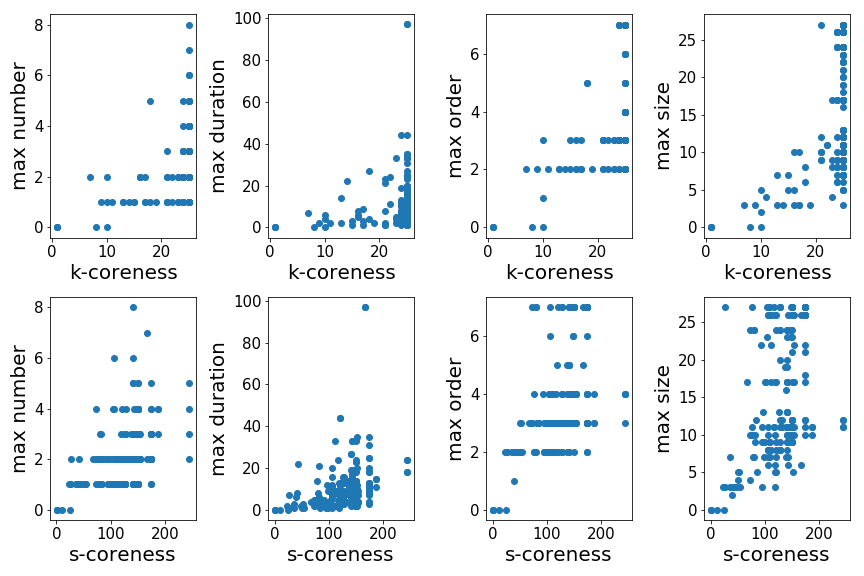}

\caption{Scatterplot of node metrics related to the span-cores it belongs to vs. static metrics, for the
High School data set (left) and the Workplace data set (right). The static metrics are
given for the top and bottom row by respectively unweighted and weighted coreness.
Each blue dot represents a node.
The span-cores-related metrics are obtained, for each node, by following over time the number,
durations, orders and sizes of the maximal span-cores it belongs to at each time step, and 
taking the maximal encountered value of these quantities in the whole timeline.}
\label{fig:scatterplots_1}
 \end{figure}

 \begin{figure}[thb]
\centering
\includegraphics[width=.48\columnwidth]{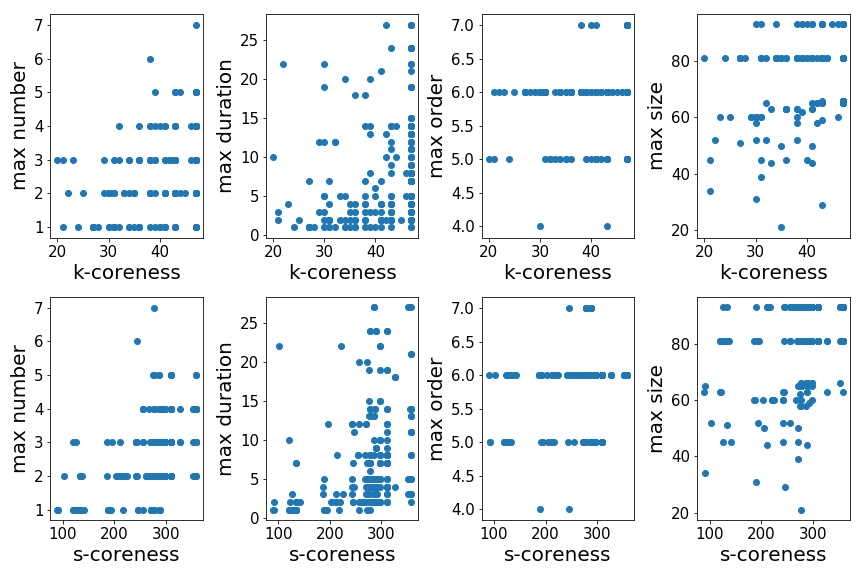}
\hspace{.1cm}
\includegraphics[width=.48\columnwidth]{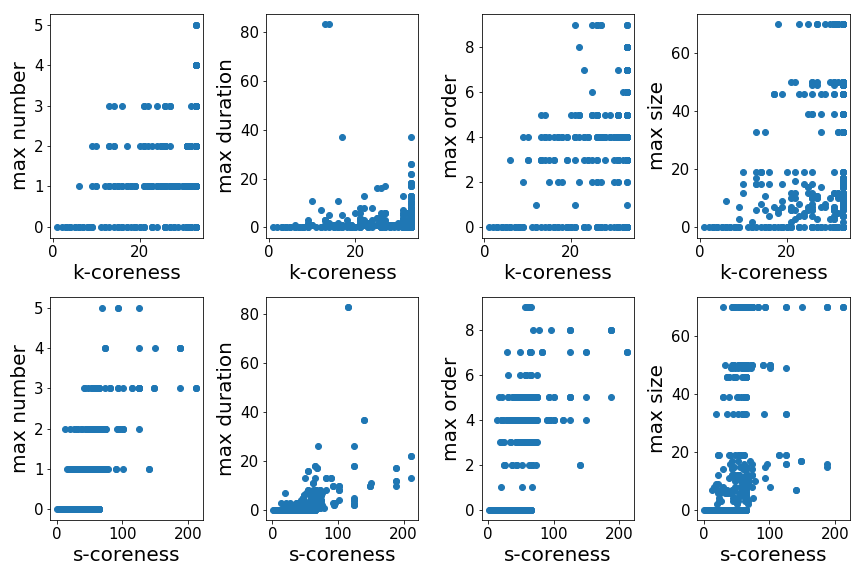}
\caption{Same as Figure \ref{fig:scatterplots_1}, for the Primary School data set (left) and the SFHH data set (right).}
\label{fig:scatterplots_2}
 \end{figure}
 
 \begin{figure}[thb]
\centering
\includegraphics[width=.48\columnwidth]{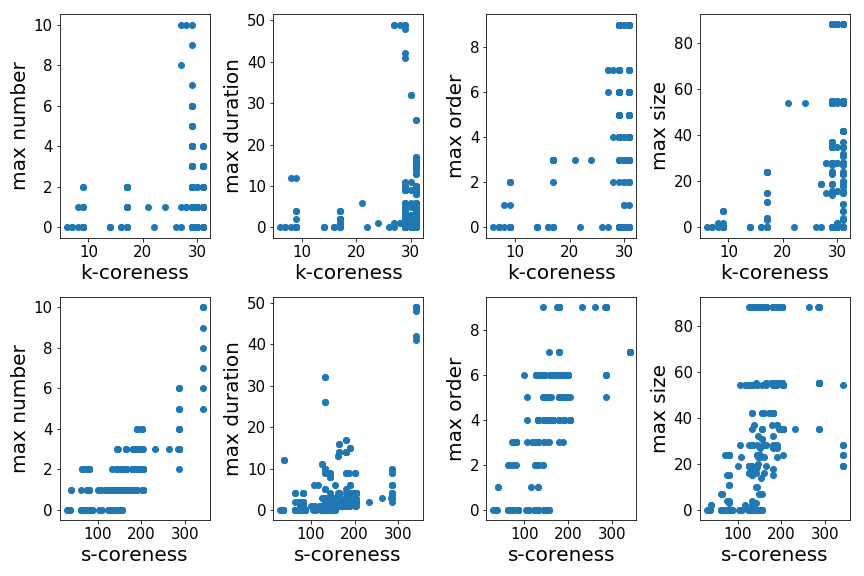}
\hspace{.1cm}
\includegraphics[width=.48\columnwidth]{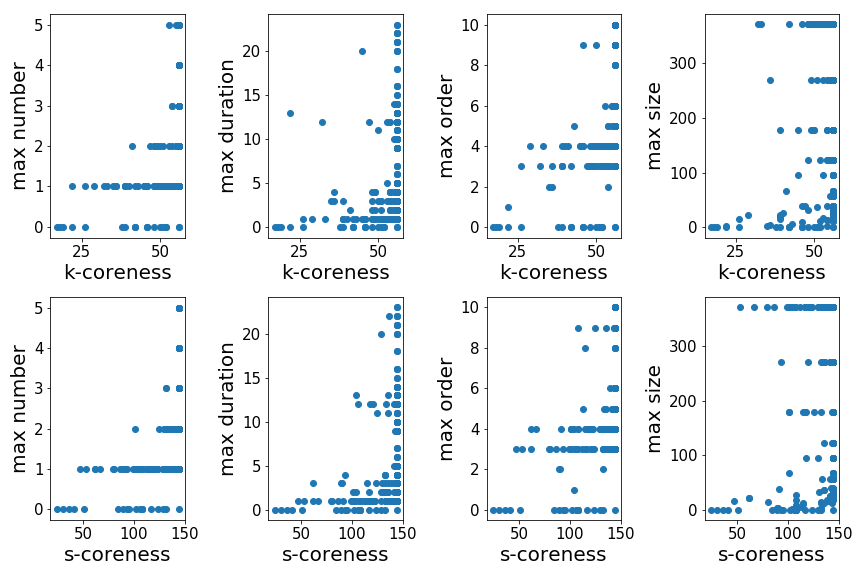}
\caption{Same as Figure \ref{fig:scatterplots_1}, for the Elementary School data set (left) and
 the Middle School data set (right).}
\label{fig:scatterplots_3}
 \end{figure}

 \begin{figure}[thb]
\centering
\includegraphics[width=.48\columnwidth]{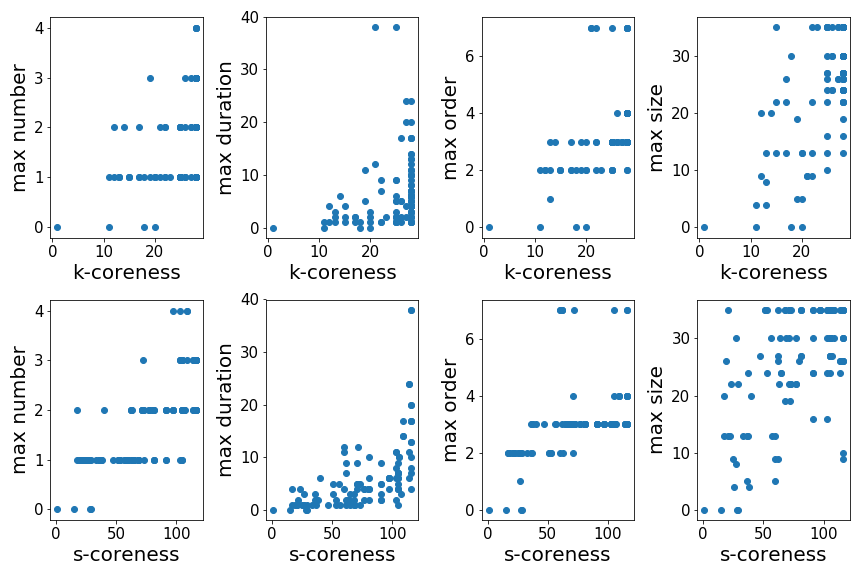}
\hspace{.1cm}
\includegraphics[width=.48\columnwidth]{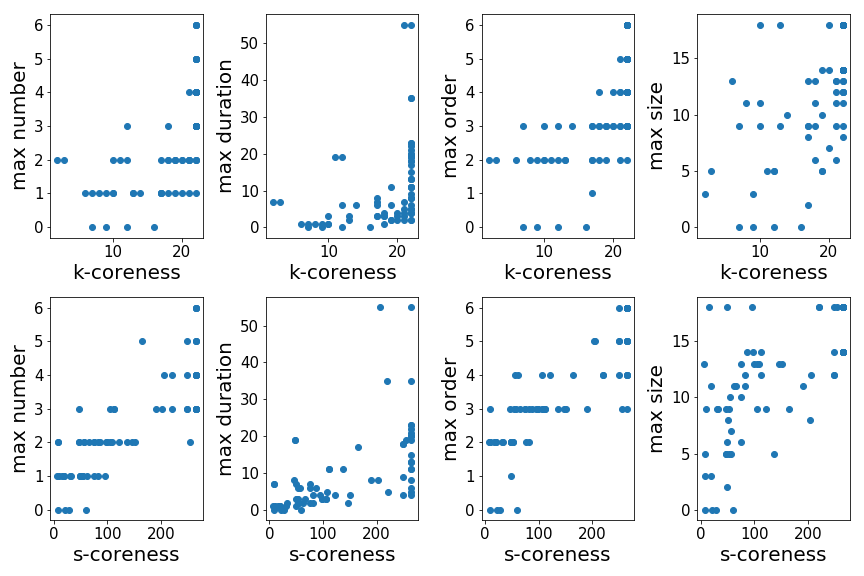}
\caption{Same as Figure \ref{fig:scatterplots_1}, for the ACM Hypertext data set (left) and
 the Hospital data set (right).}
\label{fig:scatterplots_4}
 \end{figure}

\clearpage
\newpage

\section{Properties of the (maximal) span-cores}

% I (martino) will double check this table 

\begin{table}[ht]
	\centering
	\begin{tabular}{|p{4cm}|p{2cm}|p{2cm}|p{2cm}|p{2cm}|p{2cm}|}
		\hline
		Dataset  and reference  & number of span-cores & number of maximal span-cores & average size & average duration & average order  \\
		\hline
		Primary school & 4715 & 409 & 12.74 & 3.67 & 3.31 \\
		\hline
		High school & 12514 & 456 & 6.73 & 4.91 & 2.53   \\
		\hline
		Middle school & 3024 & 281 & 13.12 & 3.47 & 2.88  \\
		\hline
		Elementary School & 4334 & 212 &  9.83 & 5 & 3.29 \\
		\hline
		SFHH Conference & 6636 & 283 & 5.86 & 3.66 & 2.71   \\
		\hline
		ACM Hypertext  & 3881 & 326 & 4.84 & 3.03 & 1.76  \\
		\hline
		Workplace & 16496 & 788 & 3.85 & 3.74 & 1.84 \\
		\hline
		Hospital & 8218 & 568 &  4.1 & 3.16 & 2.20  \\ 
		\hline
	\end{tabular}
	\caption{\label{tab:dataset_cores} Information concerning the temporal core decomposition of the datasets. The last three columns (average size, average duration and average order) are referred to the maximal span-cores only. 
	}
\end{table}

%\newpage

\section{Properties of the targeted maximal-span cores}

\begin{table}[htb]
\centering
\begin{tabular}{|p{3cm}|p{5cm}|p{1.5cm}|p{1.5cm}|p{1.5cm}|p{1.5cm}|p{1.5cm}|}
\hline
Data set & strategy & $n_{msc}$  & $\langle k \rangle$   & $\langle |\Delta| \rangle$  & $\langle n \rangle$ &  $\langle e \rangle$ \\
\hline
%Primary School & first cores in chronological order & 170 & 3.27 & 4.07 & 12.51 & 99.03  \\
\hline
Primary School &  largest order cores & 100 & 5.25 & 1.07 & 21.71 & 112.46 \\
\hline
Primary School & largest duration cores & 167 & 2.18 & 6.15 & 3.86 & 92.48 \\
\hline
%Middle School & first cores in chronological order & 231 & 2.84 & 3.52 & 13.55 & 71.35 \\
\hline
Middle School & largest order cores & 241 &  3.17 & 2.07 & 14.29 & 66.61 \\
\hline
Middle School & largest duration cores & 275 & 2.89 & 3.64 & 10.29 & 62.05 \\
\hline
%Elementary School & first cores in chronological order & 189 & 3.19 & 5.14 & 9.56 &  100.81  \\
\hline
Elementary School & largest order cores & 181 & 3.66 & 3.51 & 11.022 & 99.46 \\
\hline
Elementary School & largest duration cores & 191 & 3.32 & 5.45 & 8.76 &  104.95 \\
\hline
%ACM HT Conference & first cores in chronological order & 151 & 1.68 & 2.79  & 4.33 & 15.58  \\
\hline
ACM HT Conference & largest order cores &  110 & 2.5 & 1.92  & 4.44 &  17.07 \\
\hline
ACM HT Conference & largest duration cores & 70 & 1.17 & 8.10 & 2.32 & 26.32 \\
\hline
%SFHH Conference & first cores in chronological order & 132 & 2.97 & 3.99 & 6.47 & 100.8 \\
\hline
SFHH Conference & largest order cores & 90  & 4.05 & 1.92 & 8.92 & 99.46 \\
\hline
SFHH Conference & largest duration cores & 127 & 2.25  & 6.53 & 3.43  &  104.95 \\
\hline
%Hospital & first cores in chronological order  & 166 & 2.05 & 3.23 & 3.78 &  20.53  \\
\hline
Hospital & largest order cores & 101 & 3.49 & 1.42 & 5.07 &   23.61 \\
\hline
Hospital & largest duration cores & 15 & 1 & 24.26 & 2 &  127.40  \\
\hline
\end{tabular}
\caption{\label{tab:core_properties_interventions}
Basic properties of the maximal span-cores removed in each of the targeted strategies ($f=20\%$).
$n_{msc}$: number of maximal span-cores with all temporal edges removed;
$\langle k \rangle$: average order of the targeted maximal span-cores;
$\langle |\Delta| \rangle$: average duration of the targeted maximal span-cores;
$\langle n \rangle$: average number of nodes in the targeted maximal span-cores;
$\langle e \rangle$: average number of temporal edges removed per time step impacted by the strategy.}
\end{table}

 \clearpage
 \newpage
 
 \section{SIS results for all data sets}

% \vspace{-2mm}
 
 \begin{figure}[hb] \label{fig:th_80}
	\centerline{
		\begin{tabular}{cc}
			\includegraphics[width=0.32\textwidth, height = 0.32\textheight, keepaspectratio]{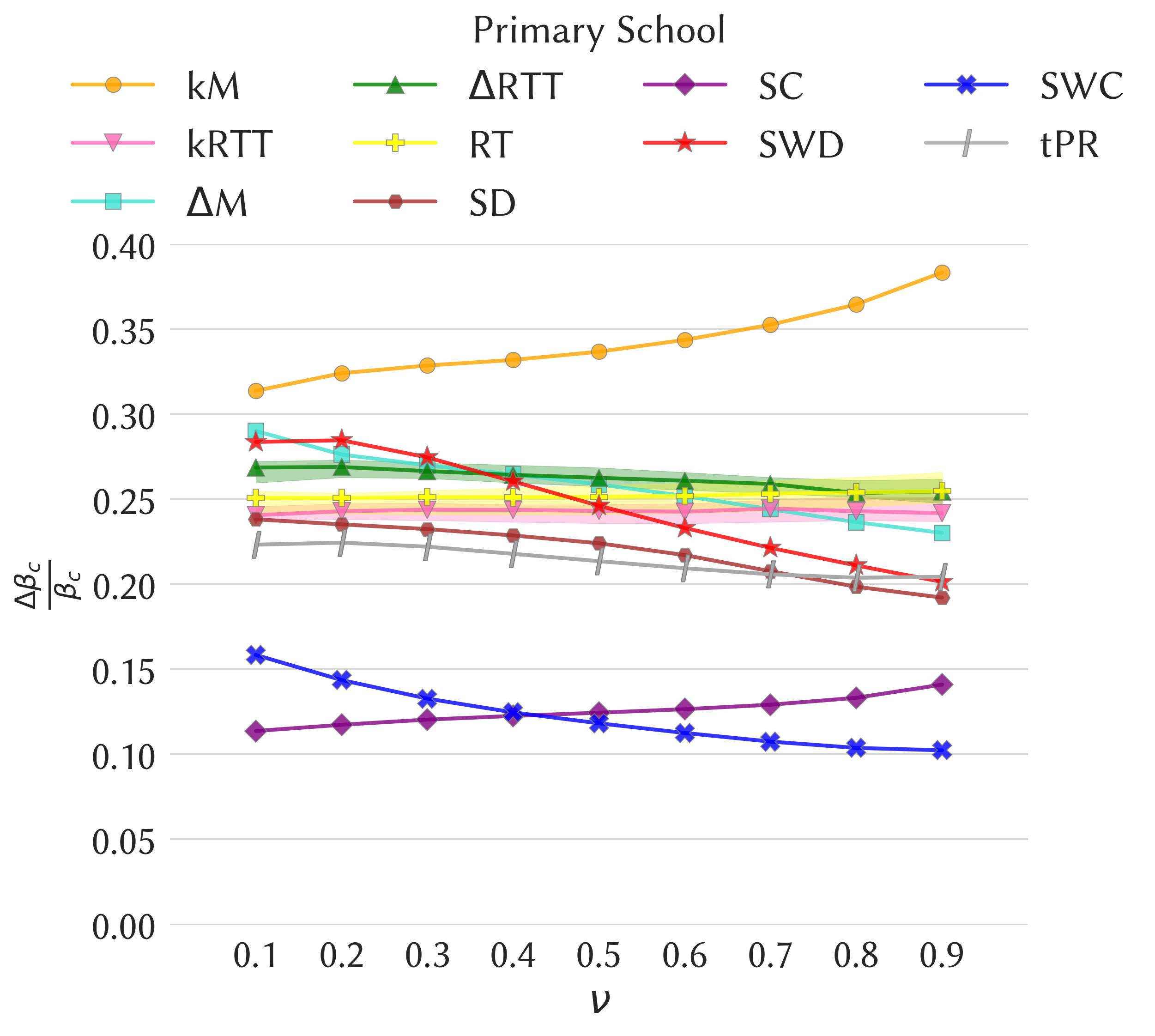} & 
			\includegraphics[width=0.32\textwidth, height = 0.32\textheight, keepaspectratio]{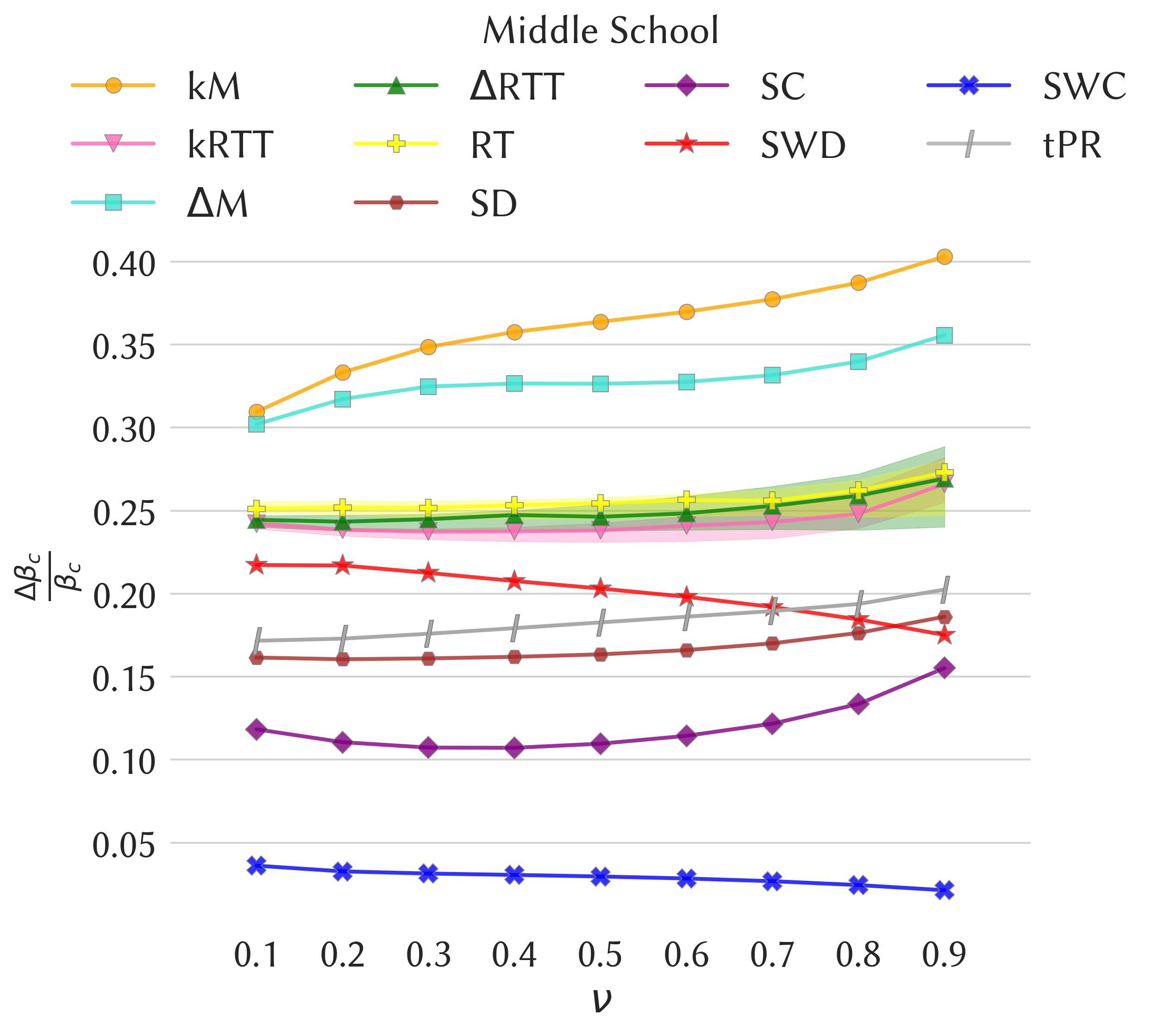} 
			\vspace{-1mm}\\
			\includegraphics[width=0.32\textwidth, height = 0.32\textheight, keepaspectratio]{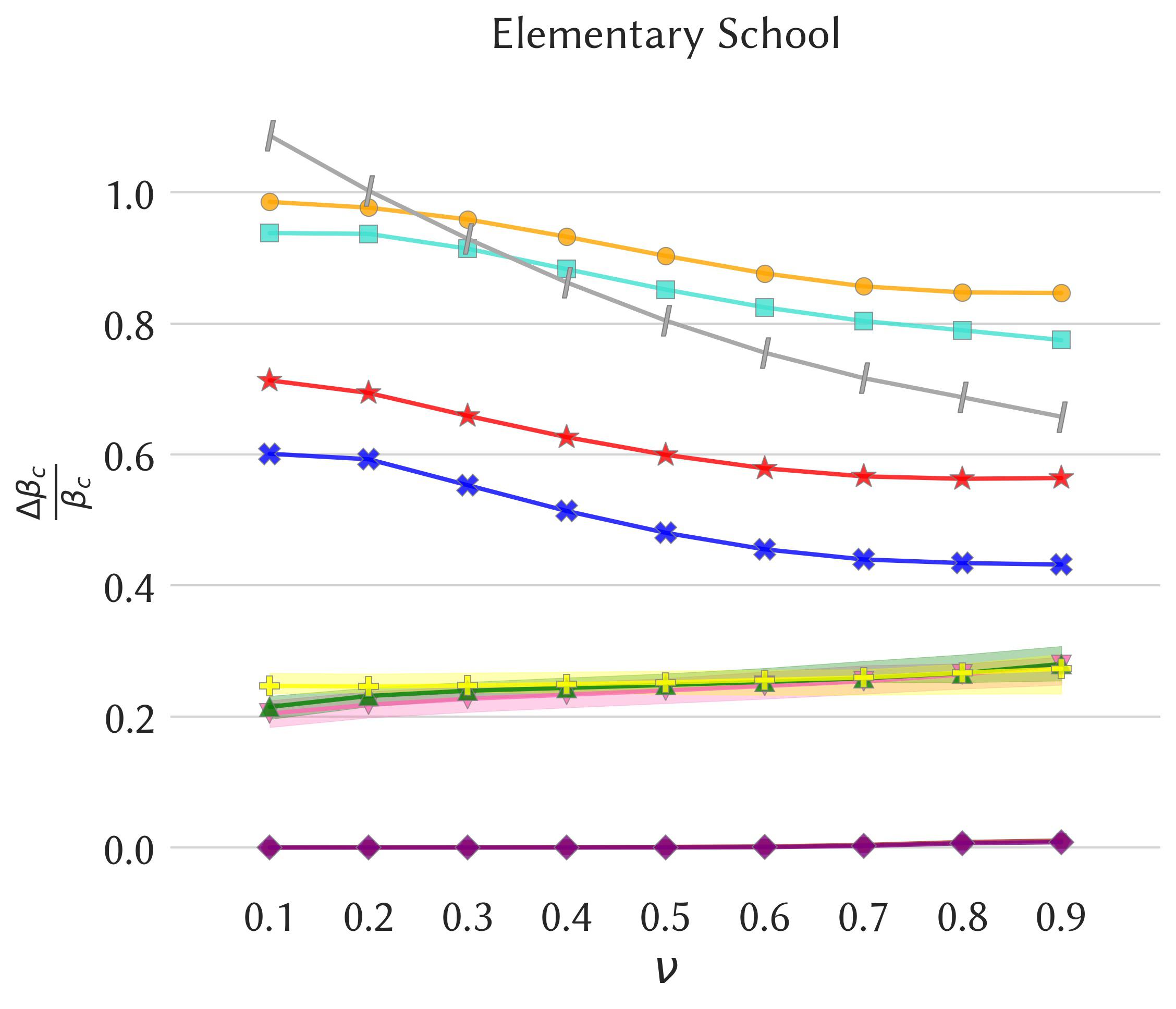} 
			&
			\includegraphics[width=0.32\textwidth, height = 0.32\textheight, keepaspectratio]{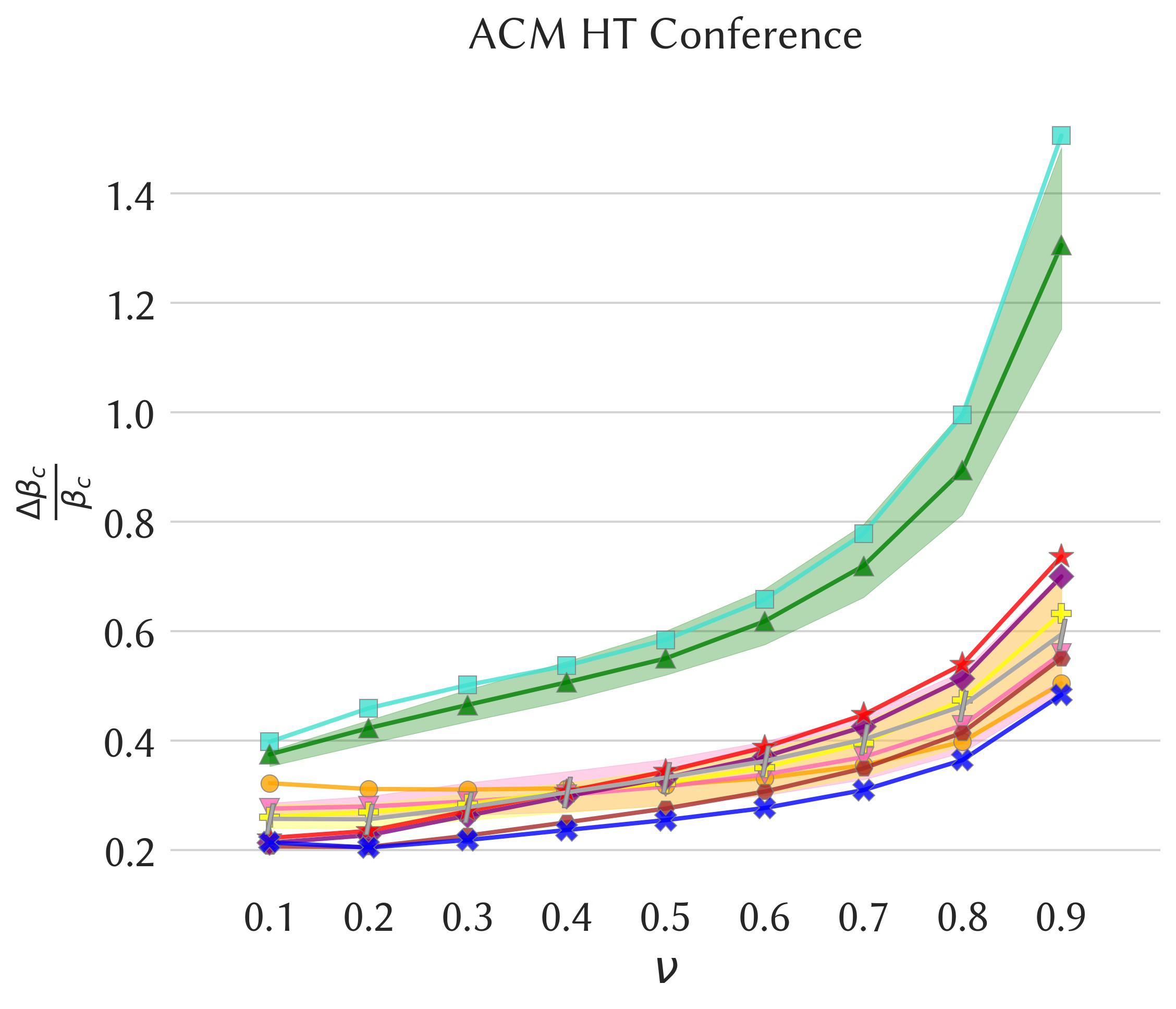} 
			\vspace{-1mm}\\
\includegraphics[width=0.32\textwidth, height = 0.32\textheight,keepaspectratio]{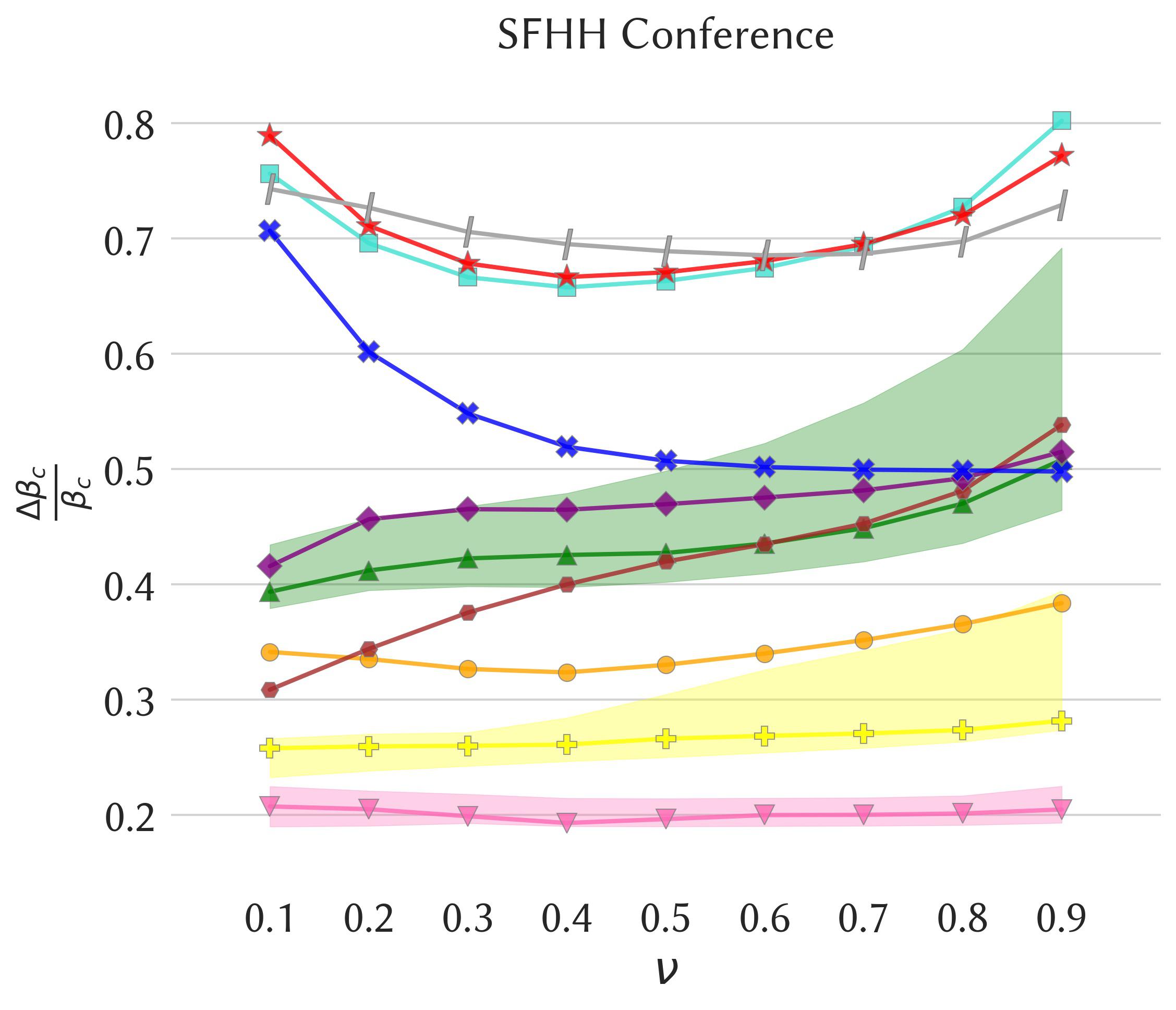}
			&
\includegraphics[width=0.32\textwidth, height = 0.32\textheight, keepaspectratio]{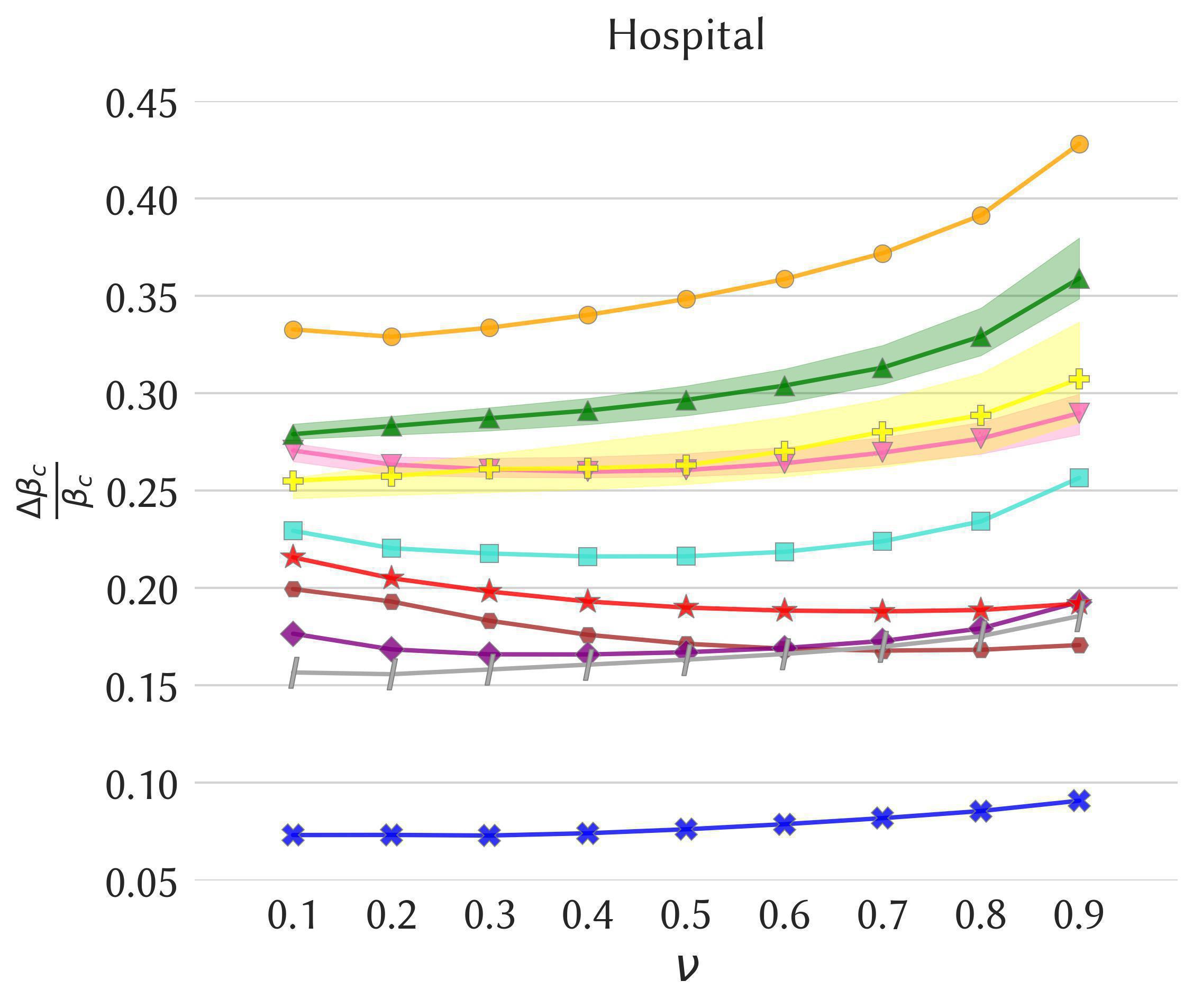} 
			\vspace{-2mm}\\
		\end{tabular}
	}
	\vspace{-2mm}
	\caption{Impact of the intervention strategies as measured by the change in the epidemic threshold of SIS processes  ($f=20\%$).
    In each panel we plot for the various strategies the relative change $\Delta \beta_c / \beta_c$
    in the epidemic threshold as a function of the recovery rate $\nu$. For each strategy
    based on random choices, we show the confidence
    interval (computed using $30$ samples) between the $5^{th}$ and $95^{th}$ percentiles as shaded areas.   }
\end{figure}

 \clearpage
 \newpage
 
 \section{SIS results for $f=10\%$}
 
 \begin{figure}[!h] \label{fig:th_90}
	\centerline{
		\begin{tabular}{cc}
			\includegraphics[width=0.25\textwidth, height = .25\textheight, keepaspectratio]{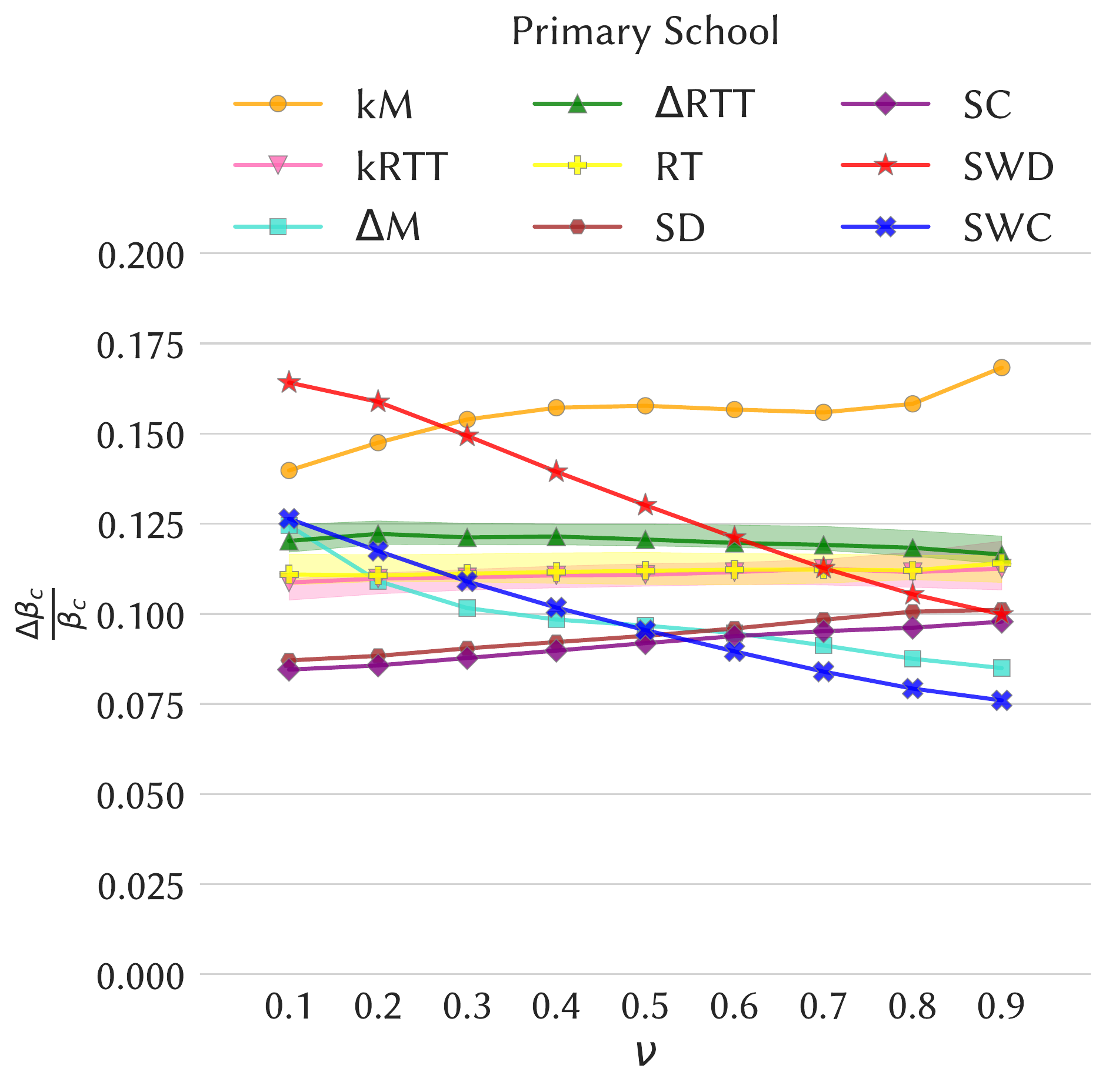} & 
			\includegraphics[width=.25\textwidth, height = .25\textheight, keepaspectratio]{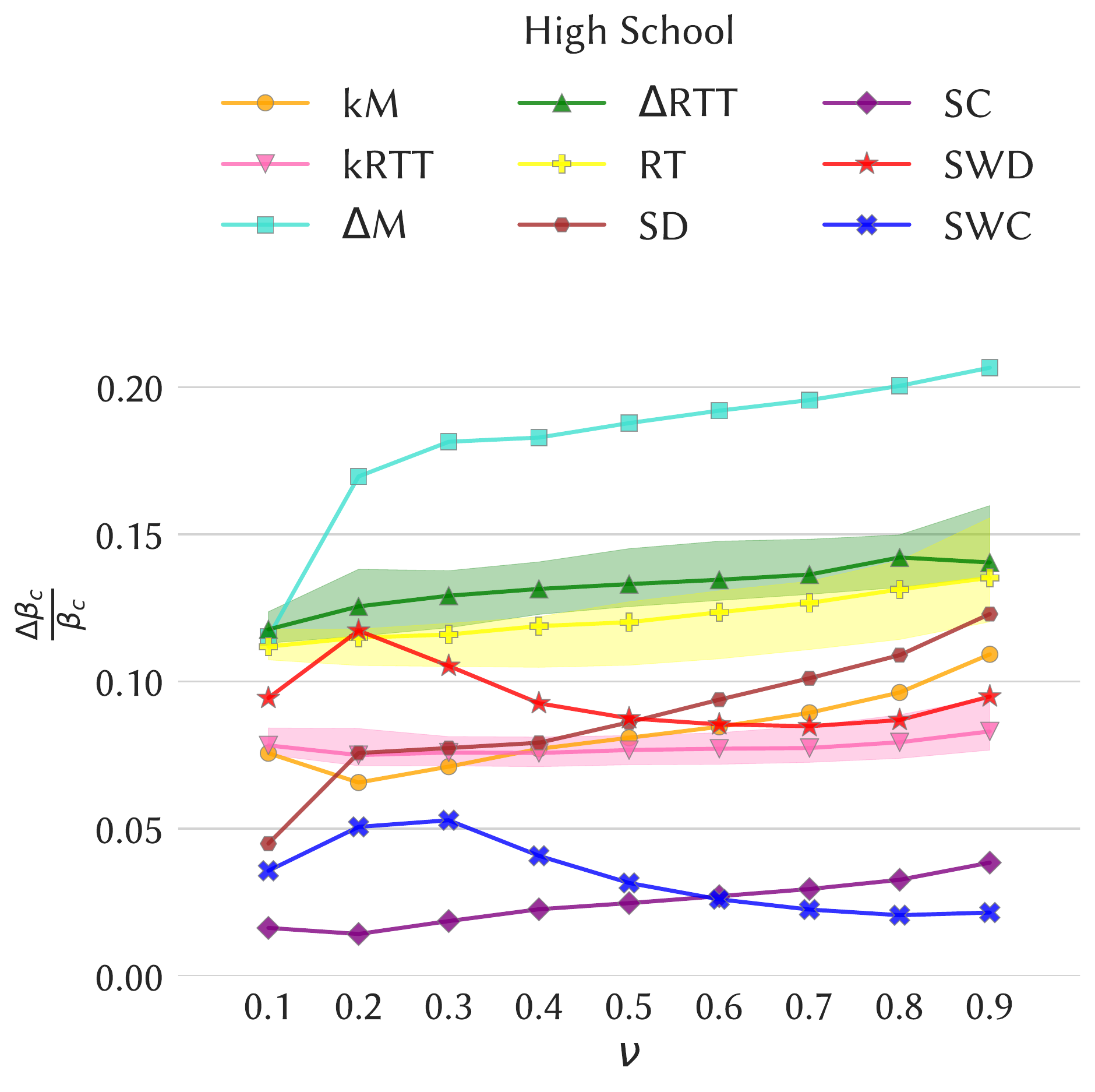} 
			\vspace{-1mm}\\
			\includegraphics[width=0.25\textwidth, height =0.25\textheight, keepaspectratio]{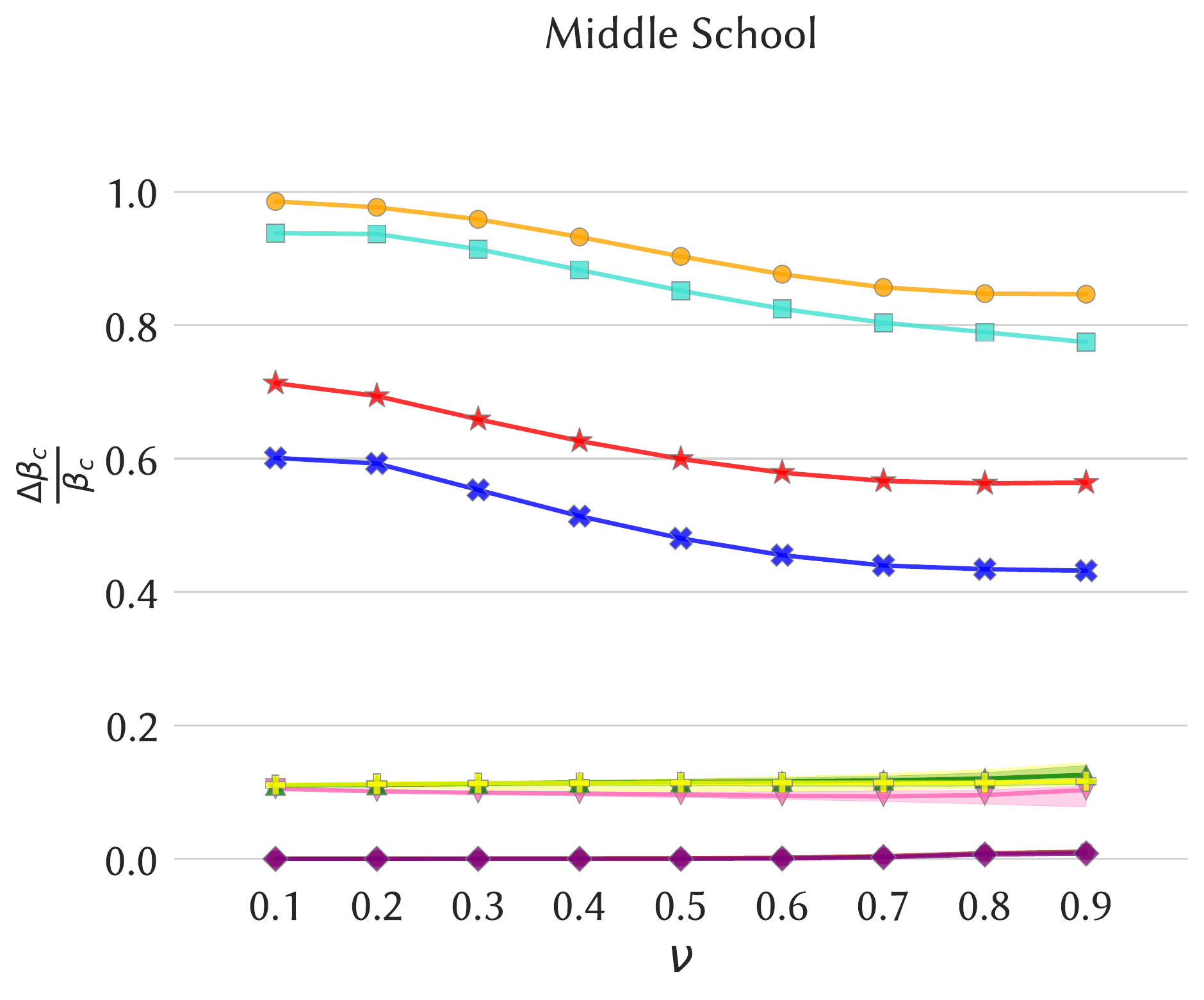} 
			&
			\includegraphics[width=0.25\textwidth, height = 0.25\textheight, keepaspectratio]{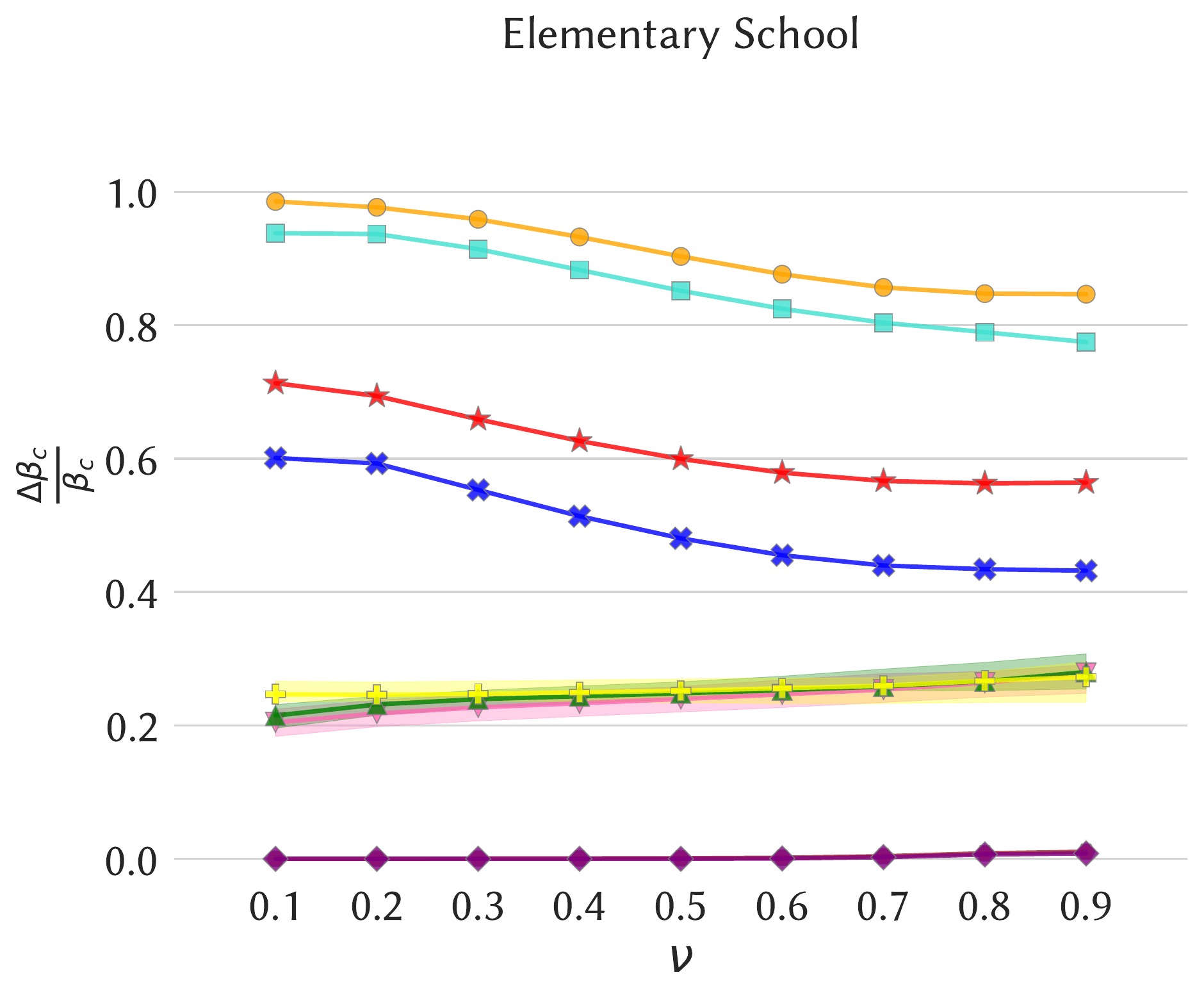} 
			\vspace{-1mm}\\
			\includegraphics[width=0.25\textwidth, height = 0.25\textheight, keepaspectratio]{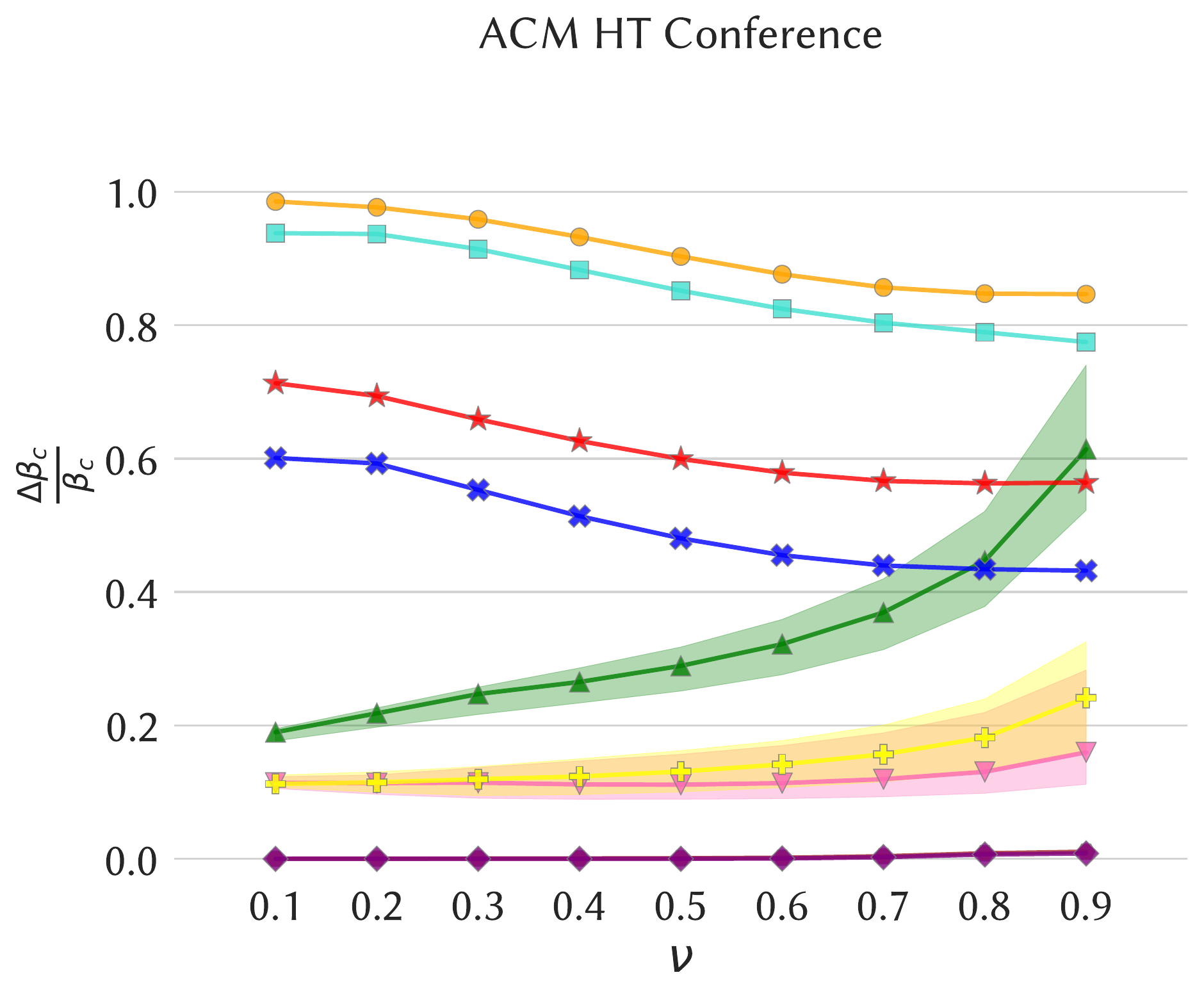} 
			&
			\includegraphics[width= 0.25\textwidth, height = 0.25\textheight, keepaspectratio]{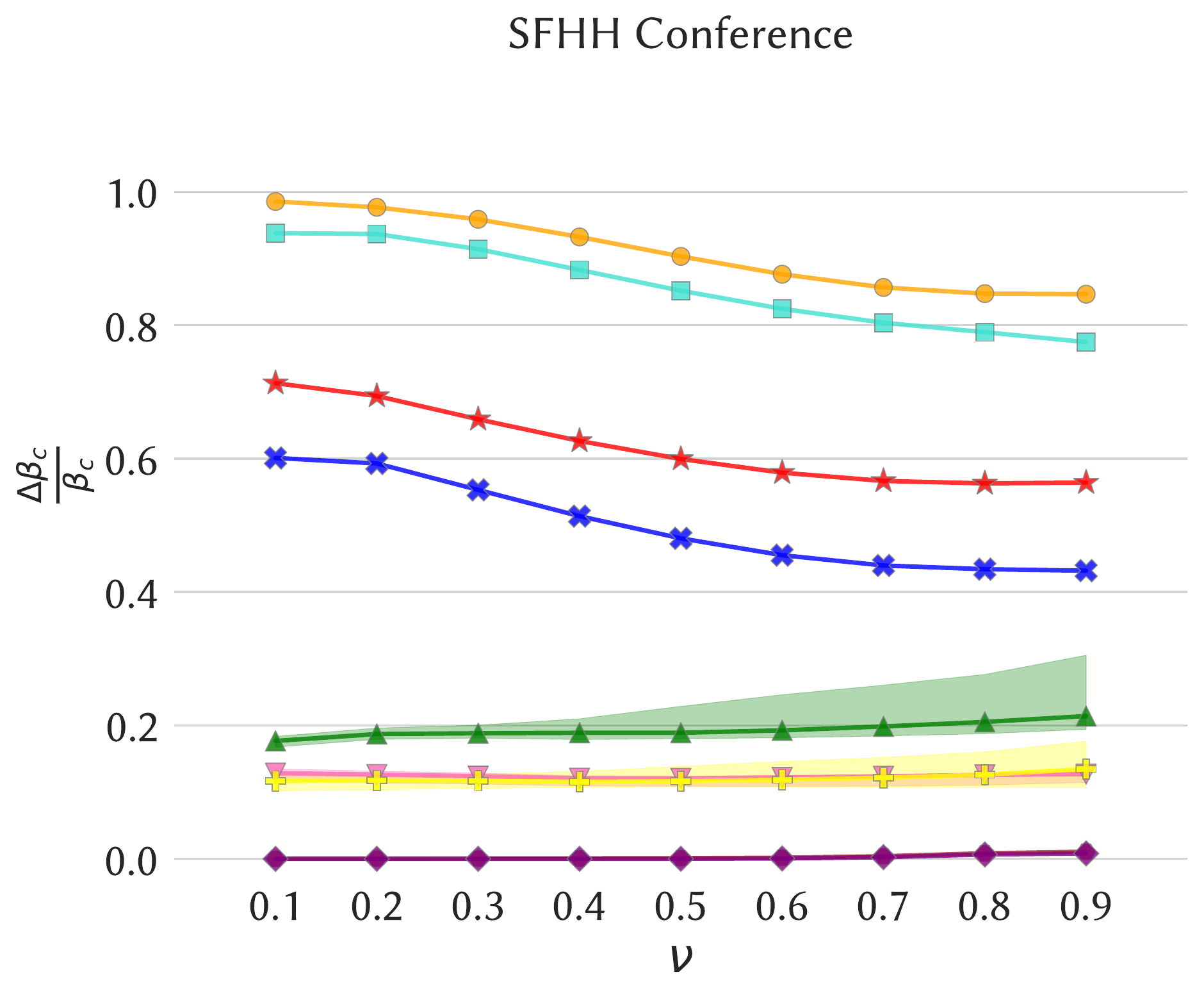}
			\vspace{-1mm}\\
			\includegraphics[width=0.25\textwidth, height = 0.25\textheight, keepaspectratio]{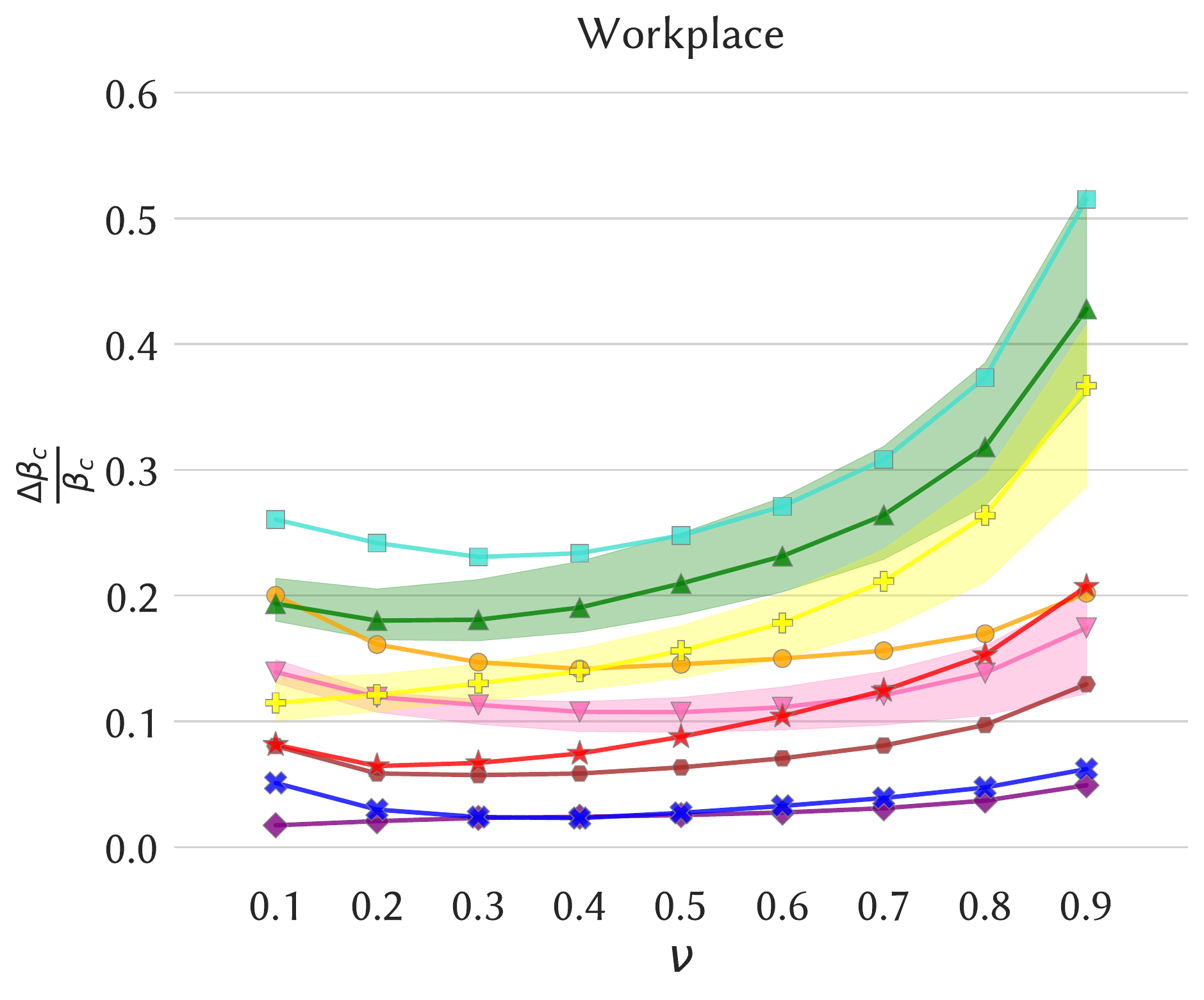} &
			\includegraphics[width=0.25\textwidth, height = 0.25\textheight, keepaspectratio]{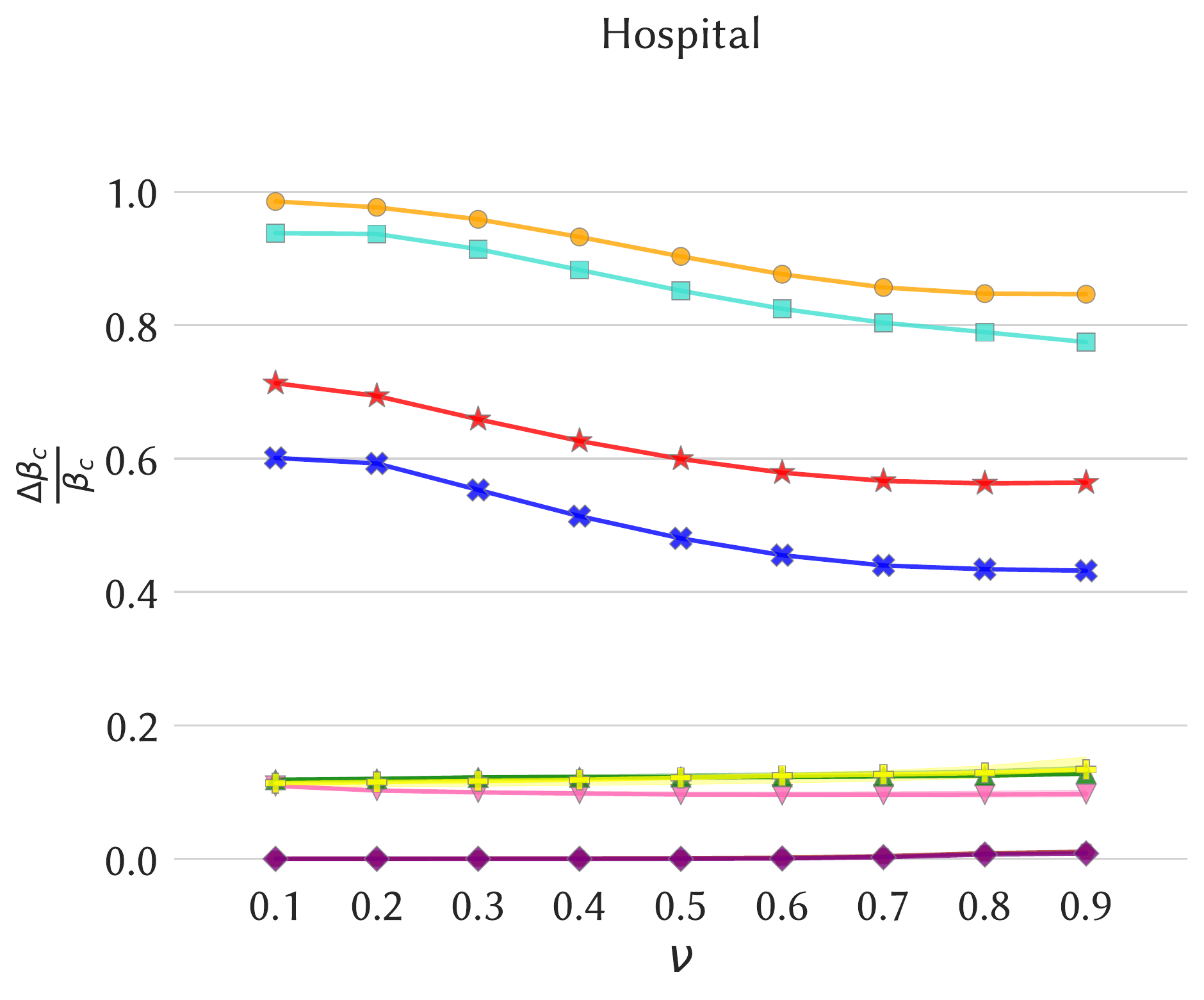} 
			\vspace{-2mm}\\
		\end{tabular}
	}
%	\vspace{1mm}
	\caption{\small Impact of the various intervention strategies as measured by the relative change $\Delta \beta_c / \beta_c$ in the epidemic threshold of SIS processes, for a fraction $f=10\%$ of temporal edges removed.
 %   In each panel we plot for the various strategies the relative change $\Delta \beta_c / \beta_c$  in the epidemic threshold as a function of the recovery rate $\nu$.
    For each strategy
    based on random choices, we show the confidence
    interval (computed using $30$ samples) between the $5^{th}$ and $95^{th}$ percentiles as shaded areas.   }
\end{figure}

\clearpage
\newpage

\section{Spread mitigation results for the SIR process}

\begin{figure}[!h]
	\centerline{
		\begin{tabular}{cc}
			\includegraphics[width=0.25\textwidth, height = .25\textheight, keepaspectratio]{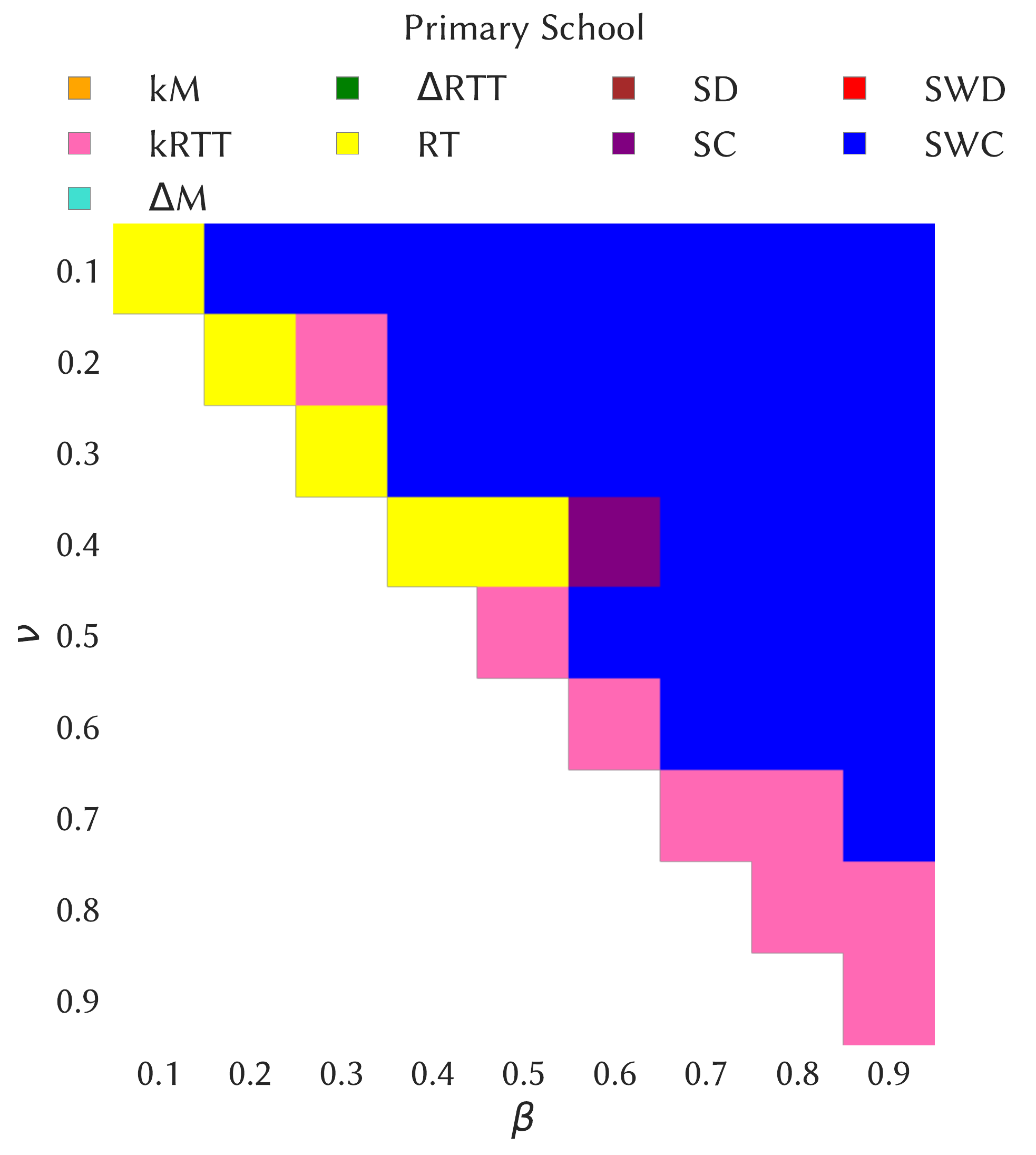} & 
			\includegraphics[width=0.25\textwidth, height = .25\textheight, keepaspectratio]{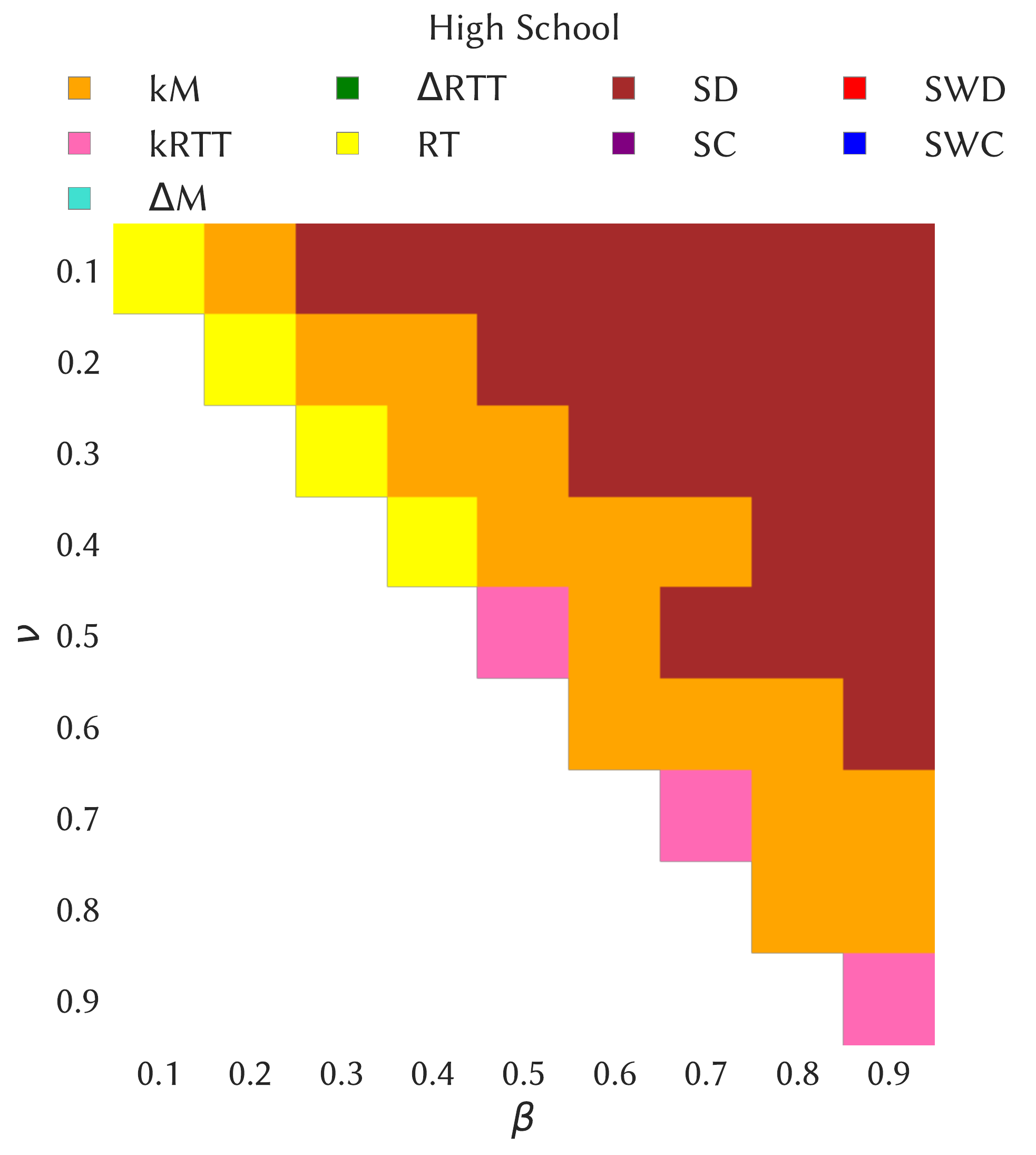} 
			\vspace{-1mm}\\
			\includegraphics[width=0.25\textwidth, height = .25\textheight, keepaspectratio]{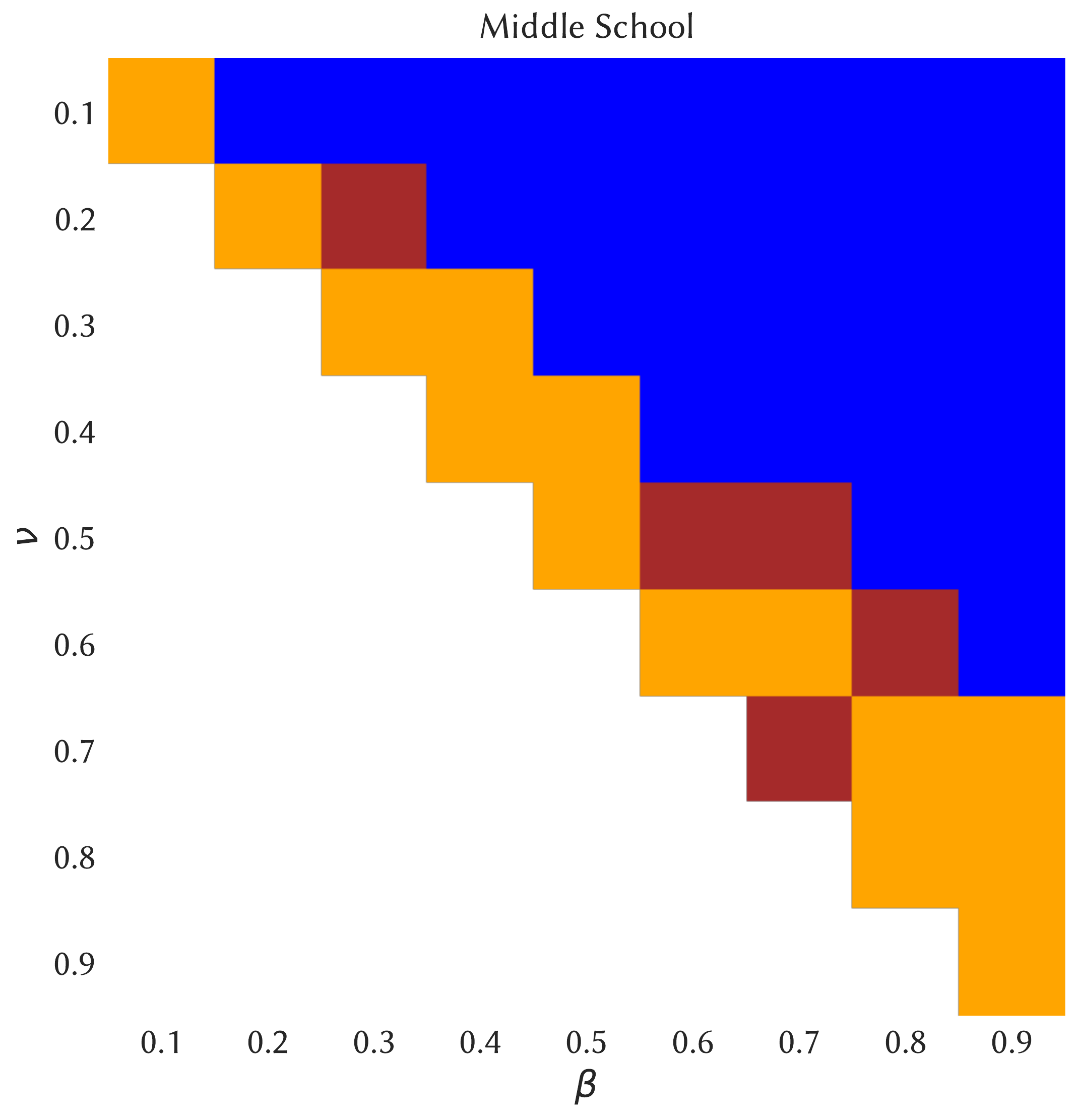} & 
			\includegraphics[width=0.25\textwidth, height = .25\textheight, keepaspectratio]{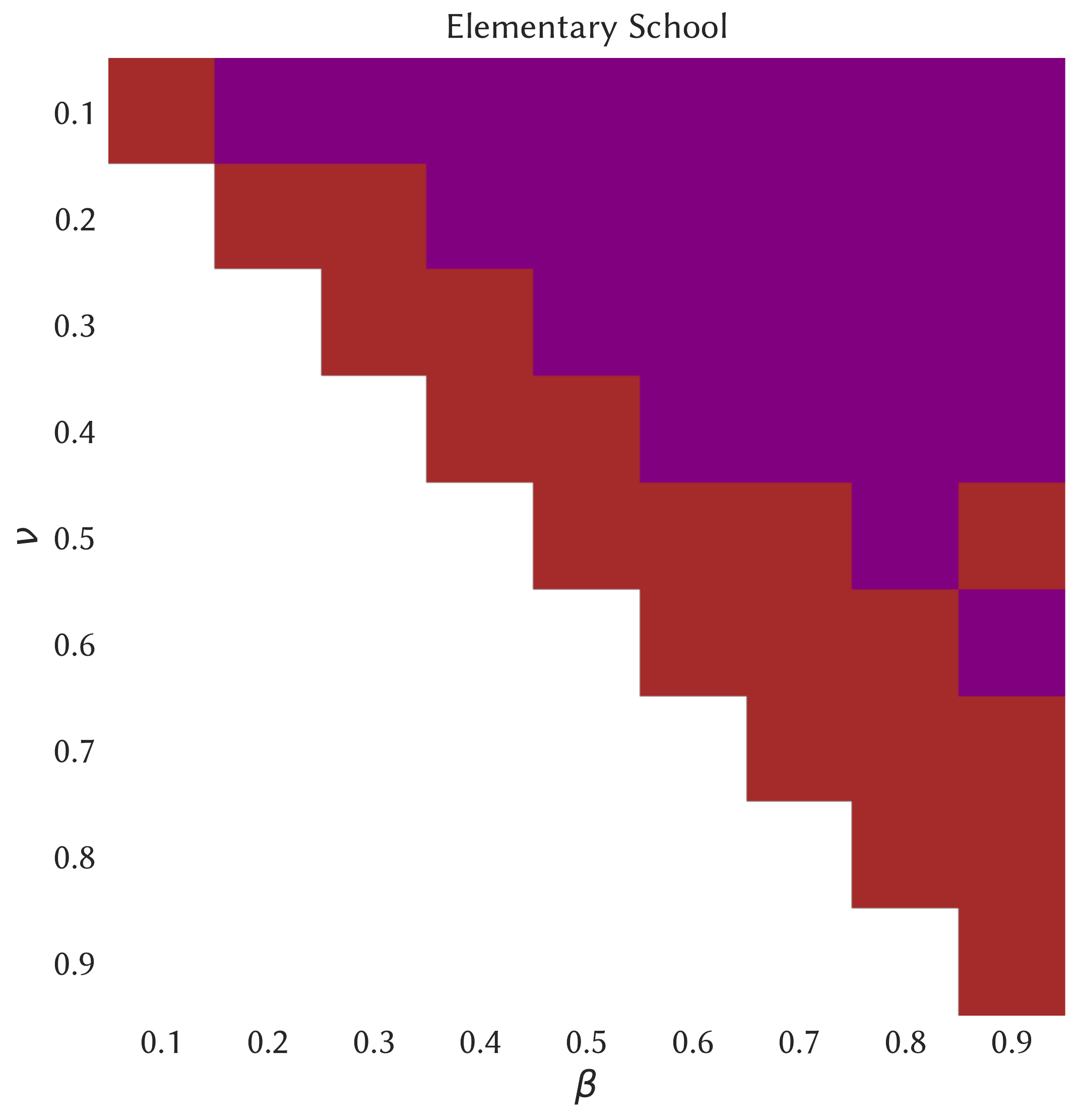} 
			\vspace{-1mm}\\
			\includegraphics[width=0.25\textwidth, height = .25\textheight, keepaspectratio]{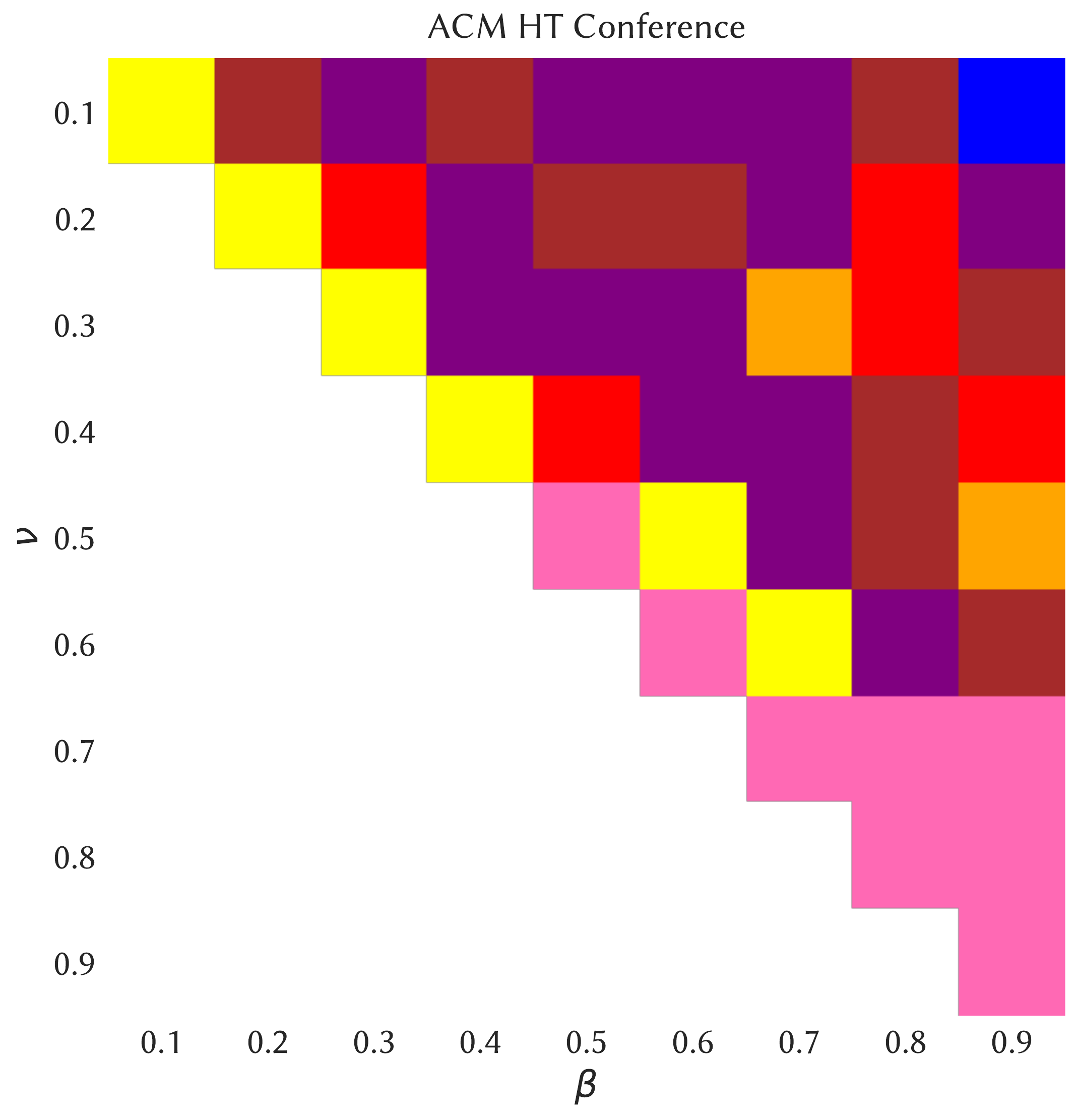} 
			& 
			\includegraphics[width=0.25\textwidth, height = .25\textheight, keepaspectratio]{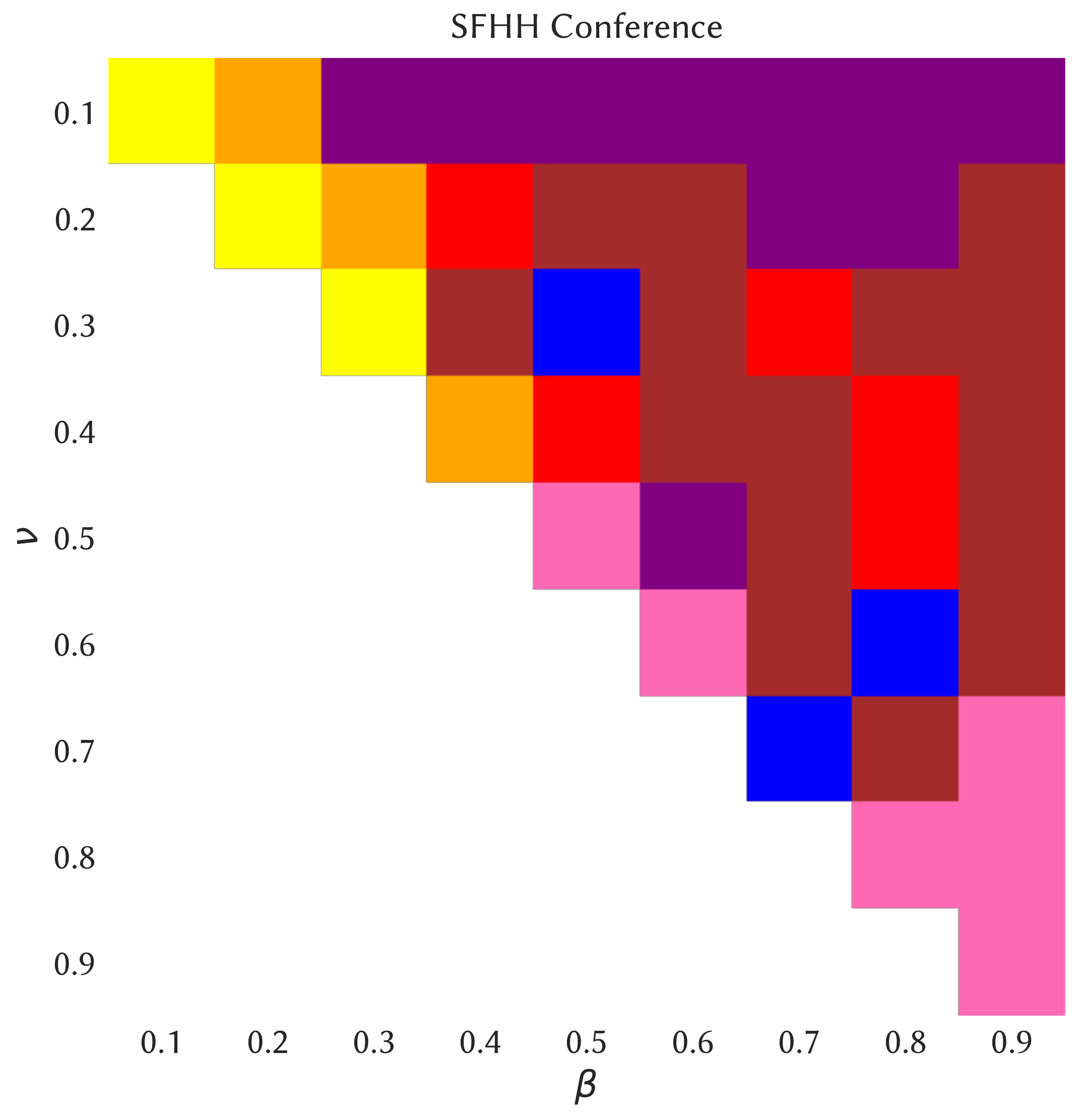} 
			\vspace{-1mm}\\
			\includegraphics[width=0.25\textwidth, height = .25\textheight, keepaspectratio]{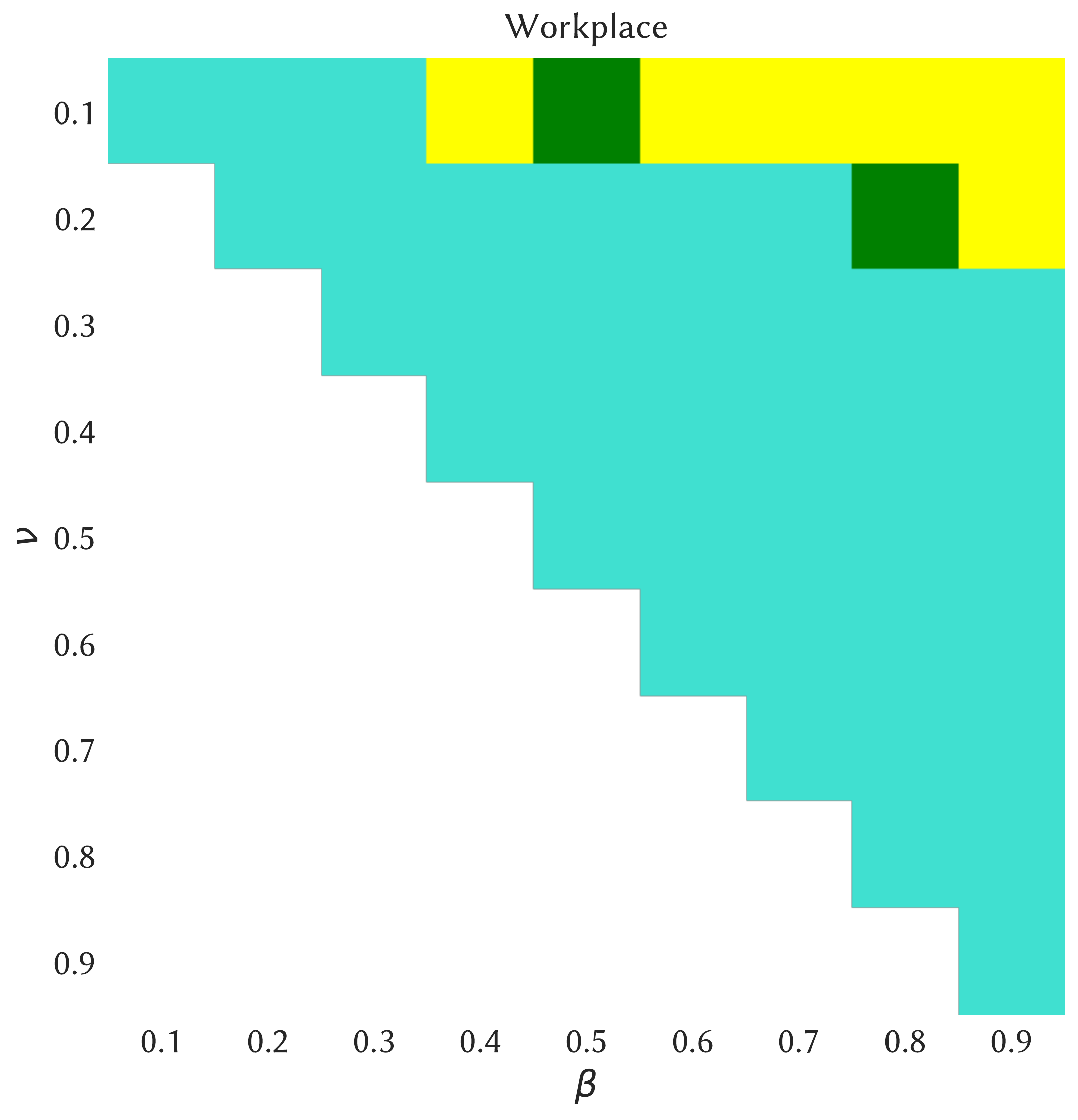} 
			&
		    \includegraphics[width=0.25\textwidth, height = .25\textheight, keepaspectratio]{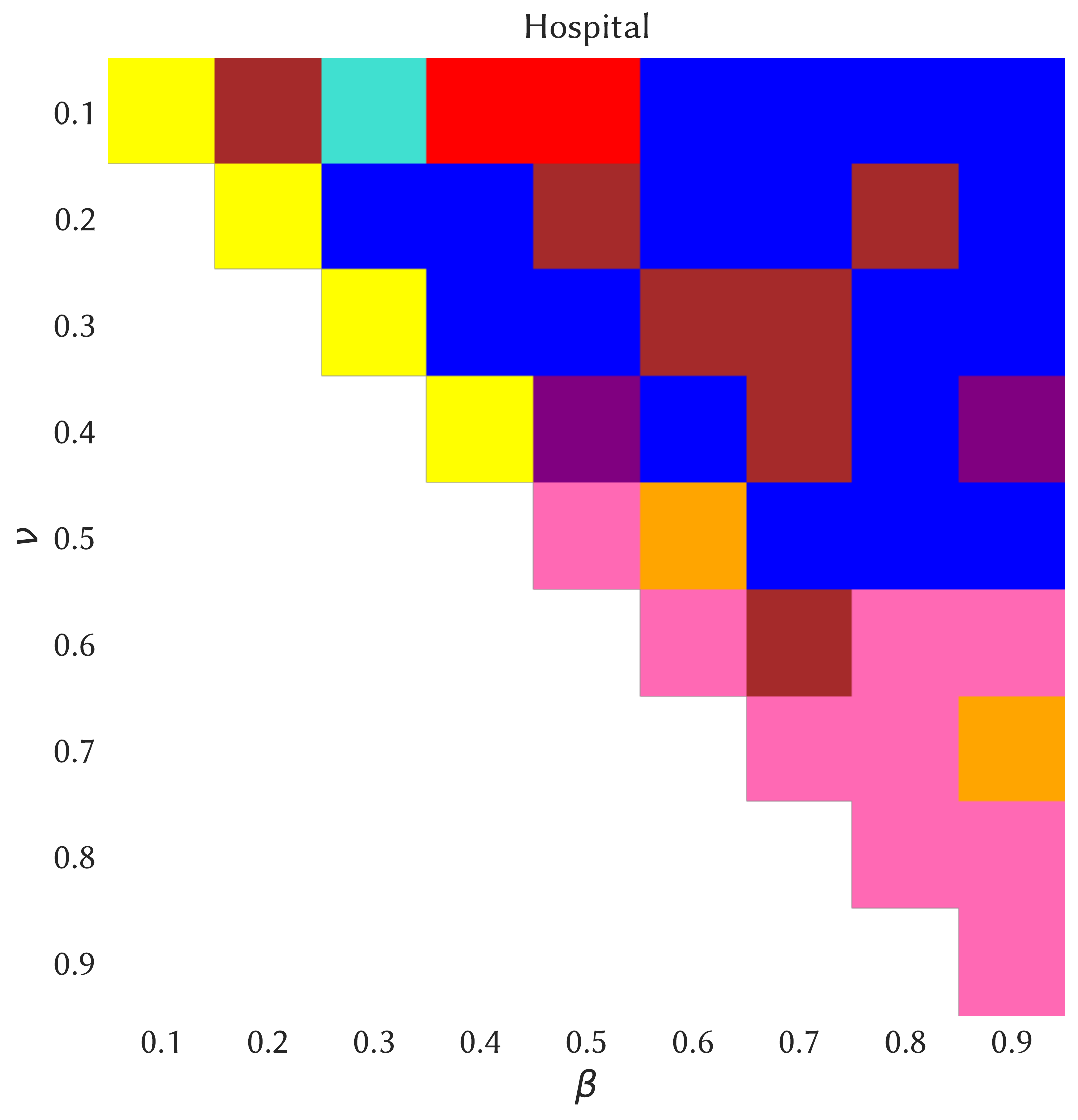} 
			\vspace{-2mm}\\
		\end{tabular}
	}
	\vspace{1mm}
	\caption{Heatmap indicating the intervention strategy leading to the lowest epidemic sizes, for each combination of spreading parameter values.
	Here $f=20\%$.
}
\end{figure}
 
\clearpage

\begin{figure}[h]
\centering
\begin{tabular}{c}
\centerline{\includegraphics[width=.5\columnwidth]{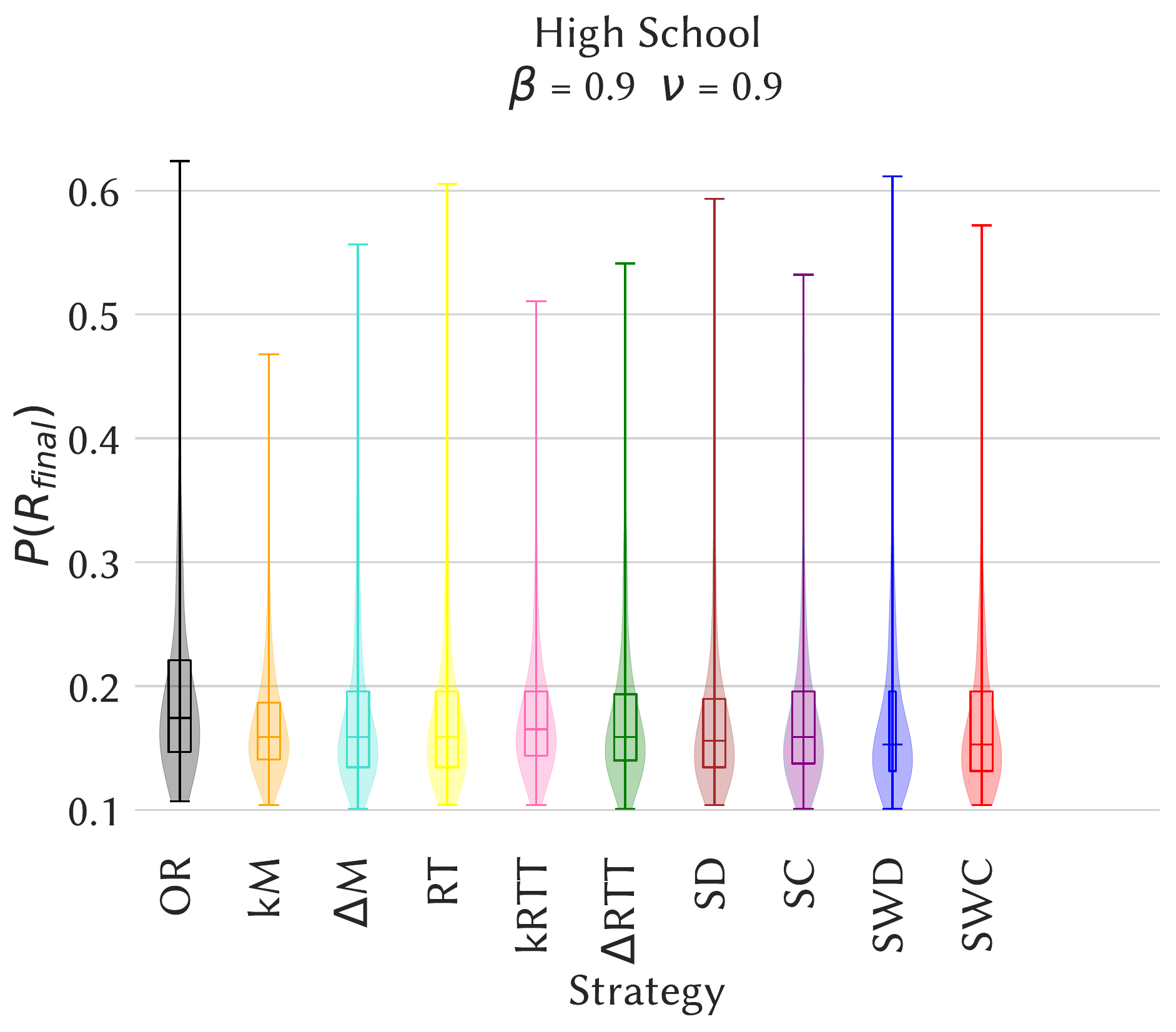}
\includegraphics[width=.5\columnwidth]{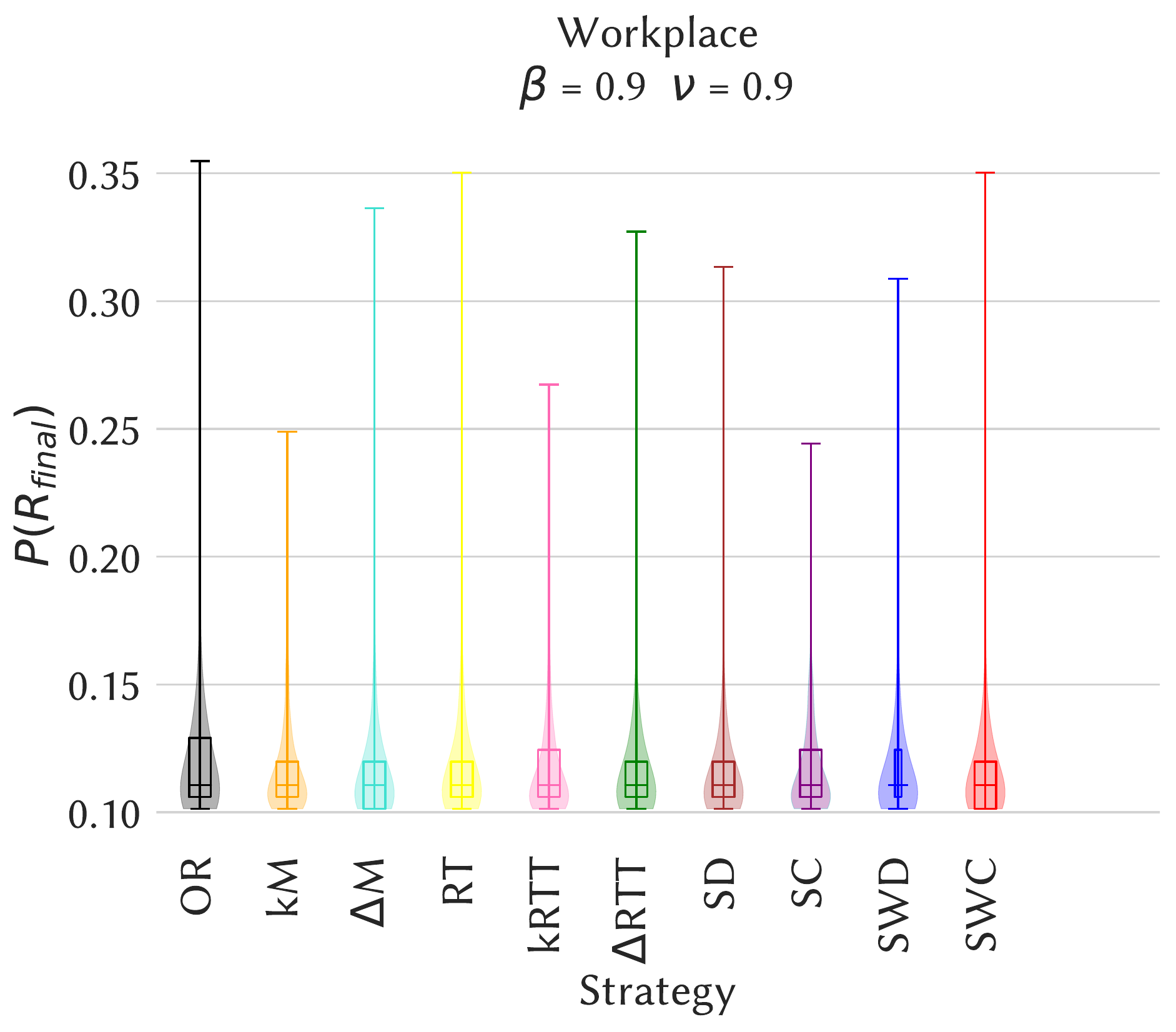}
} \vspace{-3mm}\\
\end{tabular}
\caption{Distributions of final epidemic sizes for some illustrative cases, for an SIR process on the original temporal network (OR)
and for the various mitigation strategies ($f=20\%$). Left column: High School. Right column: Workplace.
}
\label{fig:SIR_distributions}
\end{figure}

\clearpage
\newpage

 \section{Results of the seeding strategies for the SIR process, for all data sets (spread maximization)}

\begin{figure}[!h]
	\centerline{
		\begin{tabular}{cc}
			\includegraphics[width=0.3\textwidth, height = 0.3\textheight, keepaspectratio]{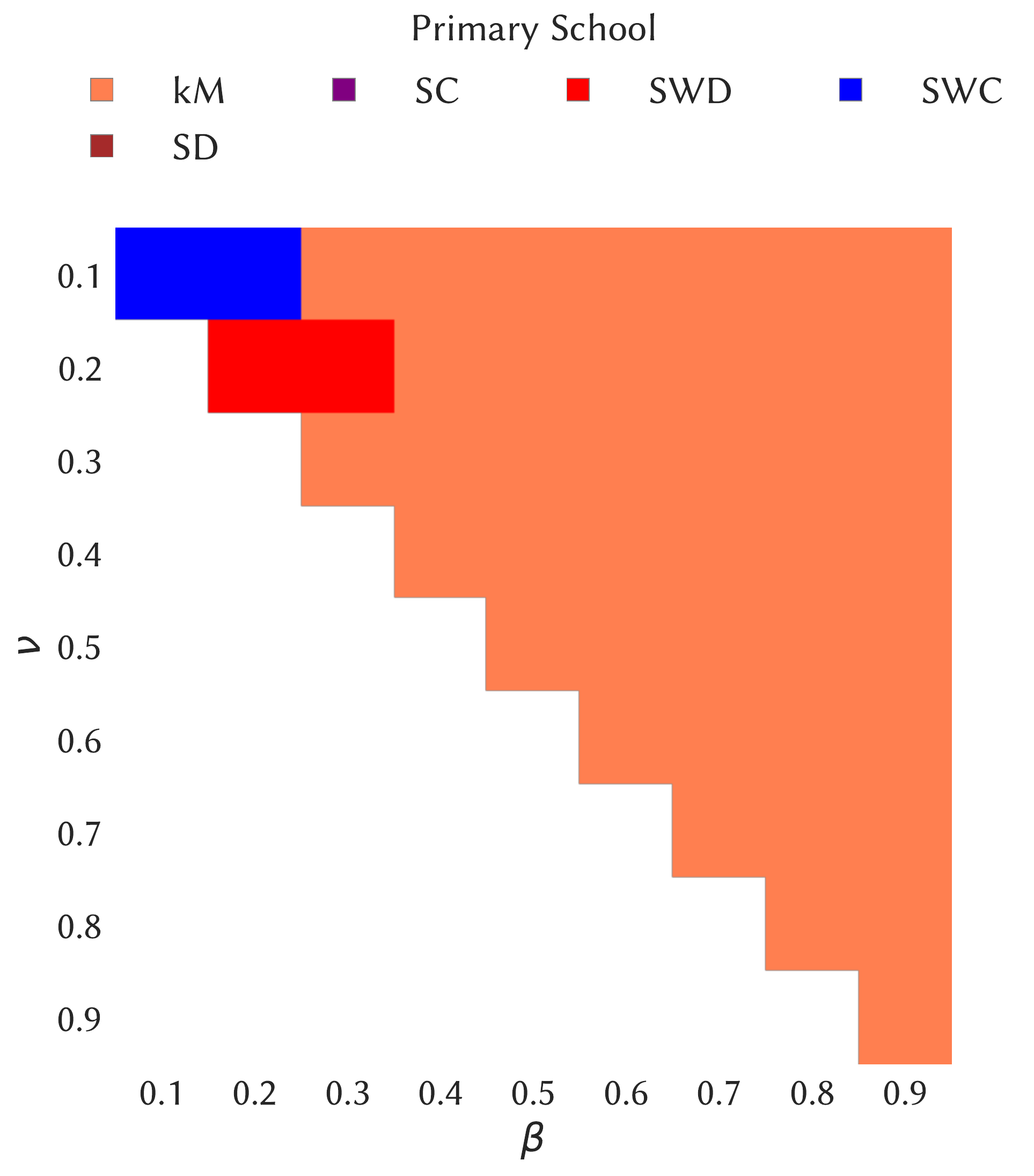} & 
			\includegraphics[width=0.3\textwidth, height = 0.3\textheight, keepaspectratio]{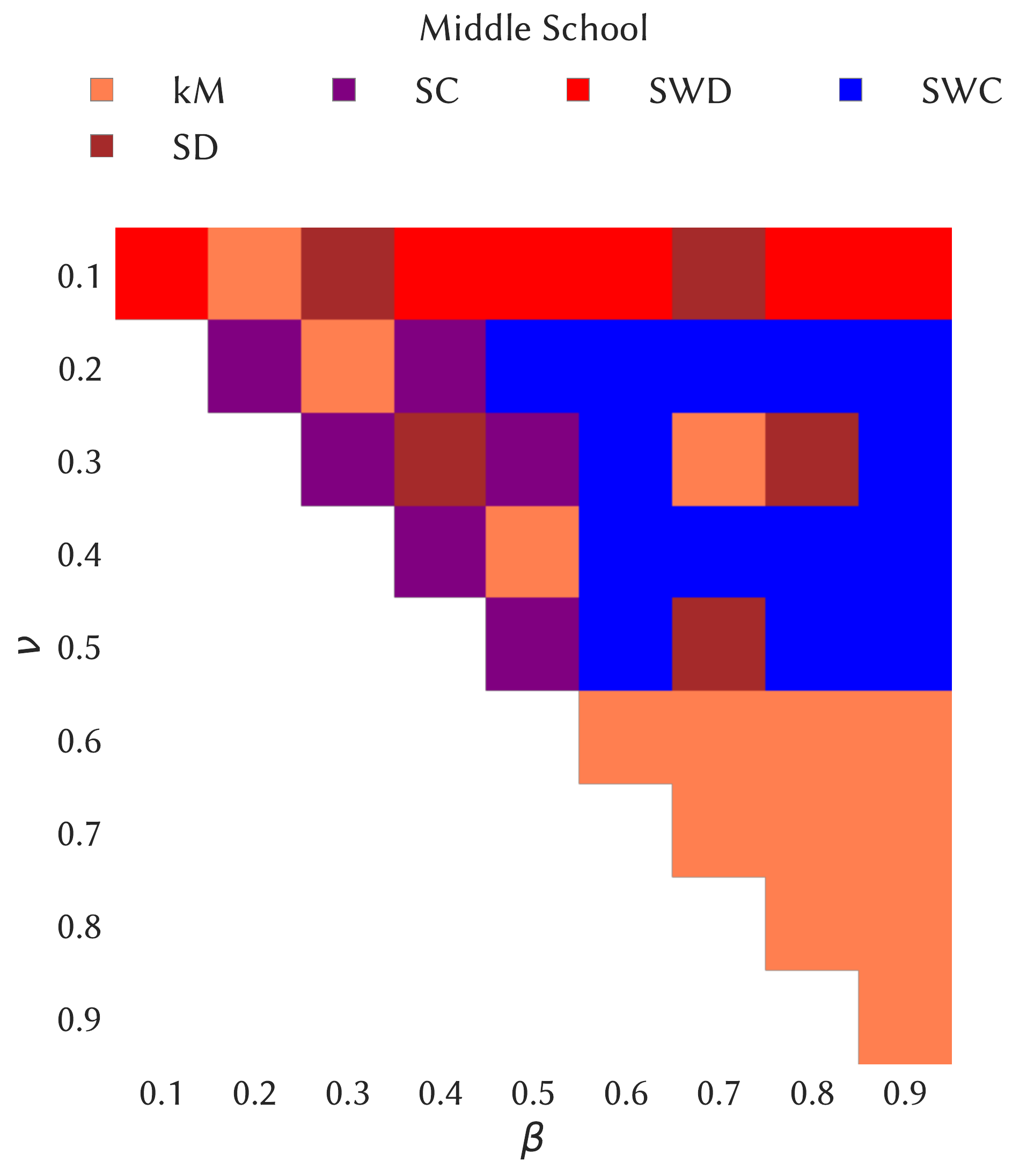} 
			\vspace{-1mm}\\
			\includegraphics[width=0.3\textwidth, height = 0.3\textheight, keepaspectratio]{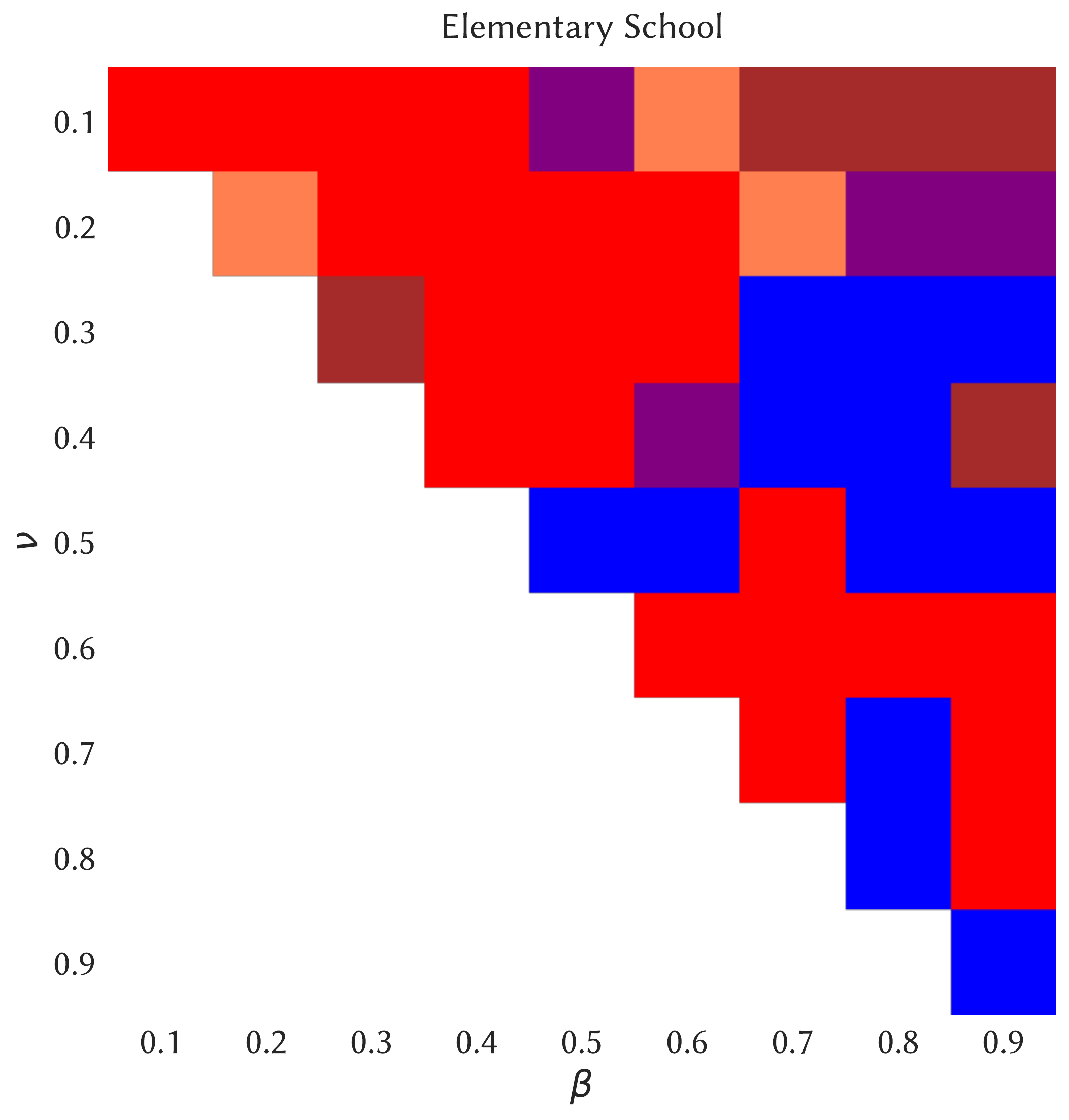} 
			& 
			\includegraphics[width=0.3\textwidth, height = 0.3\textheight, keepaspectratio]{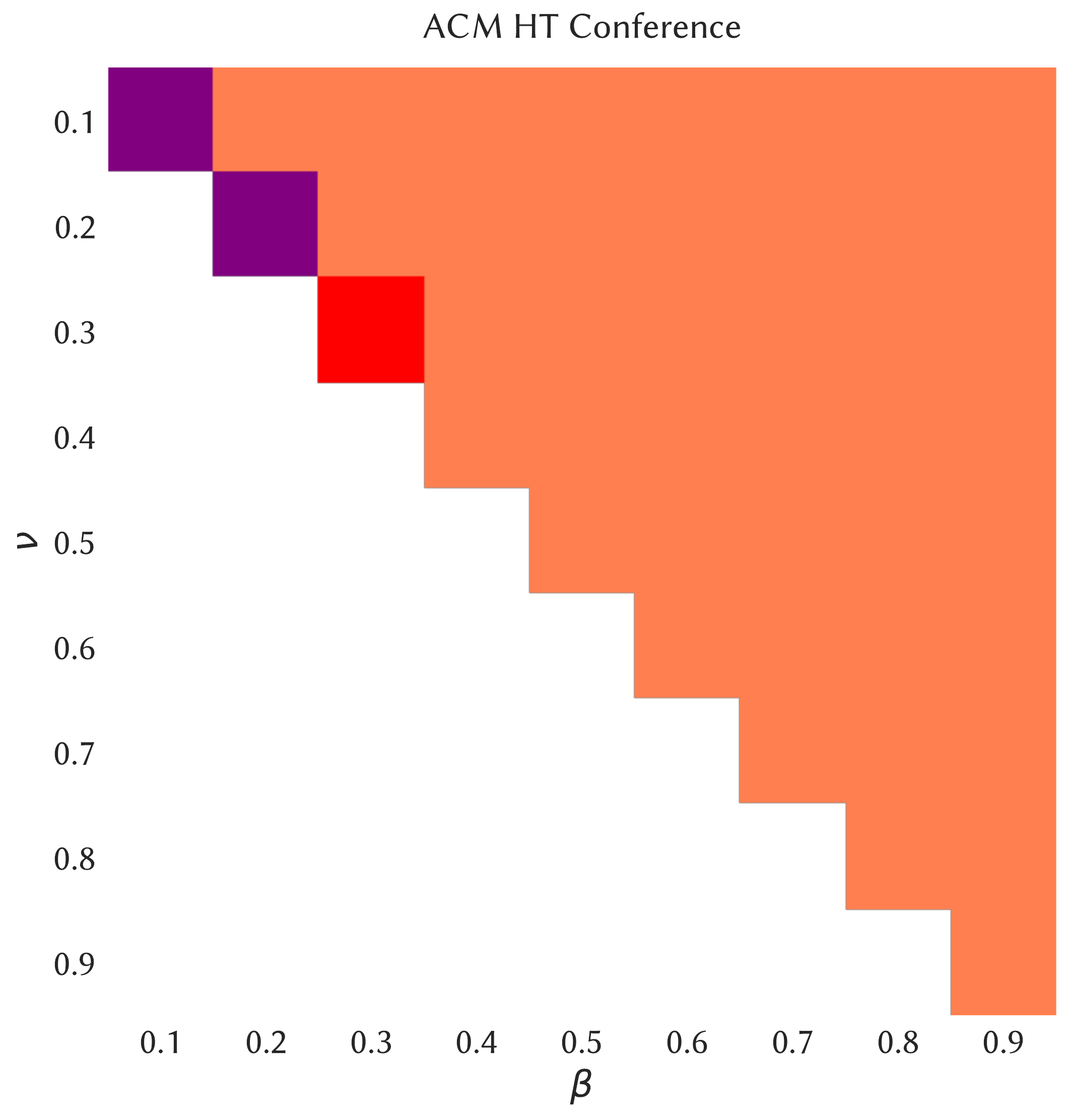} 
			\vspace{-1mm}\\
			\includegraphics[width=0.3\textwidth, height = 0.3\textheight, keepaspectratio]{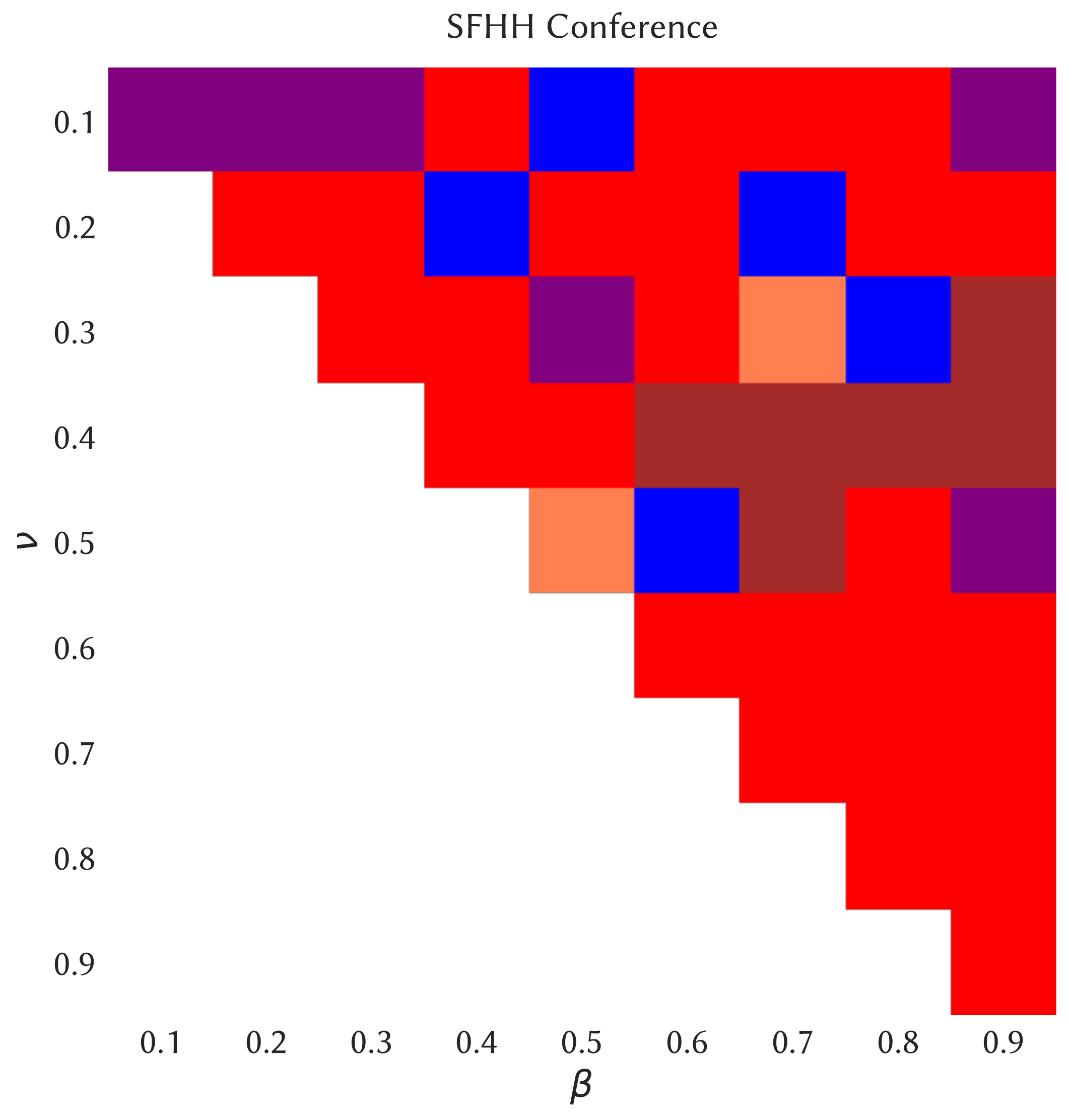} 
			&
		    \includegraphics[width=0.3\textwidth, height = 0.3\textheight, keepaspectratio]{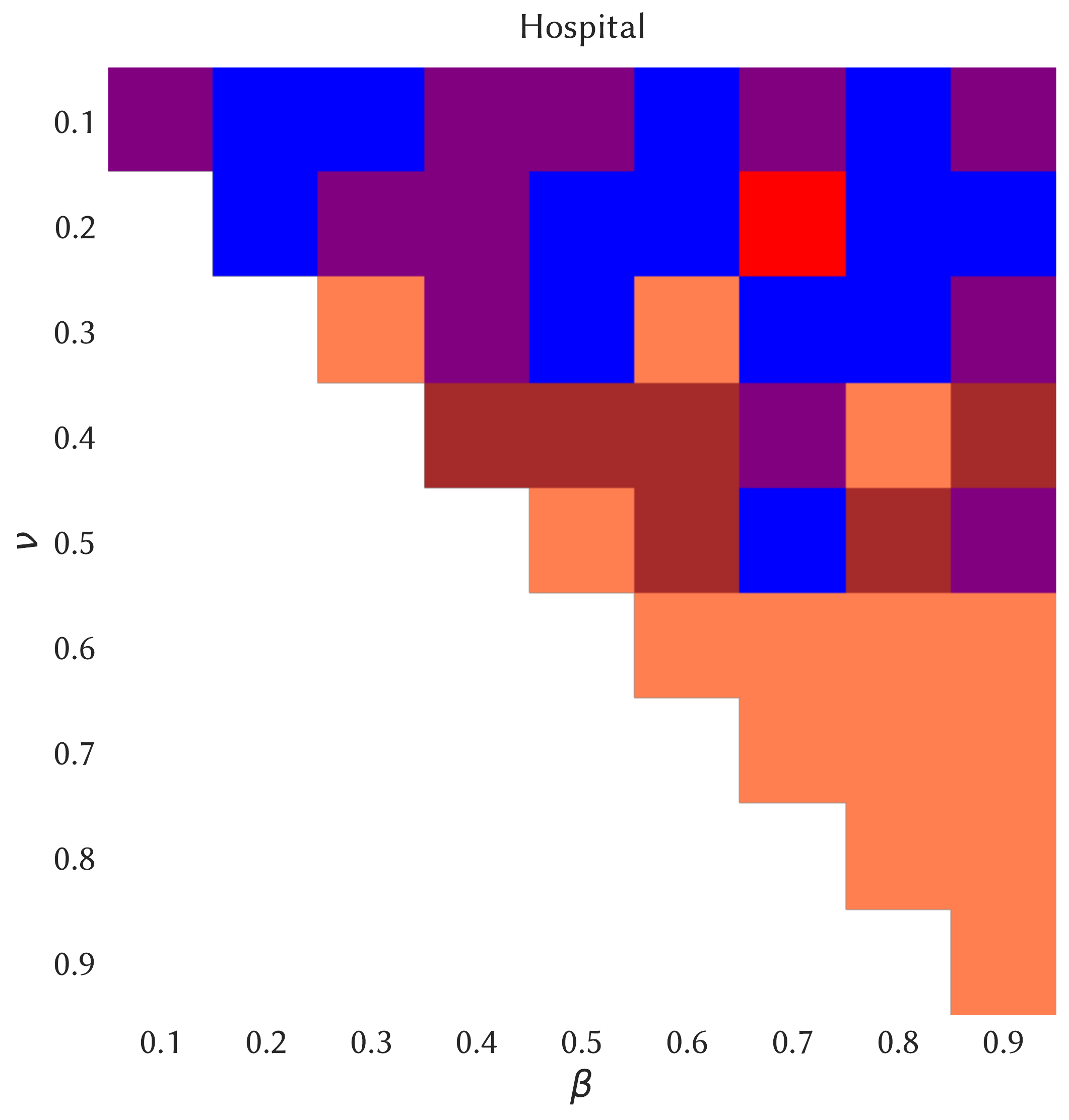} 
			\vspace{-2mm}\\
		\end{tabular}
	}
	\vspace{1mm}
	\caption{Heatmap indicating the intervention strategy leading to the largest ratio of final sizes, for each combination of 
	spreading parameter values. For the Elementary School {and SFHH Conference} data sets, no seeding strategy is consistently optimal, and we show in Figure 
	\ref{fig:SIR_distributions_elementary} that the distributions of the final epidemic sizes are in fact quite similar for the various strategies.
}
\end{figure}
 
 \clearpage

\begin{figure}[h]
\centering
\begin{tabular}{cc}
\includegraphics[width=.5\columnwidth, height = 0.3\textheight, keepaspectratio]{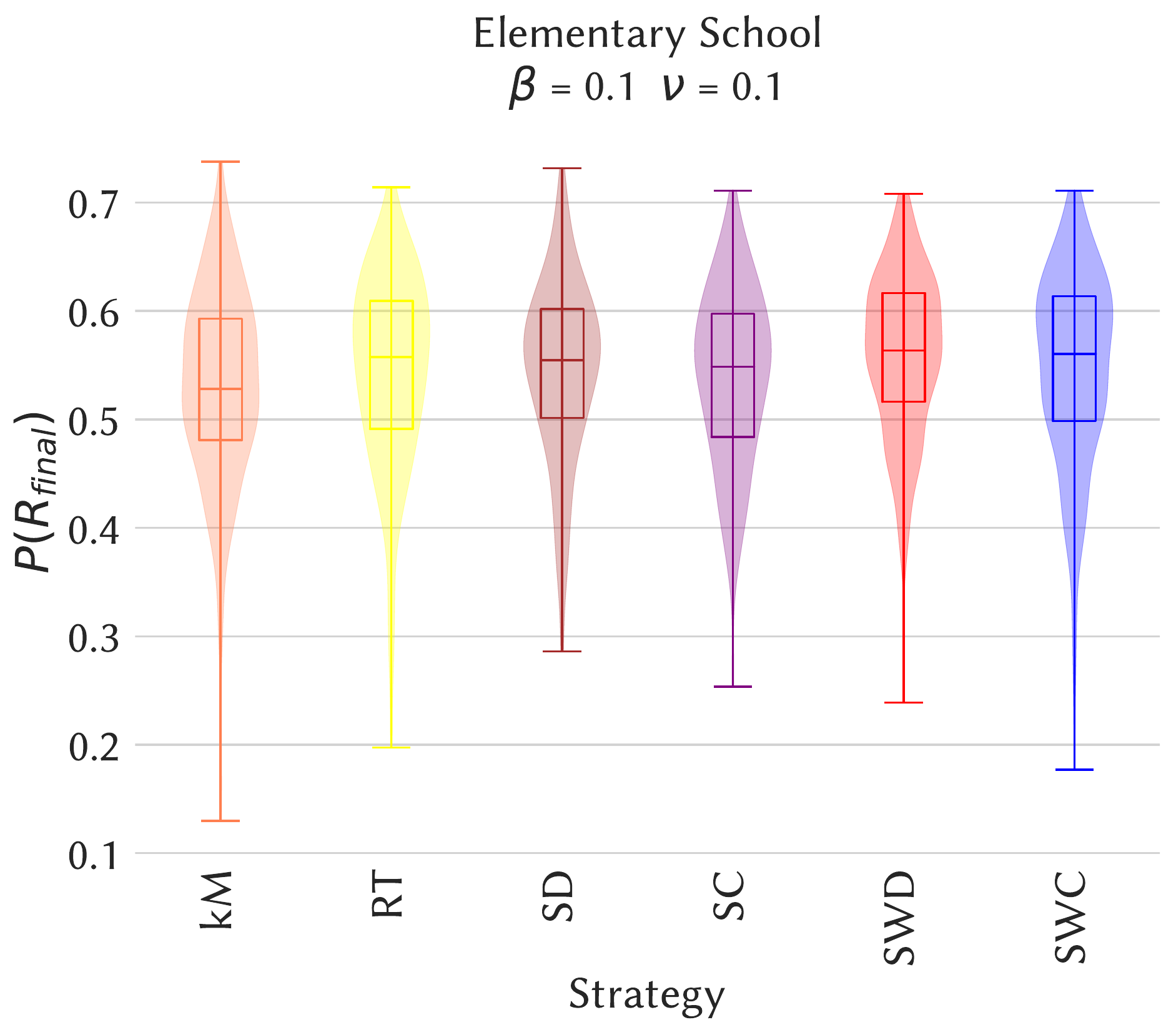} &
\includegraphics[width=.5\columnwidth, height = 0.3\textheight, keepaspectratio]{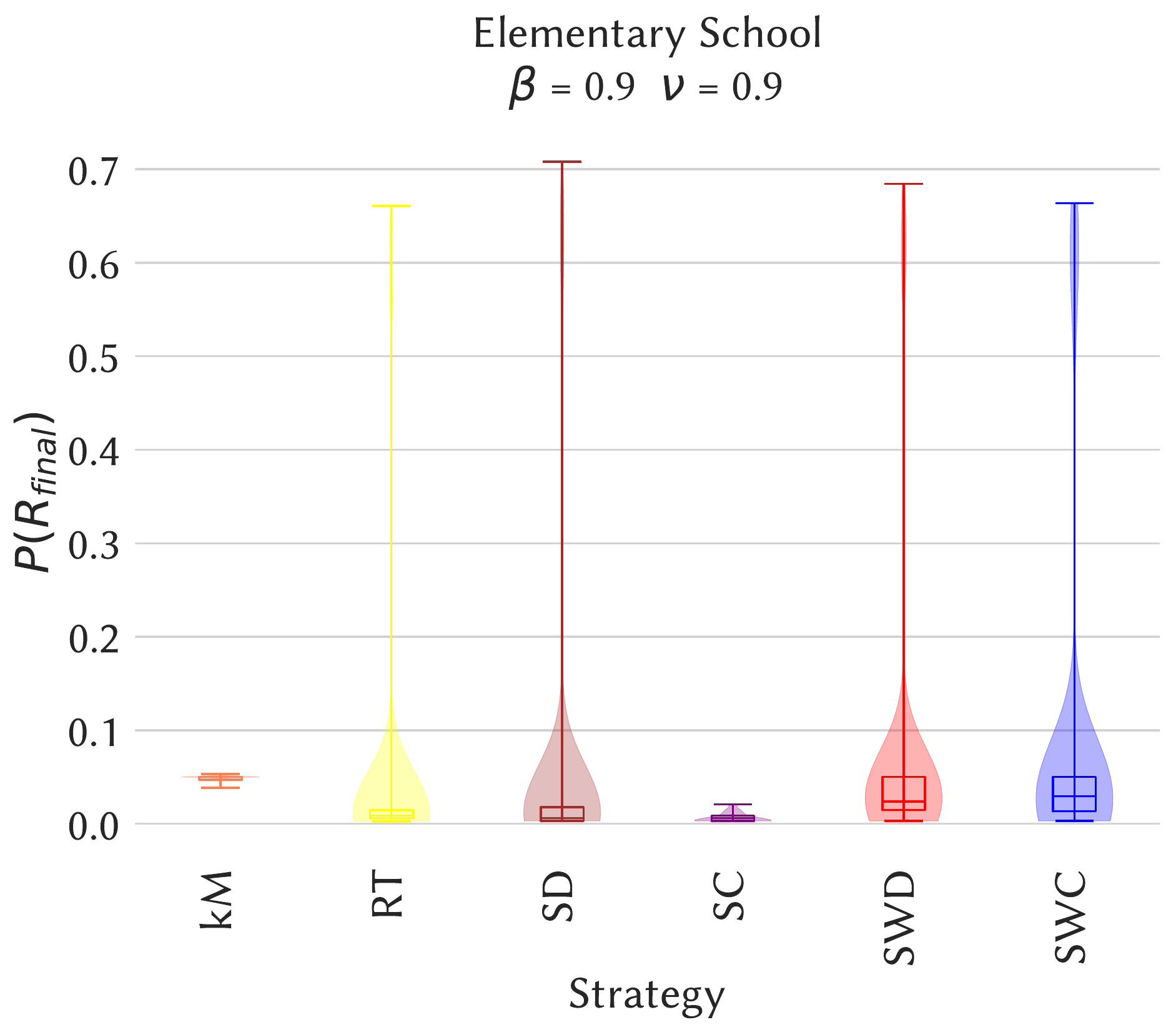} \vspace{10mm} \\ 
\includegraphics[width=.5\columnwidth, height = 0.3\textheight, keepaspectratio]{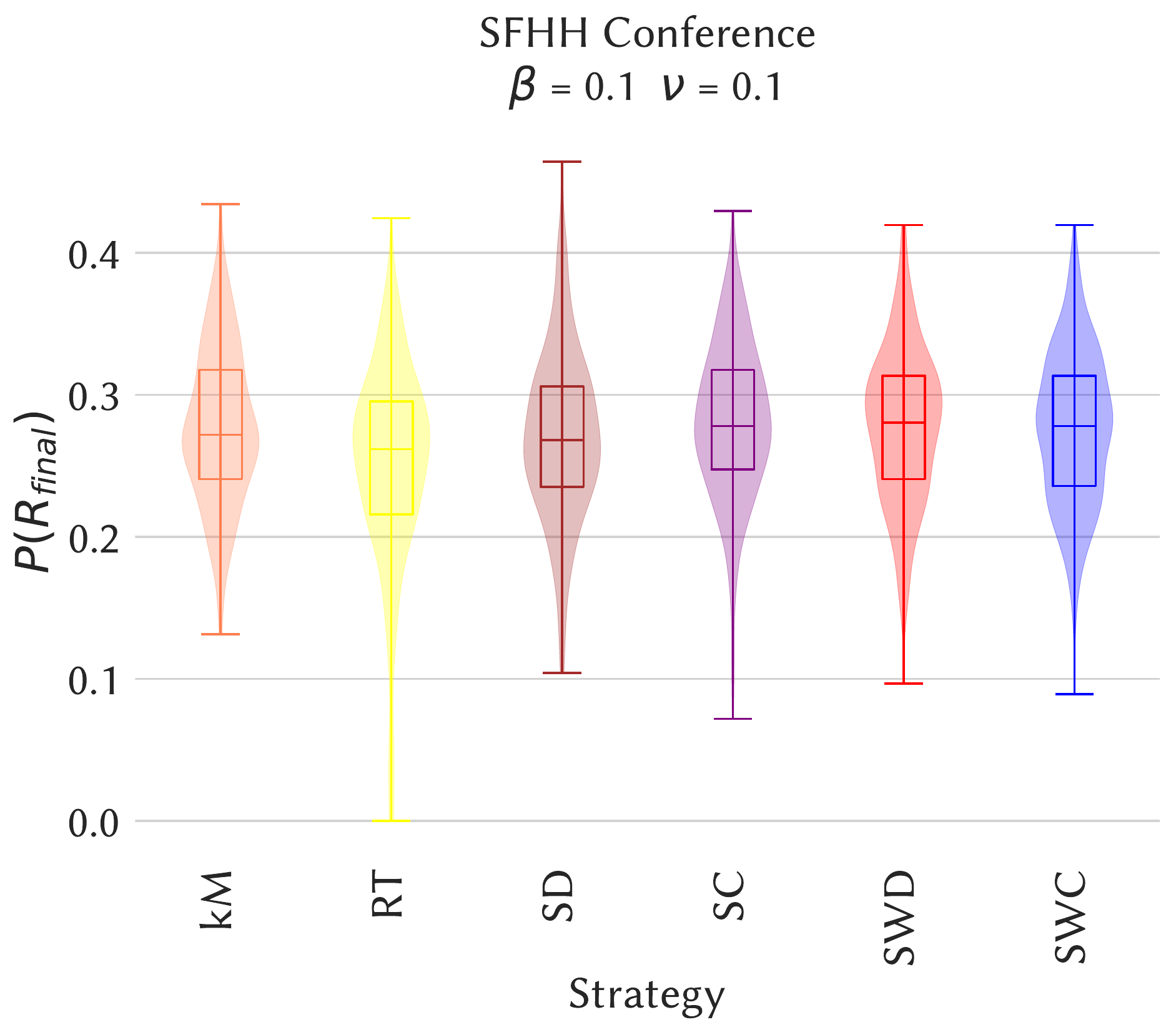} &
\includegraphics[width=.5\columnwidth, height = 0.3\textheight, keepaspectratio]{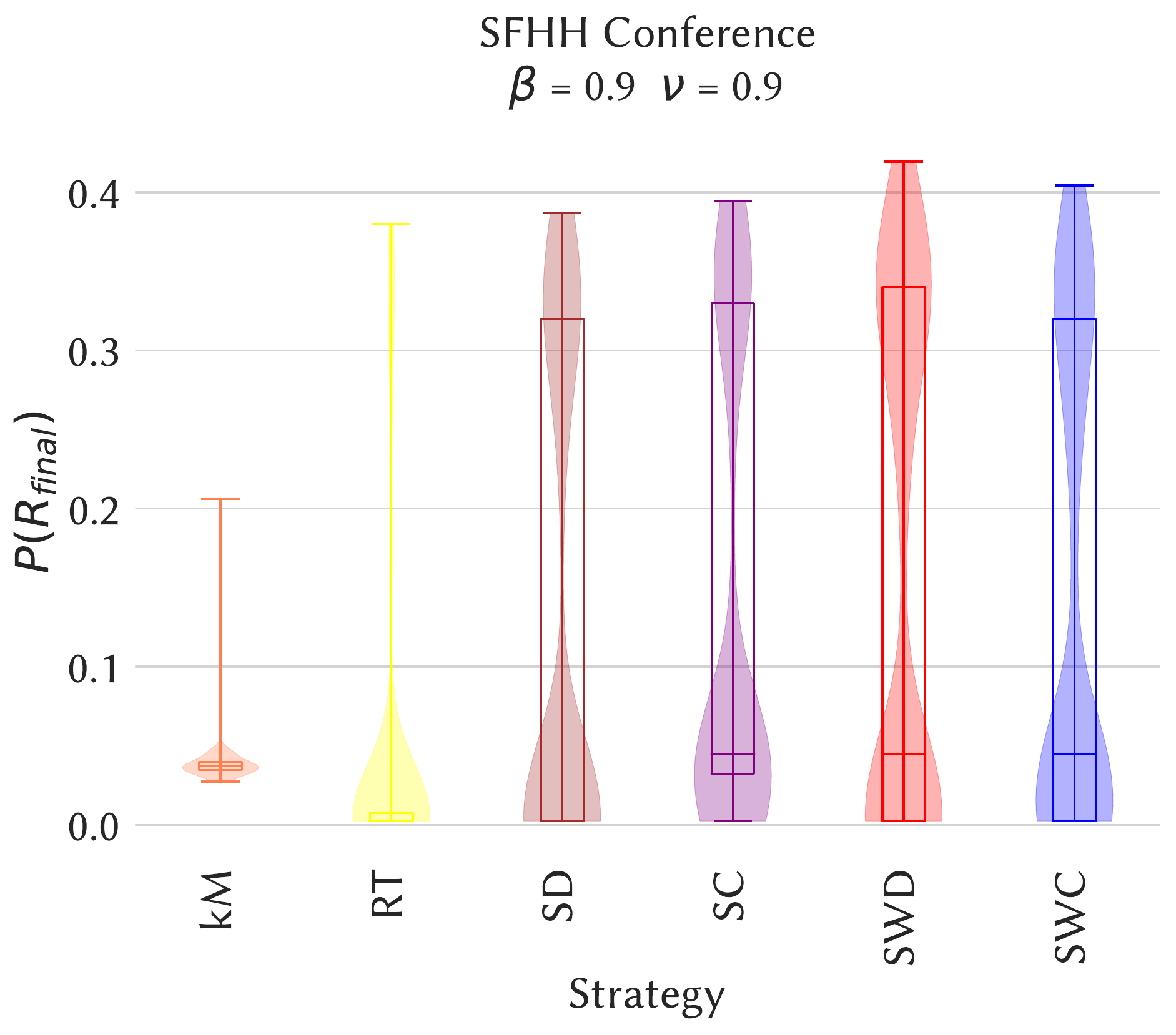}
\vspace{-3mm}\\
\end{tabular}
\caption{Distributions of final epidemic sizes in the Elementary School {and SFHH Conference} data sets for two illustrative cases of parameter values, for the
various seeding strategies considered for the SIR process.}
\label{fig:SIR_distributions_elementary}
\end{figure}

 \clearpage
 \newpage
 
\section{Effect of the order of the maximal span-cores in the SIR seeding}

\begin{table}[ht]
\begin{tabular} {*{12}{|c|}}
\hline
\multicolumn{2}{|c|}{}     & \multicolumn{2}{c|}{minimum order} & \multicolumn{2}{|c}{average order}  & \multicolumn{2}{|c|}{maximum order}\\ \hline
\multicolumn{2}{|c|}{Data set}  & ($n$,$|\Delta|$, $k$) &  $\frac{\langle R_{final} \rangle}{\langle R_{final} \rangle_{random}}$ & ($n$,$|\Delta|$, $k$) &  
$\frac{\langle R_{final} \rangle}{\langle R_{final} \rangle_{random}}$ & ($n$,$|\Delta|$, $k$) &  $\frac{\langle R_{final} \rangle}{\langle R_{final} \rangle_{random}}$\\ 
\hline \hline
\multicolumn{2}{|c|}{Primary school} & (2,14,1) &  1.39 & (10,1,3) &  5.08 & (10, 1, 7) &  33.35 \\ \hline
\multicolumn{2}{|c|}{High school}  & (4,15,1) &  1.01 & (7,1,3) &  1.1 &  (7, 1, 6 ) & 4.82  \\ \hline
\multicolumn{2}{|c|}{Middle school} & (8,10,1) & 2.23 & (11,1,3) &  4.12  & (11,1,10) & 5.17 \\ \hline
\multicolumn{2}{|c|}{Elementary school}  & (4,9,1) & 3.16 & (13,1,3) & 4.82 & (13,1,9) & 4.51  \\ \hline
\multicolumn{2}{|c|}{ACM HT Conference}  & (9,1,1) & 1.44 & (9,1,2) & 1.97 & (9,1,7) & 2.76 \\ \hline
\multicolumn{2}{|c|}{SFHH Conference}  & (4,1,1) & 1.62 & (10,2,3) & 5.56 & (10,2,9) & 5.03 \\ \hline
\multicolumn{2}{|c|}{Workplace}  & (4,1,1) & 1.38 & (8,1,2) & 1.74 & (8,1,7) & 3.46 \\ \hline
\multicolumn{2}{|c|}{Hospital}  & (4,1,1) & 1.33 & (7,2,2) & 2.39 & (7,1,6) & 3.52 \\ 
\hline
\end{tabular}
\caption{\label{tab:order_vs_spread}{Some properties (number of nodes $n$, duration $|\Delta|$, order $k$) of a maximal span-core of minimum, average and maximum order, and associated epidemic size ratio value $\frac{\langle R_{final} \rangle}{\langle R_{final} \rangle_{random}}$.}}
\end{table}

\begin{figure}[!h] \label{fig:th_90}
	\centerline{
		\begin{tabular}{cc}
			\includegraphics[width=0.3\textwidth, height = .3\textheight, keepaspectratio]{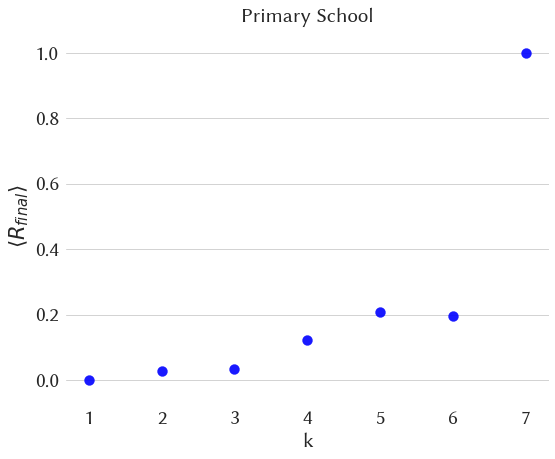} & 
			\includegraphics[width=.3\textwidth, height = .3\textheight, keepaspectratio]{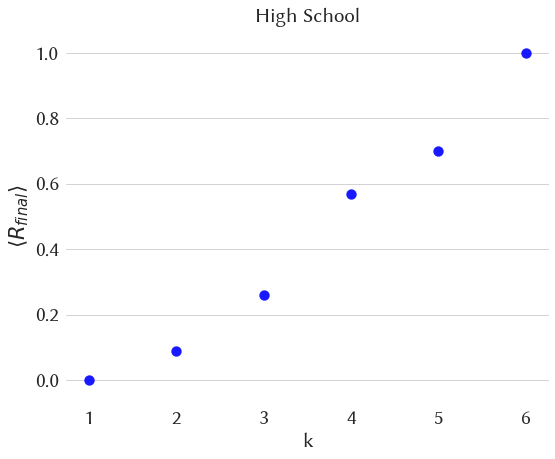} 
			\vspace{-1mm}\\
			\includegraphics[width=0.3\textwidth, height =0.3\textheight, keepaspectratio]{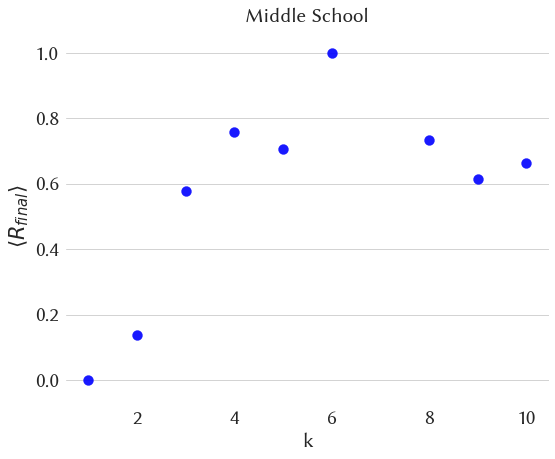} 
			&
			\includegraphics[width=0.3\textwidth, height = 0.3\textheight, keepaspectratio]{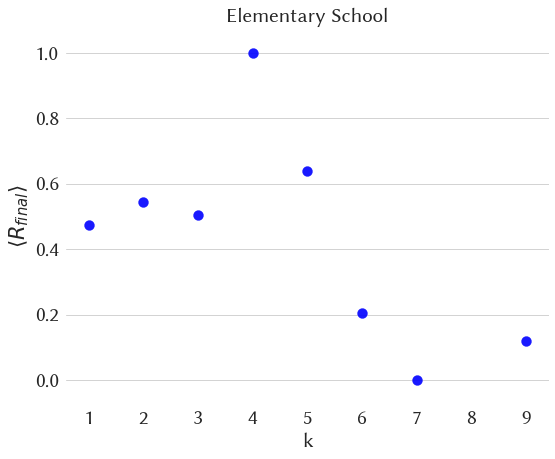} 
			\vspace{-1mm}\\
			\includegraphics[width=0.3\textwidth, height = 0.3\textheight, keepaspectratio]{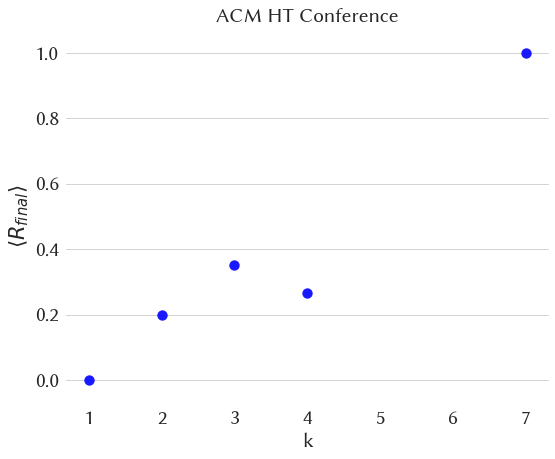} 
			&
			\includegraphics[width= 0.3\textwidth, height = 0.3\textheight, keepaspectratio]{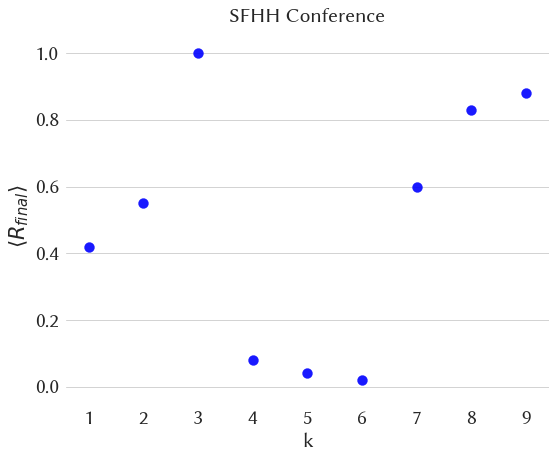}
			\vspace{-1mm}\\
			\includegraphics[width=0.3\textwidth, height = 0.3\textheight, keepaspectratio]{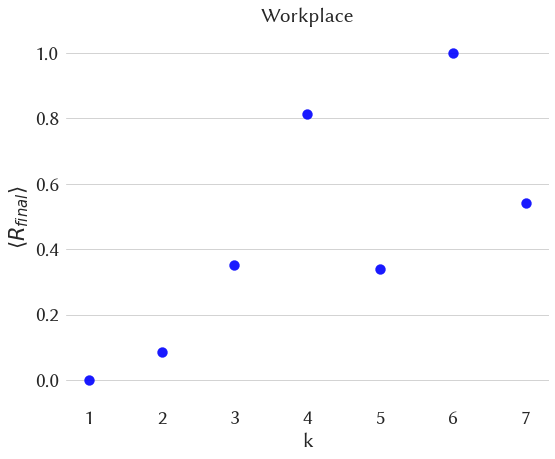} &
			\includegraphics[width=0.3\textwidth, height = 0.3\textheight, keepaspectratio]{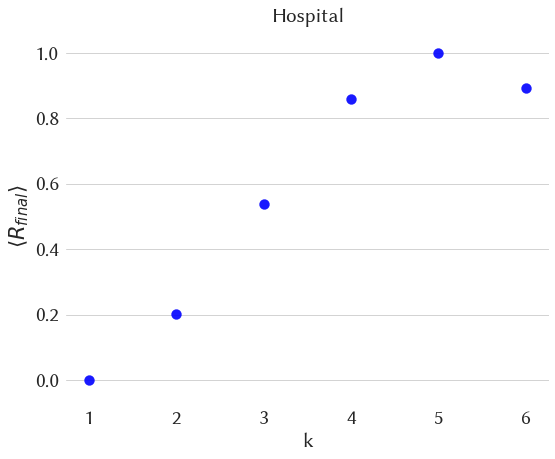} 
			\vspace{-2mm}\\
		\end{tabular}
	}
%	\vspace{1mm}
	\caption{\label{fig:order_vs_spread} { Effect of the order of the maximal span-cores in SIR processes for $\beta = 0.9$ and $\nu = 0.9$. In each panel we plot the average epidemic final size as a function of the order $k$ of the maximal span-core to which the seed belongs. 
   }}
\end{figure}

\end{document}